# Ultrafast hybrid nanocomposite scintillators: a review

V.S. Shevelev[1], A.V. Ishchenko[1*], A.S. Vanetsev[2], V. Nagirnyi[2], S.I. Omelkov[2]

[1]NANOTECH Center, Ural Federal University, Mira 19, Yekaterinburg 620002, Russia

[2]Institute of Physics, University of Tartu, W.Ostwaldi 1, Tartu 50411, Estonia

*Corresponding author. Email address: a-v-i@mail.ru (A.V. Ishchenko)

## Abstract

In recent years, demand for scintillation detectors with high time resolution (better than 100 ps) has emerged in high-energy physics and medical imaging applications. In particular, time of flight positron emission tomography (TOF-PET) can greatly benefit from increasing time resolution of scintillators, which leads to the increase of signal-to-noise ratio, decrease of patient dose, and achievement of the superior spatial resolution of PET images. Currently, extensive research of various types of materials is carried out to achieve the best time resolution. In this review, the recent progress of various approaches is summarized and scintillation compounds with the best temporal characteristics are first reviewed. The review presents the physical processes causing fast luminescence in inorganic and organic materials. Special attention is paid to nanocomposites which belong to a new perspective class of scintillating materials, consisting of a plastic matrix, inorganic nanocrystalline fillers, and organic or inorganic luminescence activators and shifters. The main features and functions of all parts of existing and prospective nanocomposite scintillators are also discussed. A number of currently created and investigated nanocomposite materials with various compounds and structures are reviewed.

**Content**

# 1 Introduction

Scintillation counting is one of the methods to detect ionizing radiation, based on an ability of some materials to convert absorbed energy into luminescence. Detectors based on scintillating materials (scintillators) are characterized by a low relative cost, simple production technology and high sensitivity, which makes them advantageous for various scientific and technological applications [1]. Currently, such detectors are used in medical diagnostics and therapy devices, high energy physics and other applications [2], [3], [4]. Rapid development of these application areas places new demands on scintillators. In particular, in many cases, the timing characteristics of scintillators have become crucial, especially when time-of-flight (TOF) radiation registration method is implemented. One of such applications is positron emission tomography (PET) and its modification TOF-PET.

Positron emission tomography allows to monitor the contribution of different molecular pathways to metabolic processes in a patient body, which makes this technique an invaluable medical imaging tool for cancer, neurodegenerative, cardiovascular and many other diseases. For diagnostics, a positron emitting radiopharmaceutical agent is injected into the body. The annihilation of the positron with an electron from a human body results in the formation of a pair of 511-keV photons simultaneously emitted in opposite directions. Typically, the detection system is designed as a ring of ionizing radiation detectors placed around a patient body [5]. All the emitted pairs are collimated along a common line called the line of response (LOR) (Figure 1). In TOF-PET, the detection of time difference between the two photons for each pair allows to localize the point on the LOR where the annihilation took place.

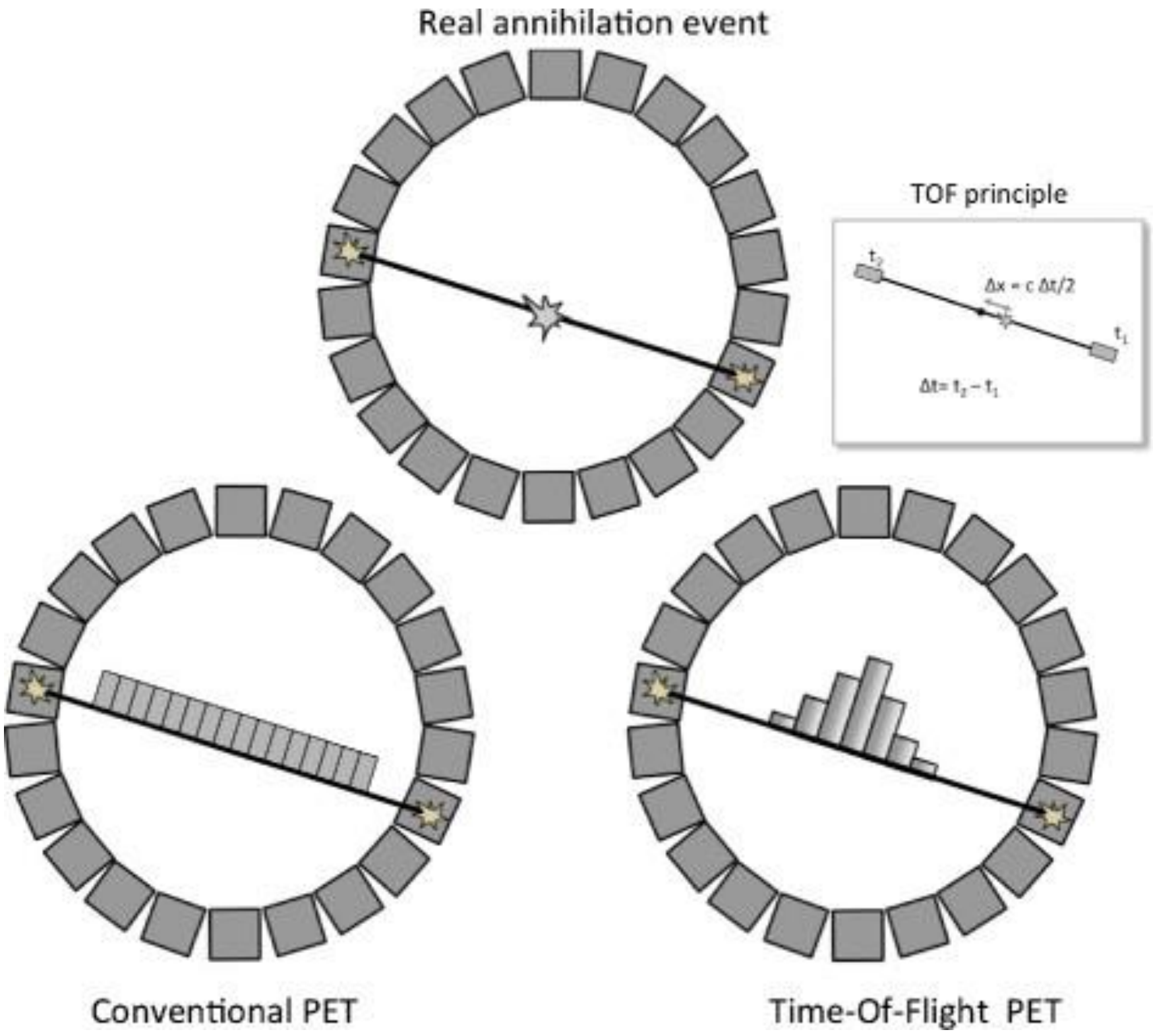

Figure 1. Main principle of the localization of a positron-electron annihilation point along the LOR by TOF-PET technique [6]. Publisher: Springer Nature, Date: Feb 16, 2016.

The accuracy of the time difference measurement is determined by the coincidence time resolution (CTR) of the detector. In modern commercially available PET scanners, the CTR value is about 310 – 400 ps [7], [8], and up to the date, the shortest CTR reported for an industrial device is 214 ps achieved with a Biograph Vision scanner [9]. It has been shown, that the attainment of the 10 ps CTR limit will allow to improve significantly the TOF-PET technique performance to bring the resolution of the mapping of radiopharmaceutical agent distribution to the millimeter scale, to improve significantly image reconstruction algorithms and to enhance the signal-to-noise ratio at least by a factor of 16 [10]. Currently, research teams from all over the world are trying to reach the 10 ps CTR. With the support of many leading scientific institutes, the “10 ps challenge” [11] has been launched with the goal to motivate the researchers to develop new fast scintillating materials. Another way to increase the sensitivity of a PET scanner is the extending of its axial length up to 2 meters for monitoring of the whole body. Such system is capable of recording 40 times more events than the state-of-art scanners [12].

Thus, in recent years, the research and development of scintillator based detectors with high time resolution is on the rise [13]. The development of scintillation materials which would satisfy new time resolution demands is an important element of this activity. The main goal of the present review is to provide an insight into the current state of modern scintillation materials and, in particular, one of the most perspective members of this family - hybrid nanocomposite scintillators. For this purpose, we have collected, analyzed and compared the latest data on timing and scintillation characteristics of the state-of-the-art nanocomposite scintillation materials.

In section 2, the main factors that determine the timing properties of a time-of-flight scanner are observed and illustrated by the example of TOF-PET. In addition, common scintillating materials used in modern medical devices and high-energy physics experiments are listed and their properties are examined.

The processes of absorption and conversion of ionizing radiation energy in inorganic and organic scintillators are discussed in section 3. The main advantages and shortcomings of scintillators based on inorganic and organic materials will be described.

In section 4, the general concept of hybrid nanocomposite materials will be presented together with the description of energy transport and conversion of electronic excitations between organic and inorganic components. The main constituents of nanocomposite materials and the primary requirements to their properties will be discussed. A comparative analysis of materials suitable for usage in fast hybrid nanocomposite scintillators will be conducted.

## 2 The principal features of modern detectors for PET applications

Let us consider the mechanism of the detection of ionizing radiation particles with high time resolution on the example of PET. All parts of the detection chain contribute to the time resolution of PET scanner: scintillation material, photodetector and analog readout circuits. A detailed description of the contribution of each factor and graphical representation of scintillator detection chain (Figure 2) is given below.

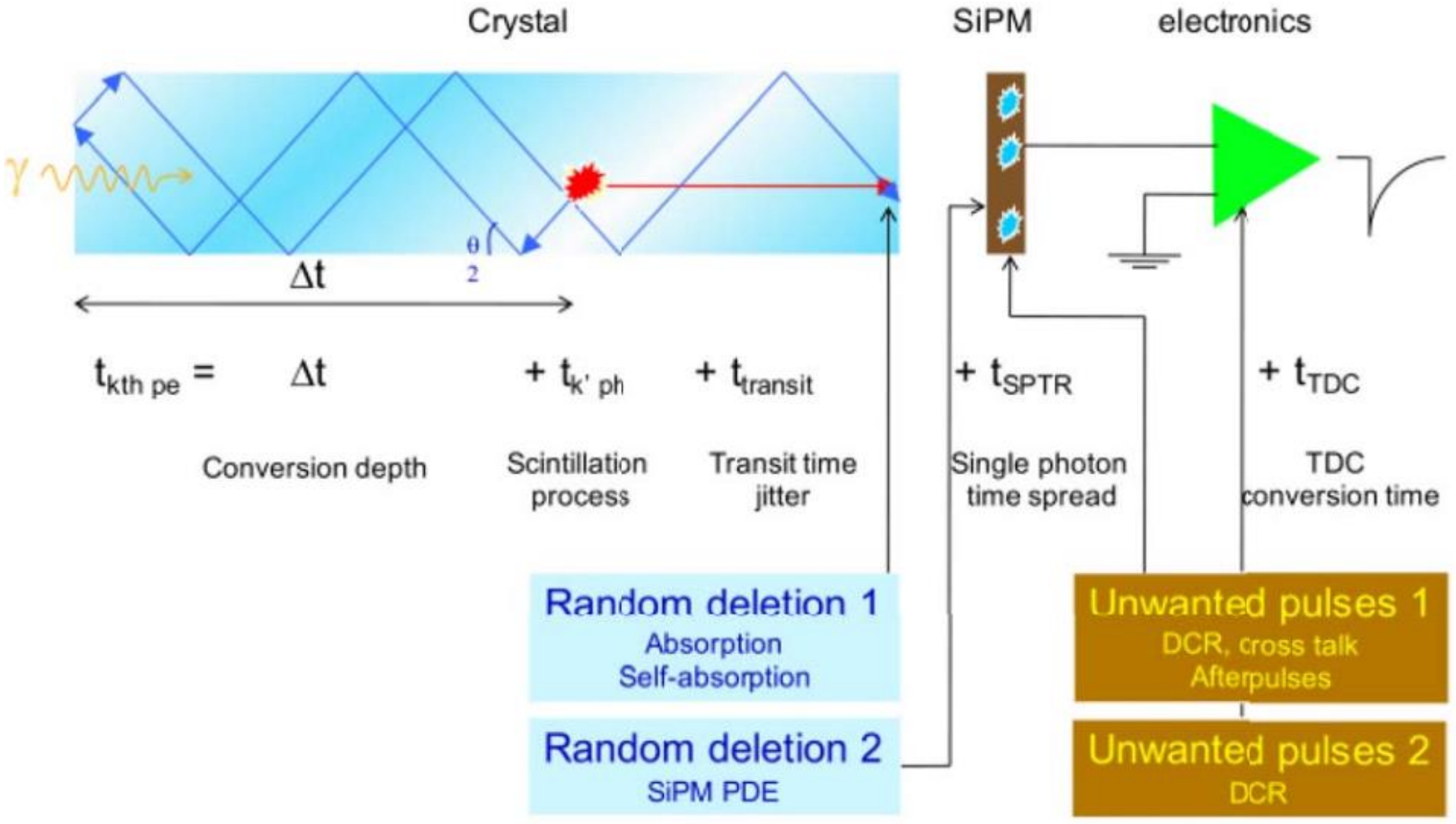

Figure 2. Scintillator detection chain showing the different contributions to the time jitter at the level of the γ conversion, scintillation mechanism, light transport in the crystal, light photoconversion, and readout electronics [10].

One of the modern solid-state photodetectors, SiPM (Figure 2), comprises an array of single photon avalanche diodes (SPAD) connected in parallel and exhibits a very short CTR. Usually, it is coupled with fast electronic readout circuits with a high bandwidth. The main parameter for photodetectors that indicates their time resolution is single photon time resolution (SPTR). SPTR for modern SiPMs typically ranges from 70 ps to 135 ps. In PET configuration for a detector pair, it is convenient to use the generalized parameter CTR which describes the timing properties of the whole detector system (SPTR, time resolution of the fast electronic readout circuits and etc.). The CTR is defined as FWHM (full-width at half maximum) of coincidence time distribution between two detectors $P_{coin}(t)$, which is a cross-correlation of each single detector timing spectrum [14]:

$$P_{coin}(t) = \int_{-\infty}^{\infty} p_{start}(t) p_{stop}(t+\tau) d\tau, \quad (1)$$

where $p_{start}(t)$ and $p_{stop}(t)$ are the timing probability densities for the start and stop detector, respectively. The best CTR value of 58 ps for a PET-compatible inorganic scintillation detection chain has been reported for a SiPM coupled with a single crystal of LSO:Ce,Ca [15].

Besides the SPTR, which quantifies the contribution of SiPM, the time parameters of scintillation materials themselves have a strong impact on the time resolution of the scintillation detection chain (Figure 2). For example, in the case of scintillator with fast luminescence, the contribution of rise time $\tau_r$ should be included in calculation along with the decay time $\tau_d$ [16]. For LSO-based bulk crystals with $\tau_r \ll \tau_d$ in the first approximation and when considering the crystal only, the CTR can be described as follows [10]:

$$CTR \sim \sqrt{\frac{\tau_r \tau_d}{N_{phe}}}, \quad (2)$$

where $N_{phe}$ is a number of photoelectrons detected by the photodetector. Within the above reasoning, it is proposed to take a fresh look at the main characteristics of the known scintillators summarized in Table 1. It shows clearly how difficult is the choice of a scintillation material for a fast timing application.

Table 1. The main characteristics of scintillation materials for fast timing applications. In case multiple decay components are manifested in the same material, a contribution of each component to the total light yield is indicated.

| Scintillator | Rise time, ps | Decay time, ns (contribution, %) | Light yield, ph/MeV | Density, g/cm$^3$ | $Z_{eff}$ |
|---|---|---|---|---|---|
| Inorganic scintillators | | | | | |
| NaI:Tl | 28200±100 [17] | 250 [18] | 37700 [19] | 3.67 [18] | 50.8 [20] |
| BGO | 30±30 [21] | 300 (90%), 60 (10%) [22]; 5.8 (1%), 28 (4%) + long [21] | 8200 [19] | 7.13 [18] | 75.2 [20] |
| LSO:Ce | 30±30 [21] | 40 [20] | 27000 [20] | 7.4 [23] | 66 [24] |
| LYSO:Ce | 68±20 [15] | 21.5 (13%), 43.8 (87%) [15] | 33800 [25] | 7.1 [23] | 65 [23] |
| $LaBr_3$:Ce | 200 (30% $Ce^{3+}$) [26] | 18 (91%), 2.5 (4%), 55 (6%) (30% Ce) [26] | 61000 [27] | 5.29 [28] | 46.9 [29] |
| $BaF_2$ (fast component) | - | 0.6 [30] | 2000 [30] | 4.88 [30] | 54 [31] |
| CsI, undoped | 9±9 [15] | 0.967 (11.5%), 5.78 (30.2%), 36.3 (58.3%) [15] | 16800 [19] | 4.51 [20] | 54 [20] |
| Organic scintillators | | | | | |
| BC-418 | 348 [15] | 1.2 [15] | 9100 [15] | 1.032 [32] | - |
| BC-422 | 32 [15] | 1.3 [15] | 7500 [15] | 1.032 [32] | - |
| EJ-232 | 350 [33] | 1.6 [33] | 8400 [33] | 1.023 [32] | - |
| EJ-232Q | 110 [33] | 0.7 [33] | 2900 [33] | 1.023 [32] | - |

Nowadays, commonly used scintillators are based on inorganic single crystals. This type of materials is characterized by high effective atomic number value ($Z_{eff}$) and high light yield (LY). One of the first scintillators used in a PET scanner is a single crystal of NaI:Tl. Despite the long rise and decay times of its luminescence (30 ns and 290 ns, respectively), this material is still used in applications requiring high light yield. However, for the applications requiring high time resolution it was replaced with much faster scintillators. New bulk scintillators with more suitable properties were developed, *e.g.*, BGO ($Bi_4Ge_3O_{12}$) [18], $BaF_2$ [30], $LaBr_3$:Ce [28]. The development of new scintillating materials continues up to now. LSO ($Lu_2SiO_5$) and LYSO ($Lu_{2(1-x)}Y_{2x}SiO_5$) with luminescence rise time of 30 and 68 ps, respectively, have been recognized as the most promising materials. Moreover, it has been shown that LSO:Ce co-doped with 0.4% Ca and LGSO:Ce ($Lu_{2x}Gd_{2-2x}SiO_5$:Ce) have shorter rise times amounting to 20 and 36 ps, respectively [34].

Organic compounds, particularly in the form of plastic scintillators, have also found their application besides inorganic single crystals [15], [33], [35]. Organic materials have a number of advantages compared with inorganic crystals. Plastic scintillators are characterized by very fast nanosecond and in some cases subnanosecond decay time and relatively high LY (Table 1, BC-422, BC-418, EJ-232Q). Recently, the CTR value of 35 ps has been reported for BC-422 combined with high frequency electronic readout and photodetectors [36]. In addition, these materials are much simpler in production than single crystals and can be obtained in large sizes and different shapes. There are several different techniques of plastic scintillator production: bulk polymerization, molding, injection molding, and extrusion [2]. This type of materials consists of several functional components. The polymeric matrix is a binding medium and commonly exhibits a low LY. For more efficient scintillation, the polymeric host is doped with an organic phosphor - a luminescent activator. The activator converts bulk electronic excitations – conduction band electrons, valence holes and excitons – into luminescence. In many cases, an additional phosphor, a wavelength shifter, is added. The role of wavelength shifter is to convert the shorter-wavelength luminescence of an activator into a longer-wavelength emission. It mitigates a relatively small Stokes shift of an activator, thus preventing luminescence reabsorption, as well as serves for better overlapping of the scintillator emission spectrum with the sensitivity range of a photodetector. Plastic scintillator is commonly manufactured in a form of strips with various cross sections and shapes or tiles with a thickness of a few millimeters and areas up to a few thousands of square meters [2]. Thanks to its unique properties, this type of scintillation materials has found wide application in high energy physics [37]. The detectors based on polystyrene stripes with 2,5-diphenyloxazole (PPO) as an activator and 1,4-bis(5-phenyl-2-oxazolyl)benzene (POPOP) as a wavelength shifter are used for neutrino investigation experiments such as MINOS [38] and MINERvA [39] at Fermilab, and OPERA [40] at CERN. Plastic scintillators are also used in preshower detectors, trigger systems, TOF measurements in the DO [41] and CDF-II [42] experiments and also in the calorimeters for the ATLAS [43] and CMS [44] experiments as well as the LHCb spectrometer [45] in CERN.

There is also an active development of a J-PET scanner based on a plastic scintillator [46]. The estimated possible CTR value of such system is 220 ps. The main drawback of plastic scintillators is a low density and low $Z_{eff}$, which significantly decreases the ionizing radiation absorption efficiency (stopping power) compared to inorganic materials. To compensate this effect, producers are forced to increase greatly the volume of plastic-based scintillators.

## 3 Energy dissipation processes in typical scintillators

### 3.1 Inorganic scintillators

Let us consider the general principles of energy dissipation processes in inorganic scintillators. After the absorption of an ionizing radiation photon, a primary electron-hole pair is born, which afterwards undergoes a series of relaxation processes preceding the photon emission, and thus determining the luminescence rise time.

The current understanding of the scintillation process dynamics was distilled to a single chart by A. Vasil'ev (see Figure 3) [47]. The five main scintillation stages have been identified: cascade, thermalization, capture, transport and recombination, and the tentative time scales have been assigned to each stage. Generally, the duration of each subsequent stage is by 2–3 orders of magnitude slower than the preceding stage, therefore they all occur in sequence with no parallel processes. However, it is a challenge to calculate the respective durations of the scintillation stages in particular materials more precisely or to measure them experimentally. Direct measurement techniques have been widely available only for the recombination stage (the scintillation lifetime) and to some extent for the transport stage (the scintillation rise time) for slow transport cases.

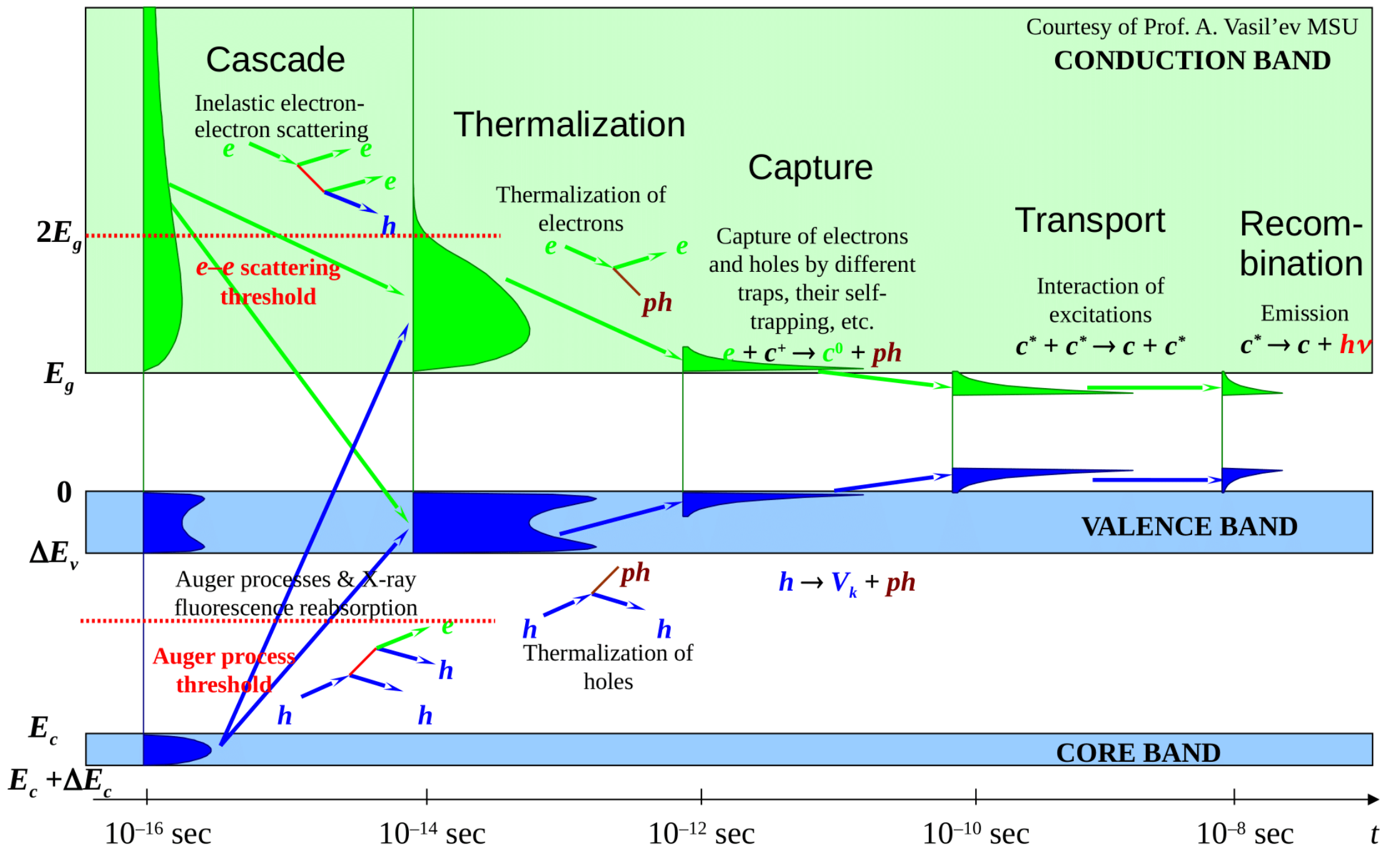

Figure 3. Electronic excitation relaxation scheme for dielectrics and broad band semiconductors [47]. Courtesy of A.N. Vasil'ev.

On the first stage, primary hot electrons and holes are formed after incident ionizing radiation quanta absorption. Then the hot electrons and holes are losing energy for the creation of secondary charge carriers during the inelastic scattering ($10^{-16}$ - $10^{-14}$ s). Only after that holes and electrons can be thermalized with phonon creation ($10^{-14}$ - $10^{-12}$ s) and captured by luminescence centers or form excitons ($10^{-12}$ - $10^{-10}$ s). The diffusion transport of thermalized charge carriers and their recombination takes generally place in time intervals longer than $10^{-10}$ s. These are the main processes for the common scintillating materials like NaI:Tl, BGO, $LaBr_3$:Ce, LSO:Ce and others. A decent amount of experimental data indicate that, in some cases, the transport stage can be slow enough to interfere with recombination, for example, in CsI:Tl demonstrating a long rise time of the order of 10 ns. Figure 4 presents time spans of various luminescence processes related to the above-described stages of the relaxation of electronic excitations. It includes known experimental data, estimates of the actual duration of scintillation stages as well as the analysis of the dynamics of ultrafast scintillation processes. The sources for the duration values are explained below.

The capture stage (Figure 3) has been investigated by different laser pump-probe techniques, but the data available are scarce due to high complexity of these methods. Femtosecond interferometry technique has been earlier developed to study the relaxation rate of free charge carriers with 100-fs time resolution in binary wide-gap materials [48], [49] and tungstates [50]. Unlike transient absorption technique, interferometry probes the change in the refractive index of the material caused by excited charge carriers and can easily distinguish trapped or self-trapped states from unbound (band) states, therefore signify the end of the capture stage and the beginning of the transport stage. The longevity of unbound charge carriers gives us the estimate of the combined duration of thermalization+capture stages. However, due to rather low excitation energy the thermalization stage is truncated as compared to the entire scintillation event. Three types of materials with different routes of charge carriers self-trapping were identified [48].

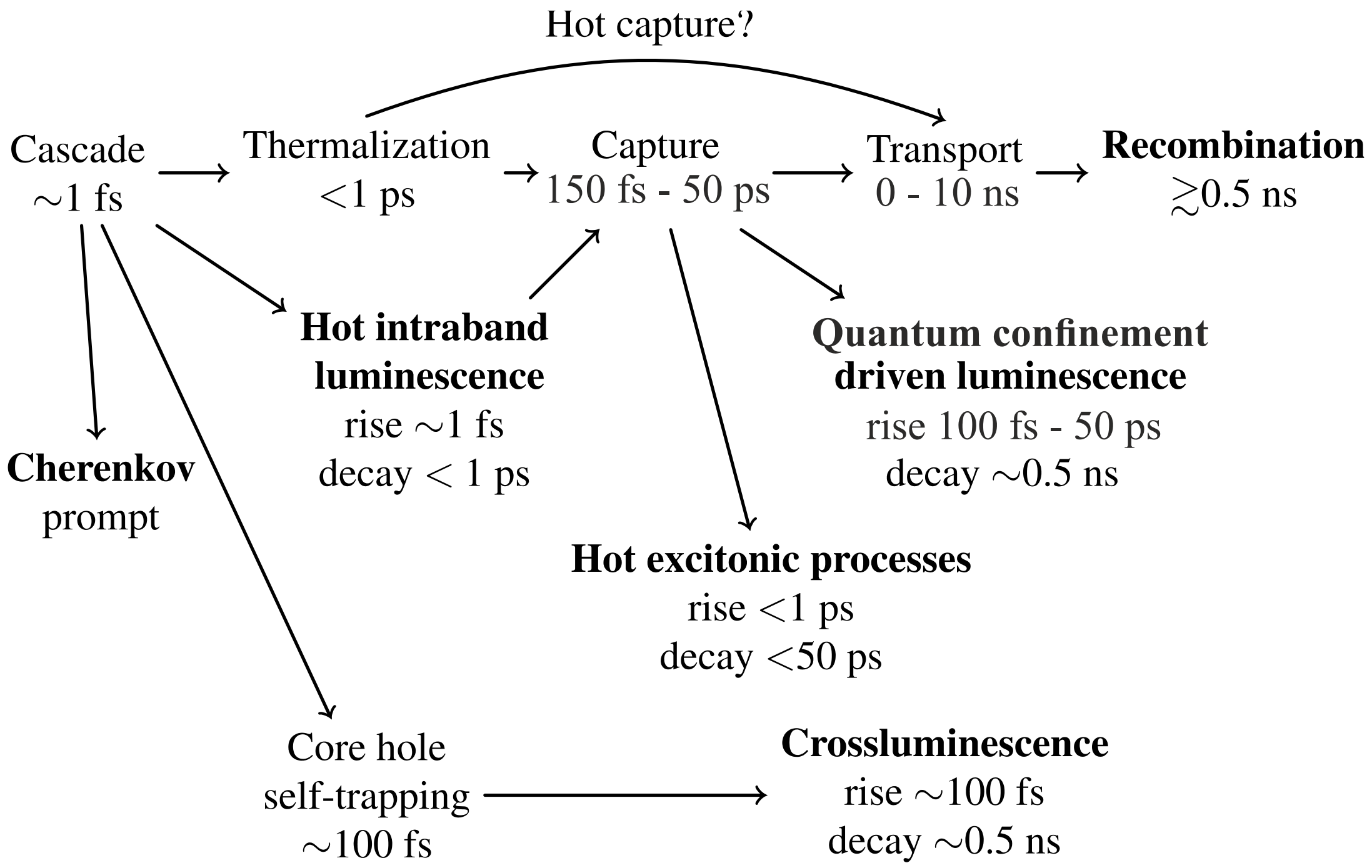

Figure 4. The scintillation stages chart and ultrafast scintillation processes in inorganics materials.

The first type ($SiO_2$, $CaWO_4$, $CdWO_4$) features fast self-trapping of excitons, which occurs within 150–200 fs after excitation pulse. In the second type (KBr, NaCl), the hole is self-trapped within 0.5 ps and subsequently captures an electron. The rate of electron trapping depends on the concentration of holes, therefore the total duration of the capture stage depends on excitation density and reaches a few picoseconds for low-density cases. In the third type ($Al_2O_3$, MgO, diamond), no self-trapping occurs, and electrons stay in the conduction band for more than 50 ps before they are trapped by defects.

The duration of the thermalization stage is currently outside the reach of any time-resolved technique, not only because of time resolution requirements, but also because the above-mentioned methods cannot distinguish high-energy band carriers from thermalized ones. However, the general analysis sets this duration to about 1 ps, and as one can see in the case of fast self-trapping of excitons, the capture stage is much faster and lies in the sub-picosecond domain. Therefore, the thermalization may sometimes occur on the same time scale as the capture, potentially allowing for a competing process like a hot capture which is a direct trapping (or self-trapping) of a hot charge carrier with subsequent release of excess energy as phonons during lattice relaxation.

There are also processes which circumvent some scintillation stages, leading to ultrafast scintillation. In quantum dots (QDs), electron-hole pairs are confined within one nanoparticle, which prevents them from escaping, thus circumventing the transport stage. As a result, the rise time of the luminescence caused by their recombination is determined by the capture stage duration, which can be below 1 ps [51]. Hot intraband luminescence (IBL) [52], [53], [54] occurs as a process competing to thermalization before the capture stage, therefore its decay time is determined by the thermalization stage duration. In some materials (e.g., $BaF_2$), holes at the uppermost core levels have too low energy for Auger recombination, so they undergo self-trapping [55] and float to the top of the core band. The recombination of valence electrons with such holes leads to cross-luminescence (CL) [56], also known as core-valence luminescence or Auger-free luminescence. Some core holes recombine instantly, others undergo self-trapping first, in which case the rise time of CL is determined by core hole self-trapping time, estimated as 200 fs in CsCl [57] and 950 fs in CsBr [58].

Several alkali halides, including KI and RbI, have an energy barrier between the free and self-trapped exciton state, which leads to rather long self-trapping – up to 2 ns at 4.2 K [59]. As a result, free and self-trapped excitons are shown to coexist at low temperatures [60], [61], [62]. At room temperature, free excitons are not luminescent, however, KI still exhibits hot luminescence of excitons during their self-trapping [52]. In some alkali halides, singlet states of self-trapped excitons are efficiently populated, which have intrinsic decay times by 3–4 orders of magnitude faster than triplet self-trapped excitons found in the majority of materials. Although such hot excitonic processes are usually strongly quenched at room temperatures, their fast intrinsic decay time allows them to compete relatively efficiently with thermal relaxation in alkali halides, which is slower than in oxides. Therefore, in KI, the yield of hot excitonic luminescence rivals that of IBL even at 297 K [52]. In materials with fast self-trapping, such processes could in principle also exist, but their decay time and yield would be much lower.

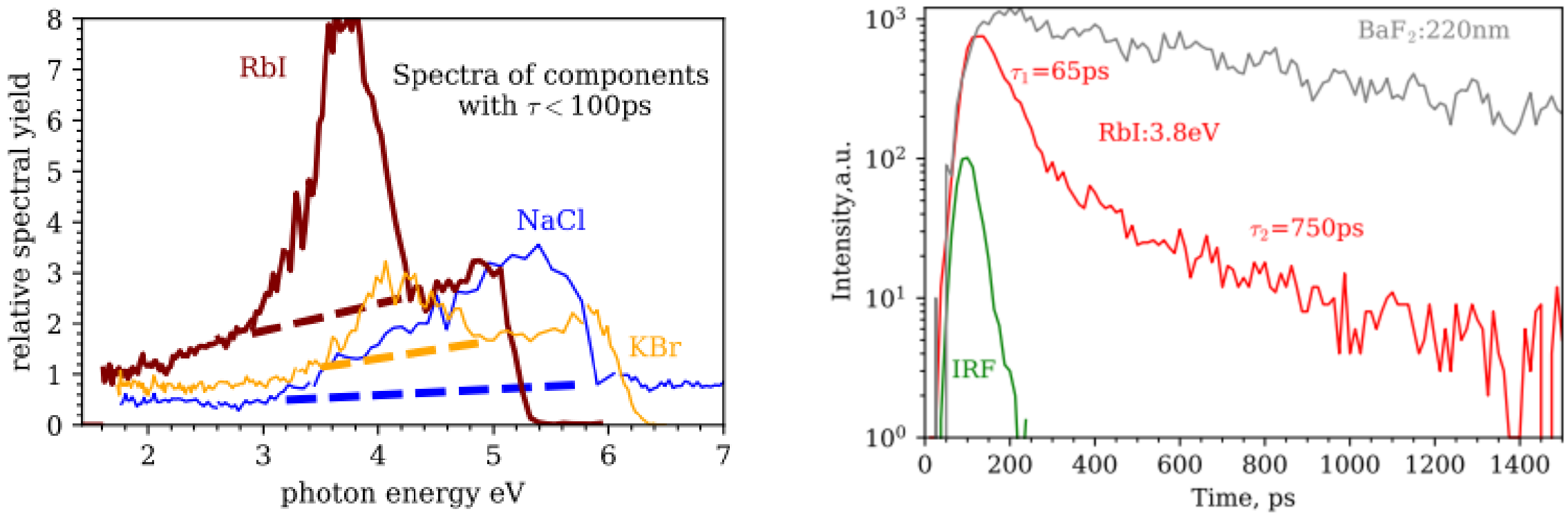

a) b)

Figure 5. The partial cathodoluminescence light yield of fast (τ<100 ps) emission components of RbI, KBr and NaCl (a). The cathodoluminescence decay curve of RbI recorded with 55 ps time resolution (FWHM) for the 3.8 eV emission band. T=297 K (this work).

As the studies [59], [60], [61], [62] were performed long time ago and mostly done at low temperatures, we have conducted pulsed cathodoluminescence experiments on RbI, NaCl and KBr to confirm the existence of these processes at room temperature. The partial spectral yield of the emission components with decay time less than 100 ps (Figure 5a) was obtained by recording decay curves for every spectral point in the range of 1.6-7.0 eV and then by applying least squares fitting to them as described in detail in [52]. The results indicate indeed the existence of additional fast emission peaks on the background of a broad structureless pedestal of IBL. The positions of the peaks are in good correspondence with the sigma-polarized singlet self-trapped exciton emission bands at liquid helium temperature reported at 3.88-3.95 eV for RbI, 4.42 eV for KBr, and 5.35-5.6 eV for NaCl [61]. The high-resolution (55 ps FWHM) cathodoluminescence decay curves (Figure 5b) were recorded for RbI at 297 K using a novel time-correlated multiphoton counting technique as described in [63] The decay constants were obtained by multi-exponential least-squares fitting. The main decay component of a singlet self-trapped exciton emission band at 3.8 eV comprising 75% of the total light yield has a decay time of 65 ps at room temperature. At liquid helium temperatures, the decay time is reported to be as long as a few nanoseconds [61], which indicates quenching by about 2 orders of magnitude. The total yield of cathodoluminescence emitted as prompt photons (including IBL and hot excitonic emission) in RbI was determined by comparison with the IBL intensity in $PbF_2$ as described in [52]. The result of 70 ph/MeV is an attractive figure for potential application of RbI for advancing the 10-ps challenge.

Finally, one should also mention Cherenkov light which is emitted when the electron velocity is greater than the speed of light in a medium and may serve as another possible source of prompt photons. In this case, the time interval between the incident γ-photon absorption and visible photon emission is no more than few ps and depends only on the travel path length of γ-photon before the first interaction. This phenomenon is observed in inorganic crystals with high refractive index, such as LSO and BGO [64]. Cherenkov light presents the fastest mechanism of the conversion of absorbed ionizing radiation energy into photons, but its yield is relatively low as well. In the common scintillating crystals used in PET, at most two tens of Cherenkov photons are produced after absorption of the incident γ-photon with 511 keV (e.g., 17 photons on average in one of the most efficient Cherenkov emitters, BGO [15]). In addition, the maximum emission wavelength lies in the UV region, while conventional photodetectors are characterized by weak sensitivity in this region and only few photons can be detected [10].

It becomes evident from the analysis of Figure 3 that the engineering of new ultrafast scintillation materials would require circumventing slow transport stage in the relaxation of at least a significant fraction of the hot charge carriers. Hot excitonic processes, though promising in some cases when bulk inorganic materials must be used, are of limited applicability due to their occurrence only in specific materials and low yield at room temperature. Finally, the above-described quantum confinement effects appear only in nanoscale objects such as quantum dots, which are not useful in scintillator applications on their own because of low stopping power of a single nanoparticle. A popular way of creating bulk scintillator out of quantum dots is embedding them into organic polymer matrix, such as plastic scintillators.

### 3.2 Plastic scintillators

The processes of absorption and conversion of ionizing radiation in organic materials are different to those in inorganic crystals. Their luminescence is due to the transition of π-electrons from excited states and, therefore, it is observed in compounds possessing unsaturated bonds in their molecular structure [1]. A generalized scheme of electronic states of such system is depicted in Figure 5. Upon excitation, a molecular electronic system transfers from the ground state $S_0$ to one of the excited singlet states $S_1$, $S_2$, $S_3$, ...$S_n$ [1] (Figure 6). Each principal level is accompanied by vibrational sublevels $S_{ij}$. Emission can occur either due to the direct singlet-singlet transition $S_n \rightarrow S_0$ or due to the transitions involving the triplet states $T_1$...$T_n$. The former mechanism is much faster and because of that it is mainly responsible for fluorescence. The radiative lifetime of $S_1$ is $10^{-8}$-$10^{-9}$ s, which is rather long compared to the period of molecular vibration (~$10^{-12}$ s). Therefore, the molecule reaches thermal equilibrium before emission and the fluorescence transition occurs from the $S_{10}$ to $S_{01}$, $S_{02}$, $S_{03}$...$S_{0n}$ vibrational sub-levels (Figure 6). In this case, luminescence decays exponentially with time:

$$I = I_0 e^{\frac{-t}{\tau}}, \quad (3)$$

where $I_0$ is the intensity at t=0 and τ is the fluorescence decay time [1].

Compounds in which direct radiative transitions dominate are much more suitable for scintillation applications than those showing triplet luminescence. Typical properties of organic scintillators are summarized in Table 1.

The second mechanism is responsible for slow luminescence components, which appear due to the population of a metastable state (T1 state in Figure 5). The possibility of the transition from the $S_1$ state to the metastable triplet state $T_1$ has been proven experimentally [65]. The transition from $T_1$ to $S_0$, characterized by a long decay time of hundreds of μs or even longer, is responsible for phosphorescence and may also cause an exponential luminescence decay. Besides, there is a possibility of return from the metastable state back to $S_1$. On the one hand, such process will shorten the decay time of the phosphorescence. On the other hand, it can cause the appearance of slow components in fluorescence, whose decay times will depend on the lifetime of metastable state and on the temperature [1].

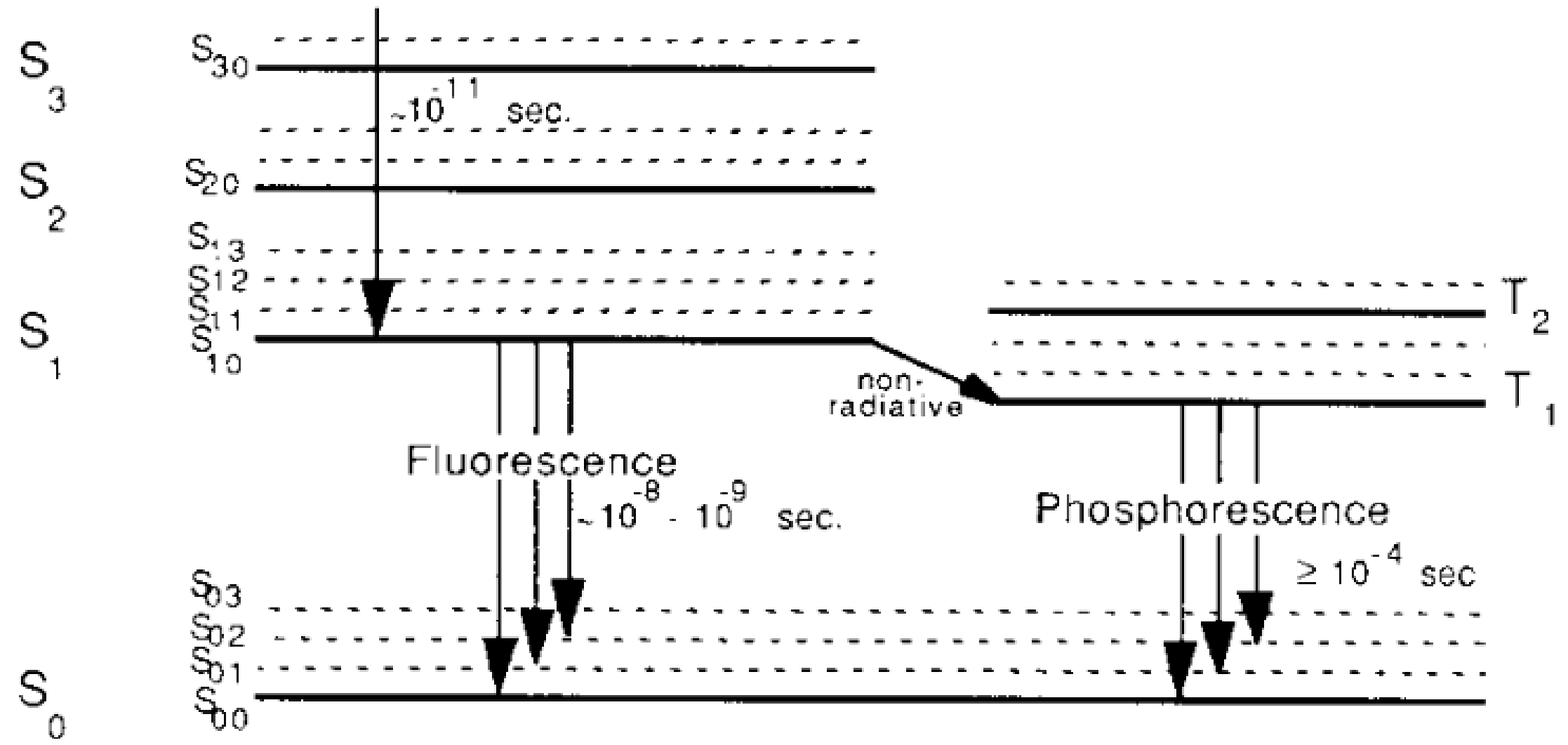

Figure 6. Electronic energy levels and transitions in organic molecules. (Reprinted with permission from Ref. [66]. Copyright with license Id: 4964150907094.)

## 4 Composite scintillators

Based on the analysis of energy dissipation processes in organic and inorganic scintillators, it is possible to make the following conclusions.

Inorganic crystals can have high ionizing radiation stopping power. Doping bulk crystals with luminescent impurities can provide high LY, but relatively slow scintillation response. The IBL and Cherenkov emission are much faster, but have negligible LY. Moreover, inorganic crystal growth is technically difficult and expensive routine driving the cost of the manufacturing of big detecting setups relatively high.

Organic compounds are much easier to produce, but in order to compensate low stopping power it is compulsory to increase the volume of organic scintillator dramatically. This in turn negatively impacts time resolution due to higher uncertainty of light travel path and transit time through the material before reaching the detector. Thus, the development of a scintillation material which could combine the advantages of both inorganic and organic scintillators is a very important and promising task to solve on the way towards the 10-ps challenge.

Nowadays there are several ways to combine the advantages of organic and inorganic scintillators aimed at applications demanding fast ionizing radiation detection. One of the possibilities is the manufacturing of a detecting cell consisting of alternating layers of inorganic single crystal and organic compound. In [36], proof-of-concept of a detecting pixel with BGO or LYSO plate as an inorganic layer and a plastic scintillator BC-422 as an organic layer is presented. This structure is characterized by high stopping power of incident ionizing radiation due its efficient absorption in the high-Z layers of inorganic compound. Plastic scintillator serves as a source of fast emission which is mainly caused by the interaction with recoil electrons formed in the high-Z layer. However, this structure has a number of drawbacks. Because of low $Z_{eff}$ value of organic compound, only a fraction of absorbed energy is converted into fast luminescence. Moreover, there are optical losses in a layered structure due to a difference in refractive indexes between inorganic and organic plates. Also, the production of inorganic plates remains expensive.

An alternative way to improve the scintillating characteristics is to dissolve quantum dots into common organic scintillating materials. Quantum dot systems are characterized by high luminescence quantum efficiency and ultrafast decay times [51]. Moreover, the spectra of such structures exhibit wide absorption and narrow emission bands whose maxima can be shifted by varying QD size. These materials could find application in new PET scanners, in particular, in TOF-PET, that take into account Compton scattering processes [67]. However, the required concentrations of selenide and sulfide based QD to reach sufficient $Z_{eff}$ values are very high (60-70%) [68]. The mismatch of refractive indices of the polymer and QDs makes the synthesis of transparent composites very challenging, and generally low production yield of QDs [69] makes large volume manufacturing difficult. One of the promising solutions is the development of polymer-based nanocomposite materials loaded with large amount of heavy inorganic nanoparticles [70]. Such material would be characterized by high efficiency of ionizing radiation absorption due to inorganic nanoparticle filler with high-$Z_{eff}$ value, and in addition, the transfer of the energy of electronic excitations to the polymer would be more efficient compared to a layered structure. Let us consider in more detail the peculiarities of material selection and synthesis routine as well as the properties of the known nanocomposite materials.

## 4.1 Concept

The idea of the development of scintillating heterostructure materials for the detection of high energy radiation is not new. First polymer-based scintillating materials with distributed filler for fast neutron detection were investigated in the beginning of the 50s of the last century [71]. In the proposed composition, polymethyl methacrylate (PMMA) matrix performed two functions: the production of recoil protons which initiate the ZnS filler scintillation, and light guiding. The investigations in the late 60s– early 70s were dedicated to the detection of radon daughter products [72] and discrimination of the detection of neutrons and γ-rays [73], [74], [75]. In these and several later works, the composite material was mostly handled as a combination of different elements, but not as a monolithic material.

Later in the 80s, the results of study of composite PMMA-based materials with 70 wt.% of filler ($BaF_2$ powder) for proton detection were presented [76]. In [77], polystyrene-based scintillating materials with $CeF_3$, $BaF_2$, or $PbF_2$ filler were proposed to use in a shashlik-type calorimeter (see Figure 7). In this configuration, emitted light was collected by wavelength-shifting fibers inserted into the detection medium. This allowed to minimize optical losses caused by strong scattering process.

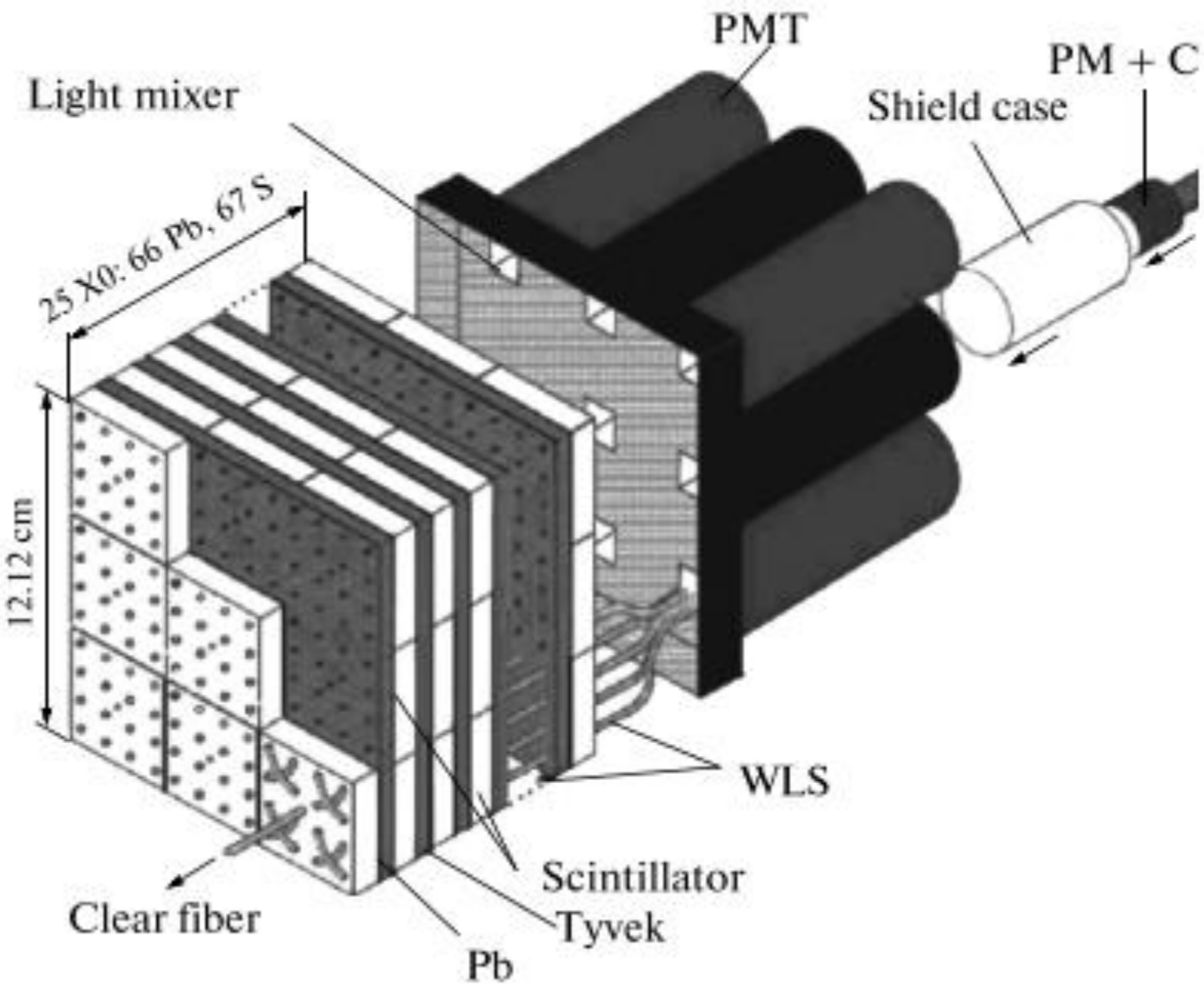

Figure 7. Shashlik-type calorimeter used in LHCb ECAL. (Reprinted with permission from Ref. [2]. Copyright with license Id: 4964150387674.)

Despite the widespread use of composite scintillators in high energy physics for a long time, the concept of scintillator heterostructure for small energies (511 keV) has only been introduced for the first time by P. Lecoq in 2008 [78] and further developed and explained in more recent papers [79], [80].

Several functional elements distinguishing nanocomposite scintillators from the above-described scintillator configurations can be highlighted: an organic matrix, a uniformly distributed heavy inorganic filler and a dissolved phosphor or several different phosphors (Figure 8). The processes of gamma radiation absorption and secondary electron creation occur in the filler particles due to their high $Z_{eff}$. Secondary electrons are escaping the particles, depositing energy in the matrix mainly by ionization. The primary phosphor acts as an activator, which converts electronic excitations of the matrix into luminescence, which usually happens by the mechanism of Förster Resonant Energy Transfer (FRET). This mechanism has characteristic times which are usually much less than the radiative lifetime of a phosphor, thus not introducing luminescence rise times. Another (secondary) phosphor can also be added perform a function of a wavelength shifter to increase the effective Stokes shift and minimize luminescence re-absorption, but the shifter usually uses radiative energy transfer from the primary phosphor leading to significant rise times The matrix in a nanocomposite material plays additional roles of a binding medium and a light guide for luminescence [81].

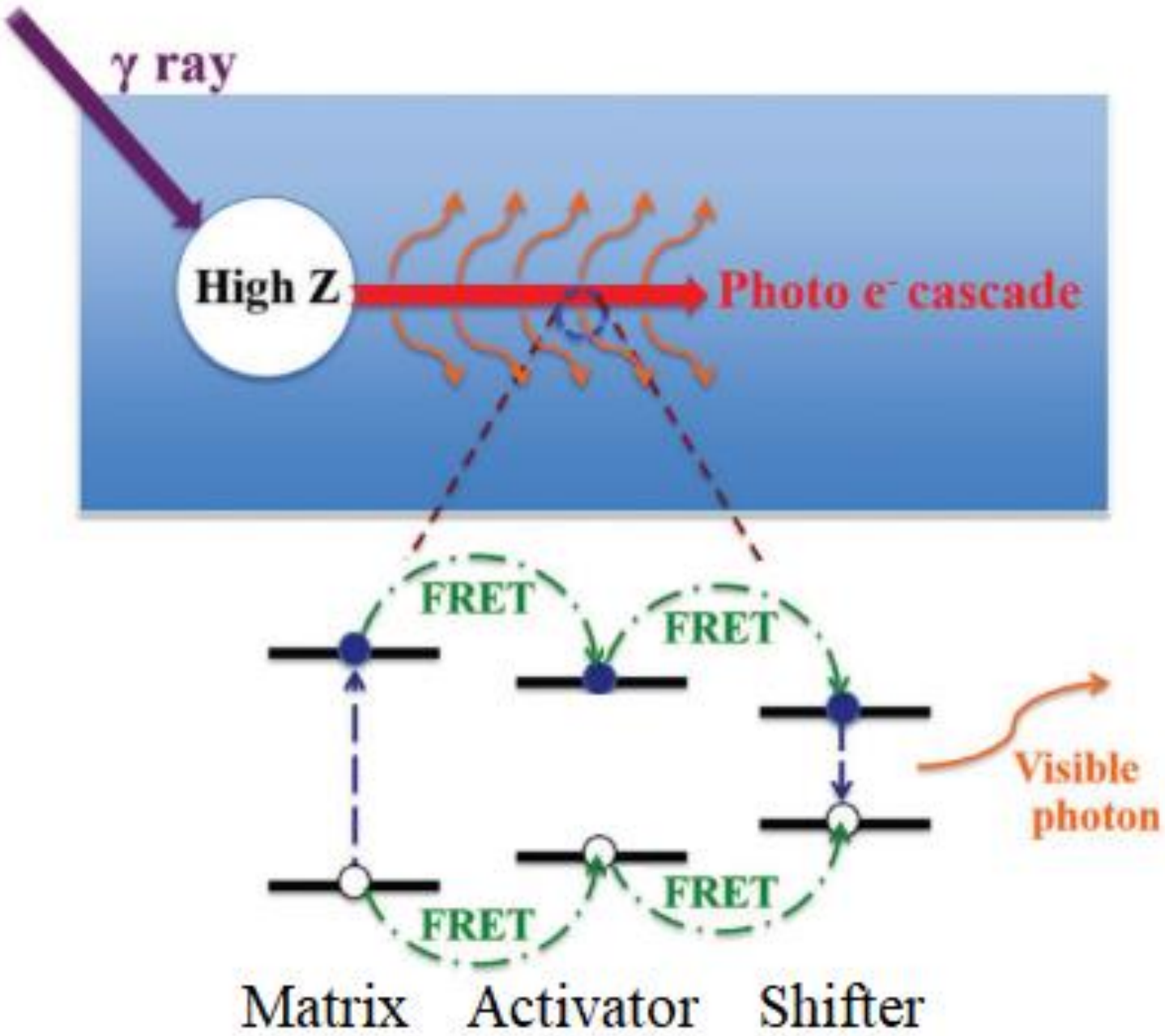

Figure 8. Structural scheme of nanocomposite scintillation material adapted from. (Reprinted with permission from Ref. [81]. Copyright with license Id: 4964080834236.)

Nanocomposite materials have several advantages compared with other similar structures with combined organic and inorganic components. Due to a filler in a form of nanocrystalline powder a large area of organic/inorganic interface can be achieved. Moreover, powder production routines are much cheaper than single crystal growth techniques. The organic phosphors or QDs as an activator allow to reach a very high conversion rates of absorbed ionizing radiation energy into luminescence.

## 4.2 Matrix

The matrix is a binding medium for the structural parts of composite materials such as phosphors and fillers. It should provide unimpeded distribution of the luminescence light and, therefore, high optical transparency in the phosphor emission range is one of the basic requirements for the matrix substance. Organic polymers are among the most suitable matrix substances in terms of the simplification of the production and postproduction treatment. Generally, organic molecules with π bonds in their structure have absorption bands in the UV spectral range and can exhibit luminescence [1]. However, the LY of such substances is usually low, so a phosphor should be used as an activator (see section 3.4).

Nowadays, there are several different polymer compositions, which can be used as organic matrixes in nanocomposite scintillation materials. The main properties of such materials are listed in Table 2.

Table 2. Properties of several polymers suitable for scintillation materials.

| Compound | Refractive index | Density, g/cm$^3$ | Absorption band, nm |
|---|---|---|---|
| 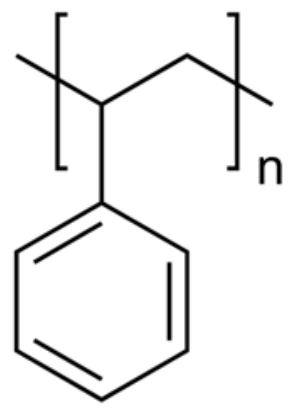 Polystyrene (PS) | 1.59 [82], 1.58 [83] | 1.04 [83] | 375 [84] |
| 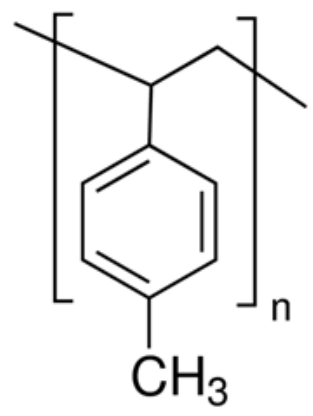 Polyvinyl toluene (PVT) | 1.6 [85] 1.55 [86] | 1.05 [86] | 390 [84] |
| 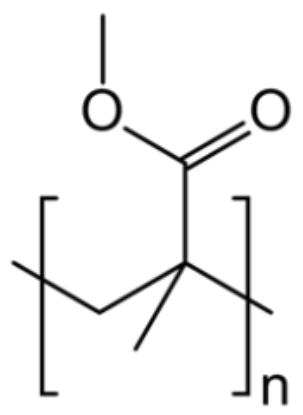 Polymethyl methacrylate (PMMA) | 1.49 [77] | 1.19 [87] | 280 [88], 300 [89] |
| 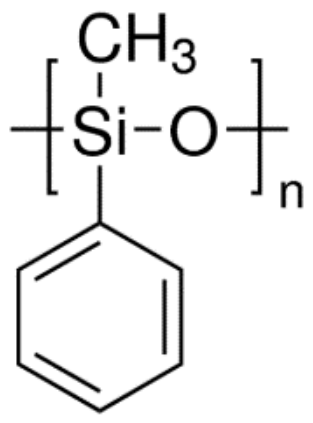 Polysiloxane (Poly(methylphenylsiloxane) | 1.49-1.58 [90] | 0.88-1,09 [91] | 400 [92] |
| 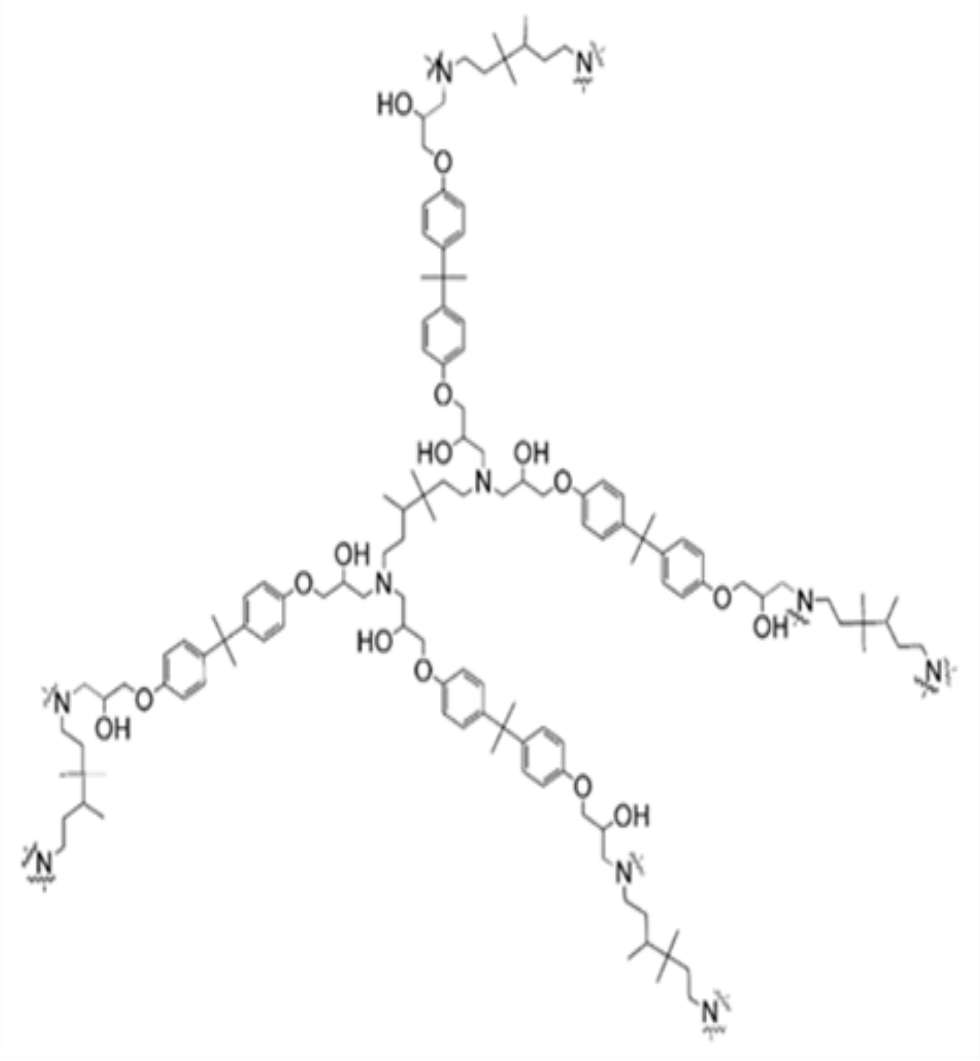 Epoxy resin | 1.48 - 1.51 [93] | 1.15 [94] | 400 [88] |

Polystyrene (PS) and its homolog polyvinyl toluene (PVT) were among the first widely used polymers. Plastic scintillators based on these polymers with additional organic activators were first described in 1949 [95]. The polymerization of vinyl group containing compounds, like styrene and vinyl toluene, results in the opening of the carbon-carbon double bond of the vinyl group $CH_2$=CH- and formation of a new σ-bond with a carbon atom of the next styrene molecule in the chain. In PS-based plastic scintillators, the absorption of ionizing radiation occurs in the polymer medium and then the excitation is transferred to the activator molecules by nonradiative Forster dipole-dipole interactions [96]. PS is characterized by high radiation and heat resistance [97] whereby this polymer is suggested for use as a polymer matrix in nanocomposite materials [98], [99], [85],[86]. Figure 9 presents transmittance spectra of polystyrene (in air) and polyvinyl toluene (in argon) [92].

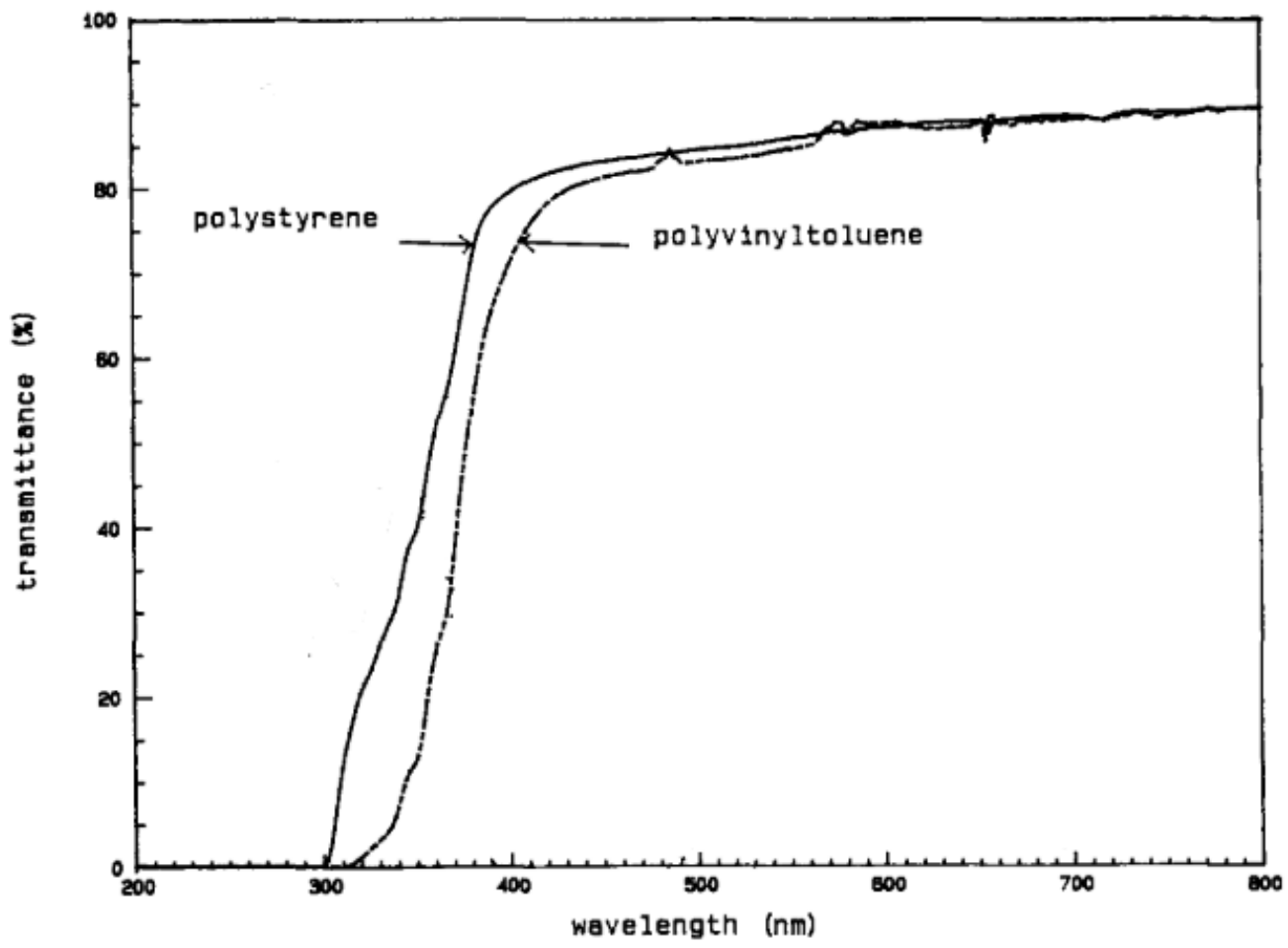

Figure 9. Transmittance spectra for polystyrene and polyvinyl toluene. (Reprinted with permission from Ref. [84]. Copyright with license Id: 5080110089005.)

Though widely used in plastic scintillators production, polystyrene has a number of shortcomings, which limit its potential as an efficient matrix for nanocomposite materials. The main drawbacks of the PS-based polymers are brittleness and have high absorption coefficient at the wavelength of activator emission. As a result PS is not suitable for large scintillators [89]. A low refractive index of polystyrene in the UV range should also be noted, as it limits the application of PS as a matrix for nanocomposite scintillators.

Another perspective compound which could be used as a matrix is polymethyl methacrylate. The main feature of this polymer is high optical transparency in a wide spectral region extending to UV (300-800 nm) (see Figure 10) [88]. PMMA does not have unsaturated C-C bonds in its structure and, thus, unlike other polymers, does not exhibit luminescence. However, this compound has several shortcomings. Due to the absence of π-bonds, the efficiency of energy transfer from the matrix to phosphor is much weaker compared to PS and PVT [83]. Moreover, due to an aliphatic structure, PMMA is a very poor solvent for most of the organic phosphors and, therefore, the maximal achievable activator concentration is rather low.

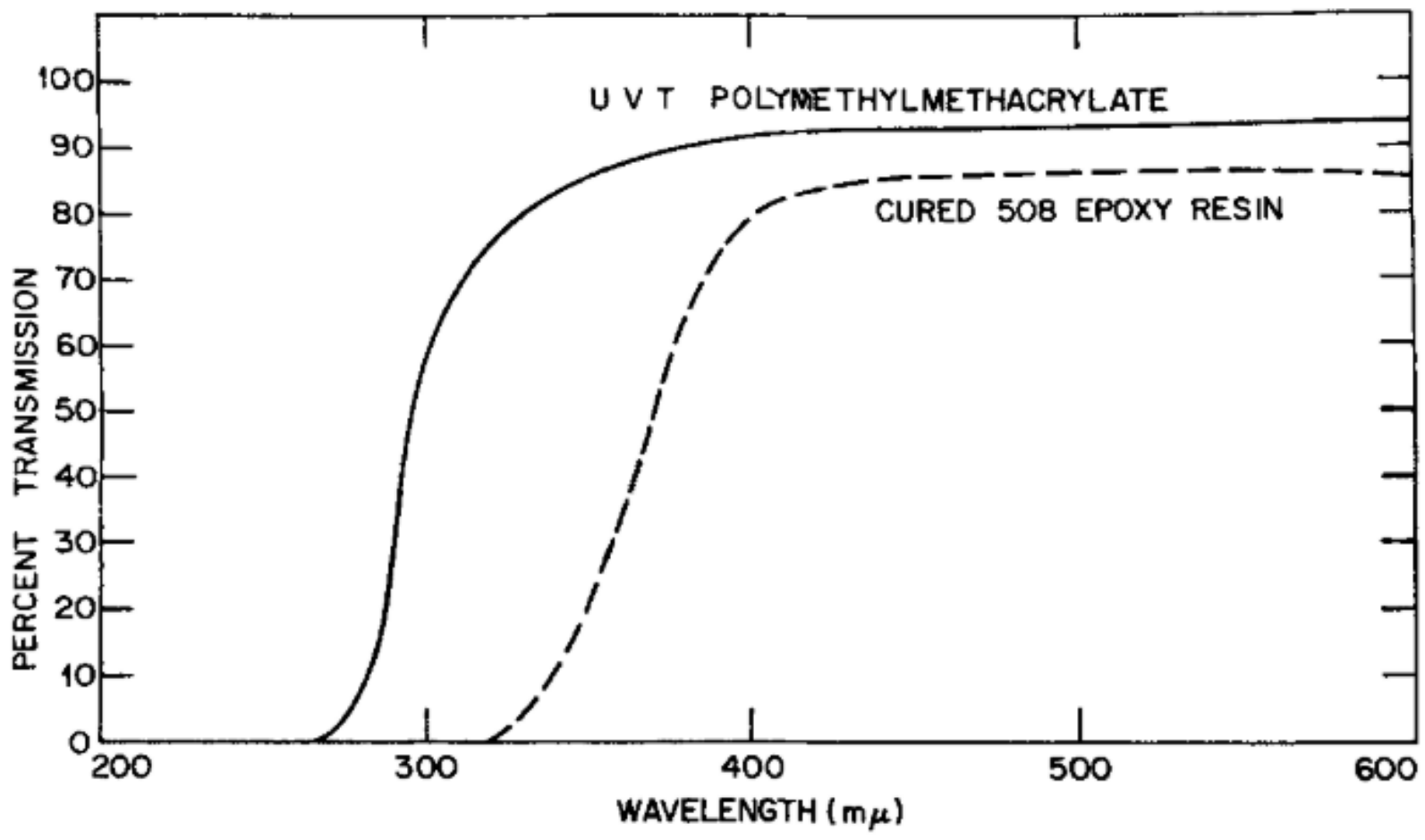

Figure 10. Typical transmittance spectra of PMMA and Epoxy Resin 508 (diglycidylethers of bisphenol A based). (Reprinted with permission from Ref. [88]. Copyright with license Id: 4964081451915.)

Radiation resistance is another important parameter for the basic materials of scintillator matrix. Incident ionizing radiation not only excites luminescence, but also breaks chemical bonds, producing harmful color centers and deteriorating optical characteristics of a polymeric matrix. Polysiloxane may be considered as one of the most irradiation resistant polymers (Table 2). An exemplary structure of polysiloxane, based on vinyl-terminated diphenylsiloxane-dimethylsiloxane copolymer and hydride-terminated methylhydrosiloxane-phenylmethylsiloxane cross-linker, is presented at Figure 11.

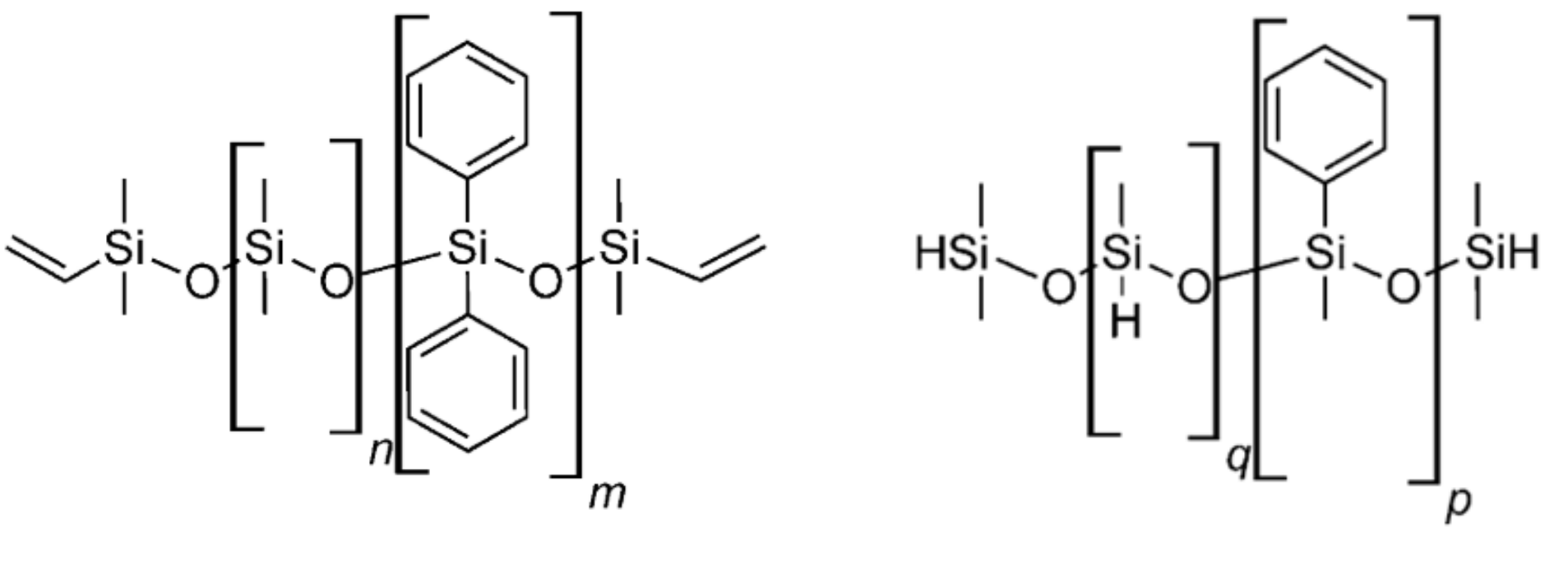

Figure 11. Structure of a polysiloxane components a) base resin and b) cross-linker.

It was historically the first polymer compound proposed as a matrix for composite scintillators [92]. Another advantage of this material over PS, PVT or PMMA is high stability of optical properties in a wide temperature range. Its unique physical properties are determined by high flexibility of chains because of long Si-O bonds situated at large angles. It should be also noted, that because of Si atoms substituting carbon in organic polymers, polysiloxanes are characterized by a higher $Z_{eff}$ value making them much more sensitive to γ-rays. The main disadvantage of these compounds is low solubility of commonly used phosphors. For a widespread PPO activator, the highest attainable concentration is only 1-1.5 wt.% [100]. In addition, polysiloxanes exhibit significant absorption in the UV range compared to other compounds (see Figure 12).

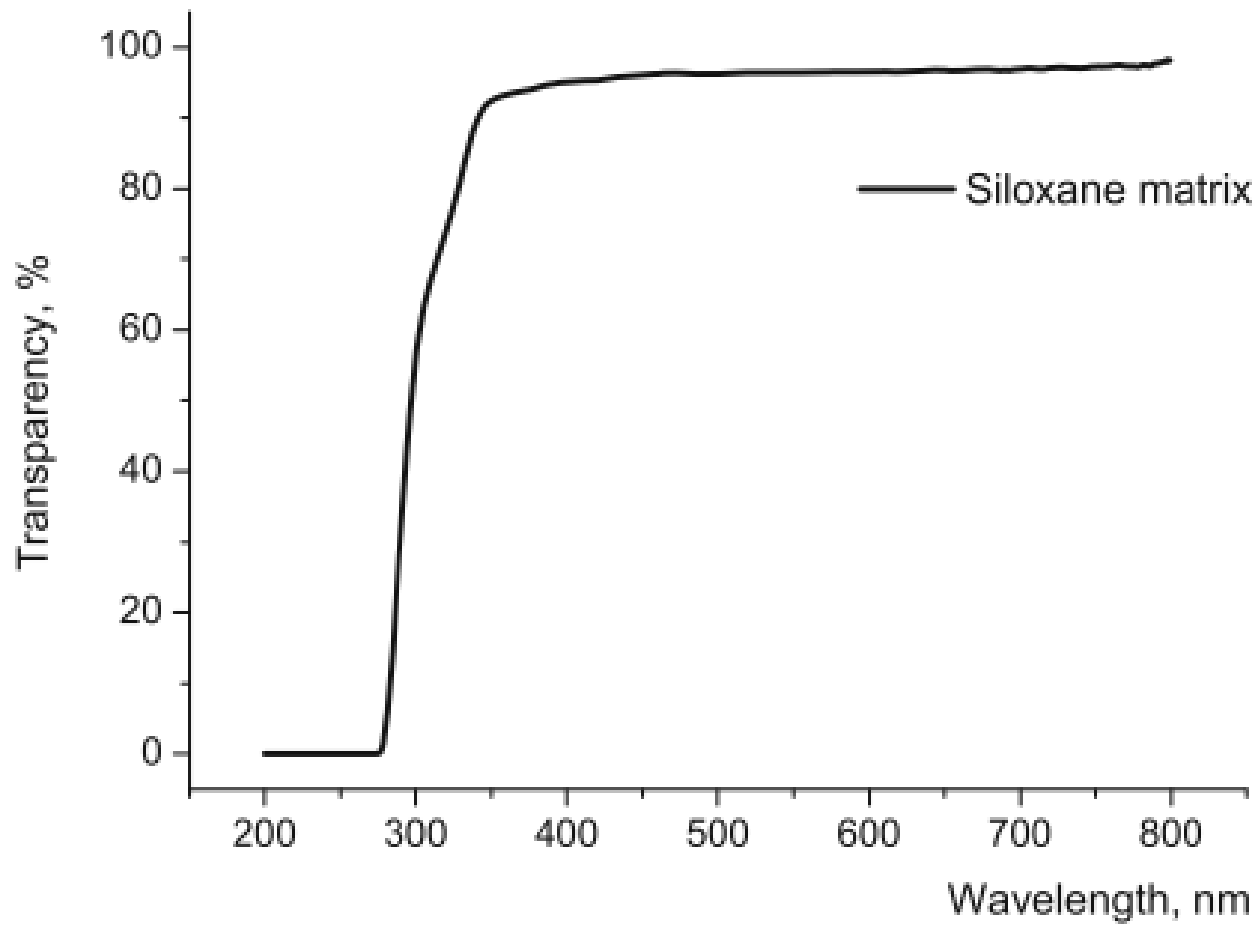

Figure 12. Transmittance spectrum of polymethylphenylsiloxane. (Reprinted with permission from Ref. [101]. Copyright with license Id: 4970050233103.)

Another perspective polymers are epoxy resins which can also be used as organic matrixes. This class of compounds is attractive due to a simple preparation routine. Such resin consists typically of two parts – prepolymer (2,2-Bis[4-(glycidyloxy)phenyl]propane) and hardener (3,3,4-trimethylhexane-1,6-diamine) (Figure 13) [83].

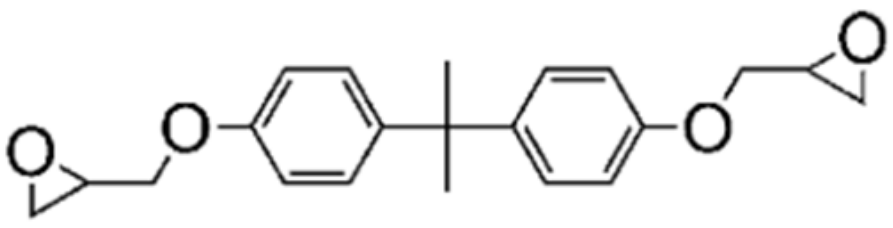

a)

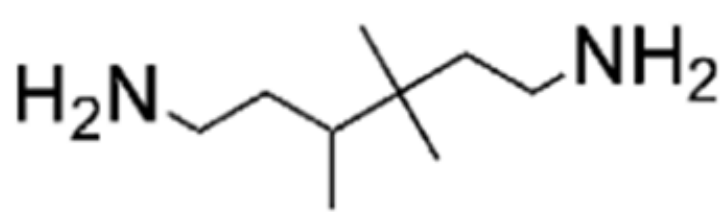

б)

Figure 13. Structure of epoxy resin components: prepolymer - 2,2-Bis[4-(glycidyloxy)phenyl]propane (a) and hardener - 3,3,4-trimethylhexane-1,6-diamine (b)

Polymerization occurs at room temperature after the mixing of the two parts. Phosphors are usually dissolved in prepolymer prior to mixing [102], [93], [103]. Highly cross-linked chains of prepolymer and hardener molecules present another advantage of polyepoxides. Due to that, these compounds are characterized by high temperature and mechanical stability. The drawbacks of polyepoxides are low solubility and poor transmittance in the UV range (Figure 10) [88].

Polyurethane (PU) is one more class of promising compounds for organic matrix. Polyurethanes are synthesized by exothermic reaction between polyols, compounds with at least two OH groups in one molecule, and isocyanates, compounds with more than one N=C=O group in one molecule (diisocyanats, polyisocyanates). Physical properties of resulting polymer can be predefined by altering the polymerization routine. The size of polymer chain, molecular weight and degree of hydrogen bonding may be increased by means of chain-extenders. Due to a variety of initial compounds, PU with defined physical properties can be produced, whereby they have widely found application as elastomers [104], mechanoluminescence polymer materials [105], fibers [106], and luminescent foams [107]. At the moment, there are many ways of producing optically clear polyurethanes with high transmittance in the UV range, which is one of the most important properties of organic matrix for scintillation material. For that purpose, aliphatic and cycloaliphatic isocyanates such as 4,4'-methylene bis(cyclohexylisocyanate) [108] are used. A typical PU transmittance spectrum is presented in Figure 14.

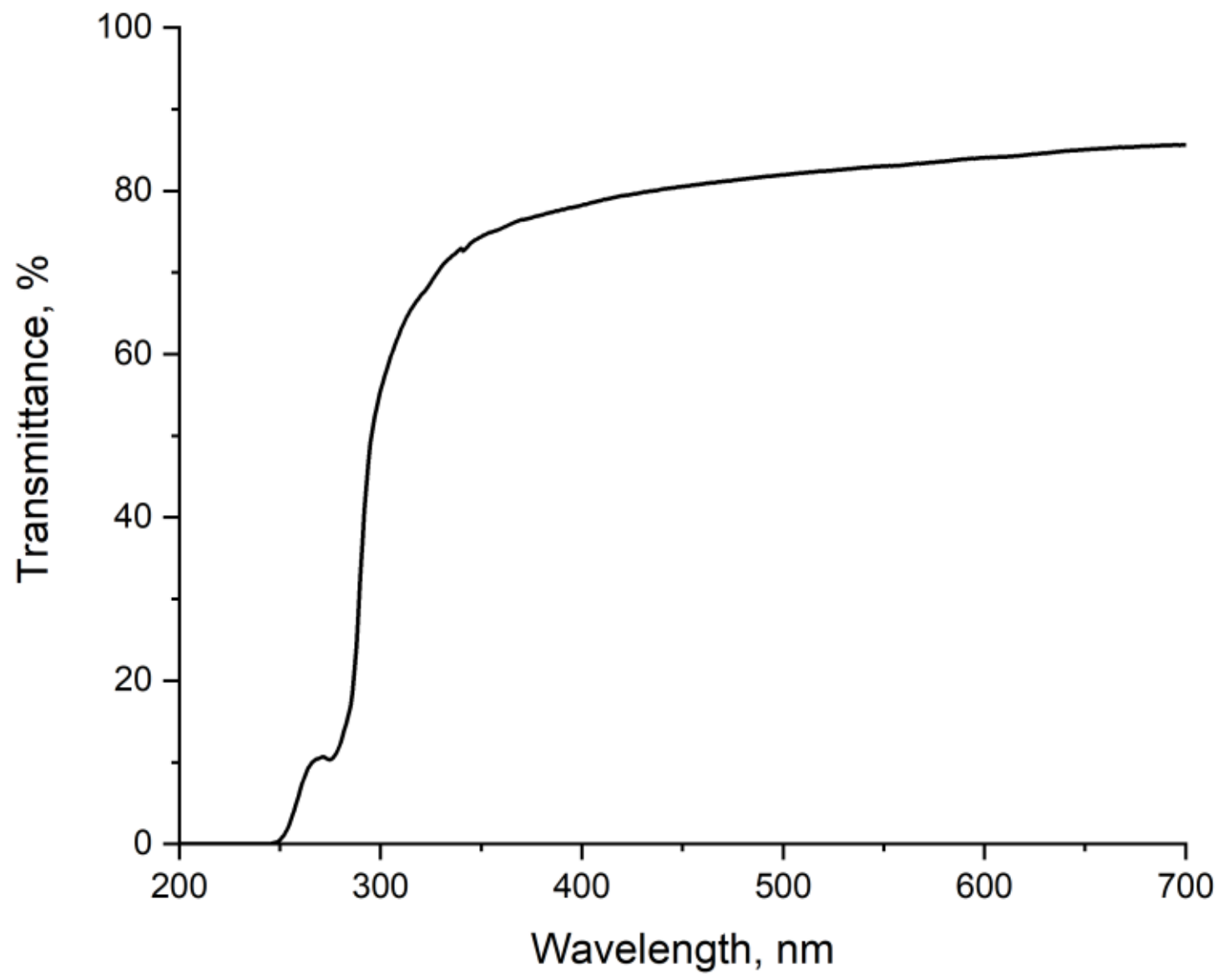

Figure 14.Typical PU transmittance spectrum. (Reprinted with permission from Ref. [109]. Copyright with license Id: 4964090133073.)

It is also important that PU is able to contain fillers in high concentrations [110]. Presently, there are many different nanocomposite PU-based materials. Filler injection improves such relevant physical properties of a polymer material as hardness, flexibility, thermal stability, electric conductivity, etc. The fillers are typically nanoparticles, carbon nanotubes, nanofibers or graphene [110]. The development of filler injection techniques allows to significantly improve the properties of the resulting materials. There are on-going investigations of novel efficient PU-based scintillators doped with various organic phosphors [106], [111], [112], [113], [114] and rare-earth elements [107], [115]. The flexibility of chemical composition of the matrix allows to securely stabilize particles with high active surface area. Combined with high optical clearance, it allows to use these polymer materials as binder media for luminescence nanoparticles [116], [117] and QDs [118], [119], [120], [121], [122], [123], [124]. Up to now, there have been no reports on PU-based scintillating compounds, however their above-listed properties make PU a perspective matrix for nanocomposite scintillators [125].

Based on the discussion above, one can conclude that the existing polymer scintillation materials possess along with advantages also a number of shortcomings which limit their applicability as nanocomposite matrixes. Therefore, the search for new optimal matrix materials is still a crucial task.

### 4.3 Filler

Transparent organic material of a host compound consists mainly of light elements with low atomic number such as H (Z = 1), C (Z = 6), N (Z = 7), O (Z = 8). As a result, organic compounds have a small value of the effective atomic number $Z_{eff}$ and interact weakly with high-energy radiation. Therefore, to register X-rays and gamma rays with organic materials, large volume detectors are required. The large dimensions have a negative impact on the time and energy resolution due to the increase in the light travel path inside the detector. The efficiency of the absorption of ionizing radiation by organic compounds can be increased by introducing an inorganic filler. In the structure of the nanocomposite material under consideration, the filler material with high effective atomic number acts as an absorber of the energy of ionizing radiation. The main luminescence mechanisms of inorganic compounds involve, the relaxation and capture of electronic excitations by luminescence centers, the formation of excitons and emitting light typically within the time intervals of $10^{-9}$ – $10^{-7}$ s (see Table 1). These time characteristics do not satisfy the demands of novel tasks related to the implementation of time-of-flight techniques as described in Introduction, therefore, it is necessary to use in a composite material with complex additives including an efficient filler and faster activator. The purpose of the inorganic filler is not the fluorescence, but the transformation of the absorbed gamma photon into electronic excitations, which will then be transferred through nonradiative mechanisms to the activator. The role of the activator is to transform the captured electronic excitation into a prompt luminescence photon. In the described energy transfer process, the luminescence characteristics of the filler will not affect the decay time of the composite scintillator.

In addition to the atomic number of filler elements, the size and morphology of its particles play an important role. As the particle size decreases, the specific surface contact area of the filler and host increases. Due to this, the processes of electronic excitation transfer become more efficient.

When choosing a filler material, it is also necessary to consider its effect on the optical characteristics of the material. In a heterogeneous structure, additional scattering will appear at the particle-host interface. This process can be quantitatively described using the Rayleigh scattering formula for spherical particles [126]:

$$T = \frac{I}{I_0} = exp\left(\frac{-32\pi^4\phi_p x r^3 {n_m}^4}{\lambda^4}\left(\frac{\left(\frac{n_p}{n_m}\right)^2-1}{\left(\frac{n_p}{n_m}\right)^2+2}\right)^2\right), \quad (4)$$

where $I$ is the intensity of the transmitted light; $I_0$ is the intensity of the incident light; $\phi_p$ is the volume fraction of nanoparticles; $r$ is the particle radius; $x$ is the optical path length; $\lambda$ is the wavelength of light; $n_p$ and $n_m$ are the refractive indices of the filler (nanoparticles) and matrix, respectively.

According to this formula, the degree of light scattering inside the nanocomposite material depends on the difference between the refractive indices of the matrix and the filler. Organic compounds are usually characterized by a small range of $n_m$ values: 1.3-1.7 [126]. For inorganic single crystals, on the contrary, the range of refractive indices is larger. The closer the values of refractive indices of the filler and matrix are, the lower optical losses will be. In [93], due to good coincidence of refractive indices ($n_m$ = 1.476, $n_p$ = 1.464 at the wavelength of 585 nm), it was possible to obtain a transparent nanocomposite based on an epoxy resin matrix and $BaF_2$:Ce particles (see Figure 15). However, the limitation caused by the differences in refractive indices can be circumvented. It was shown in [86] that for a sufficiently small size of filler particles (a few nanometers), the effect of the difference in the values of refractive indices becomes insignificant. This also makes the use of smaller nanoparticles preferable.

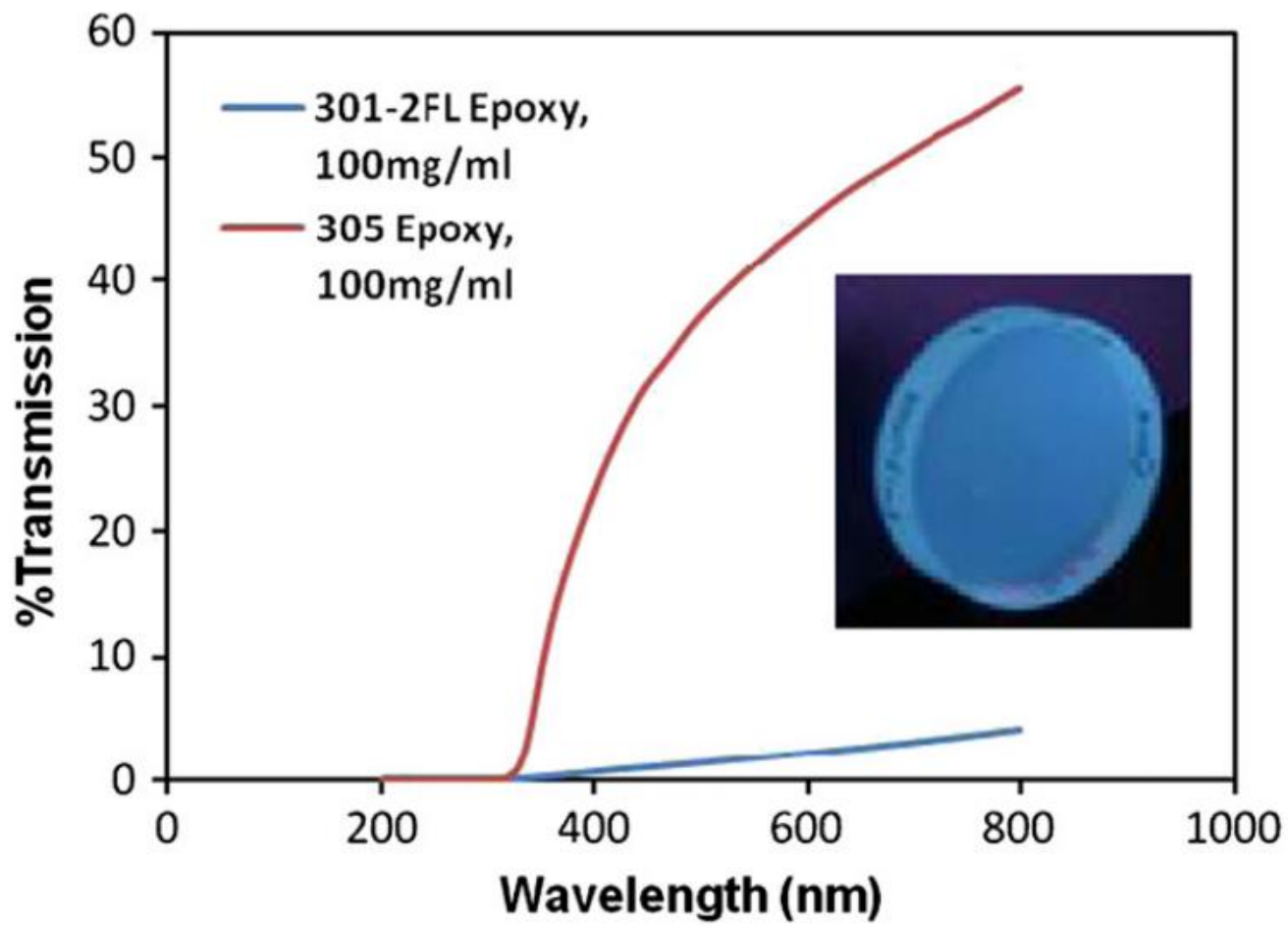

Figure 15. Transmission spectra of nanocomposite samples based on epoxy resin 301-2FL Epoxy ($n_m$ = 1.476) and 305 Epoxy ($n_m$ = 1.512) with $BaF_2$:Ce particles ($n_p$ = 1.464) (refractive indices are given for a light wavelength of 589 nm). (Reprinted with permission from Ref. [93]. Copyright with license Id: 4964090279399.)

Various synthesis methods allow obtaining nanoparticles in small size. These are mainly chemical methods based on the formation of nanoparticles in liquid media [127], [128] and physical methods based on the extraction of particles from sintered targets by laser ablation [129] or electron pulsed evaporation [130].

The main problem associated with the use of nanoparticles is their agglomeration. Due to the large specific surface area, the interaction between such particles is strong due to the van der Waals forces and large surface energy, which facilitates the formation of agglomerates [131]. The agglomeration causes the increase of the filler particle size, which in turn, according to formula (4), leads to the increase of the optical losses associated with Rayleigh scattering. In this regard, one of the important tasks is to prevent particle agglomeration and ensure their uniform distribution in the scintillator matrix. To solve this problem, as a rule, it is necessary to modify the surface of the particles, usually with the aid of a surfactant or a capping agent, which form a protective shell of molecules on the surface of nanoparticles and can also take part in polymerization.

A molecule of a surfactant consists of two components: a polar group that binds to the surface of nanoparticle and a hydrophobic tail that can participate in the polymerization process or form a bond with a polymer molecule. Anchor polar groups include amines ($-NH_2$) [132], [110], thiols (-SH) [133], [134], [135] and others (carboxyl groups -COOH [136], sulfonic acid radicals (sulfonile) $-SO_2OH$ [137], phosphonates -PO ($OH_2$) [138]). A long saturated or unsaturated hydrocarbon chain serves as a hydrophobic tail [139]. A protective shell of organic molecules can be either created on the already synthesized NPs [140] or formed during synthesis [141]. One of the ways to form a protective shell is the dispergation of nanoparticles in a solution containing the appropriate compound. In this way, it is possible to create a protective shell of, e.g., oleic acids OAc molecules [142], [143], [103], [93] or OA [144]. Inorganic nanocrystals covered with the molecules of this compound become readily dispersible in weakly polar solvents, such as hexane, toluene, chloroform and monomers such as vinyl toluene. On the other hand, OAc molecules are weakly involved in polymerization processes. As a result, NPs with an olein shell are displaced by polymerizing molecules. After the completion of the polymerization processes, the filler particles are expelled onto the sample surface. To overcome this problem, other more suitable surfactants may be used. For example, the substitution of the OAc with BMEP (bis(2-(methacryloyloxy)ethyl)phosphate) made it possible to create nanocomposites based on the PVT matrix with a high transmittance of 83.7% (at a wavelength 550 nm) uniformly loaded with 20 wt% $HfO_2$ [144], 69.6% (for the 500 nm band) with 40 wt% $Gd_2O_3$ [145], 79.5% (for the 550 nm band) with 80 wt% $YbF_3$ [86] (see Figure 16).

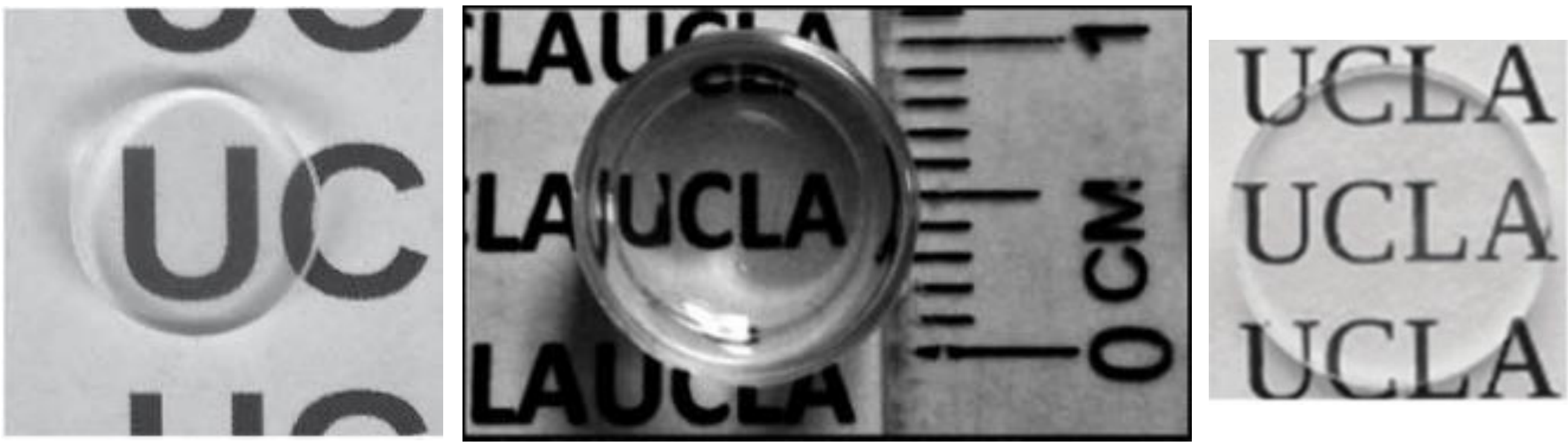

a) b) c)

Figure 16. The appearance of nanocomposites based on a PVT matrix with various fillers with a modified surface: a) 20 wt. % $HfO_2$, b) 40 wt. % $Gd_2O_3$, c) 80 wt% $YbF_3$. (Reprinted with permission from Refs. [144], [145], [86]. Copyright with license Id: 4964090492640, 1122811-1, 1122811-2)

The most promising materials for the role of fillers are fluorides, for example, $BaF_2$ and $YbF_3$, which have a sufficiently high $Z_{eff}$ value, as well as refractive indices close to those polymers. The problem of modifying the surface of particles for better incorporation into polymer chains in order to increase the concentration of the filler and improve the optical and scintillation properties remains still crucial.

### 4.4 Phosphor

The main task of the phosphor, or luminescence dopant, is to convert the electronic excitations created by the filler particles into luminescence photons. The temporal characteristics of a nanocomposite material are largely determined by this very substance. Thus, a high rate of the luminescence process is a key condition in choosing an activator. Other important characteristics are high light yield and good solubility in the matrix. The felicitous combination of these features in the organic phosphors, which are widely used in industrial plastic scintillators, makes them promising activators for ultrafast nanocomposite scintillators.

Within the whole class of organic phosphors, the systems consisting of chains of benzene rings combined by single bonds are characterized by the highest scintillation efficiency [1]. In addition, it has been shown that the substitution of phenyl groups with oxazole and 1,3,4-oxadiazole groups increases the solubility of substances without deteriorating their luminescence [146]. To increase the Stokes shift in plastic scintillators, secondary phosphors (shifters, or spectrum shifters) with a luminescence band in the longer wavelength region can be used [146]. The main characteristics of several popular phosphors are given in Table 3.

Table 3. Parameters of several industrial organic phosphors (solvents in which measurements were carried out are indicated in brackets)

| Phosphors | Luminescence band maximum (average value), nm | Decay constant, ns | Solubility, g $L^{-1}$ |
|---|---|---|---|
| 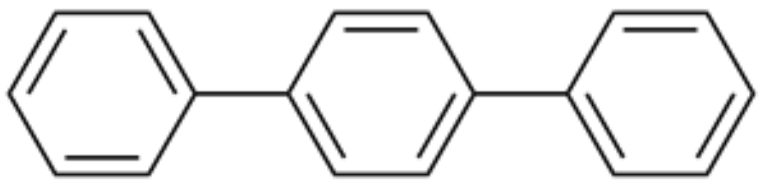 p-Terphenyl (p-TP) | 360 (toluene) [146] | 4.7 ±2.0 (xylene)[146] 6.6±1.8 (cyclohexylbenzene) [146], 2.9 (toluene) [146] | 8 (in toluene at 25 °C) [146] |
| 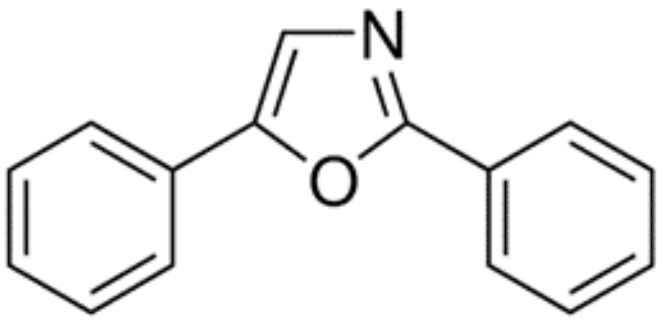 2,5-Diphenyloxazole (PPO) | 394 (toluene) [146] | 4.8±1.5 (xylene) [146] 5.8±1.5 (cyclohexylbenzene) [146] ≤3.0 (toluene) [146] | 270 (in toluene at 25 °C) [146] |
| 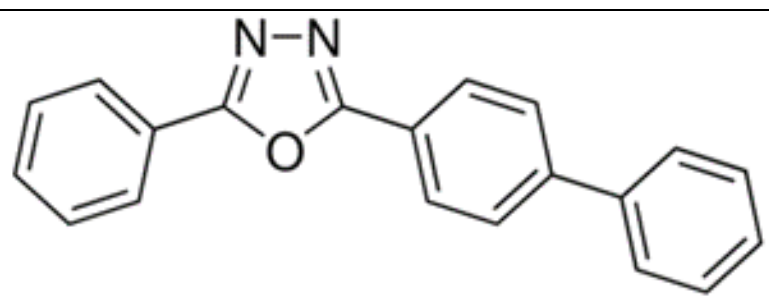 2-Phenyl-5-(4-biphenylyl)-1,3,4-oxadiazole (PBD) | 388 (toluene) [146] | 1.2 (toluene) [147] 2.8 (toluene) [146] | >20 (in toluene at 25 °C) [146] |
| 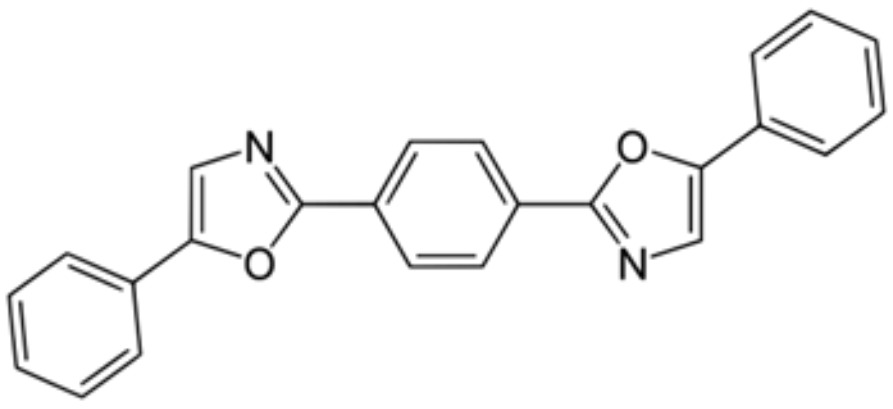 1,4-Bis(5-phenyl-2-oxazolyl)benzene (POPOP) | 444 (toluene) [146] | 1.5 (toluene) [147] | 2.2 (in toluene at 20 °C) [147] |
| 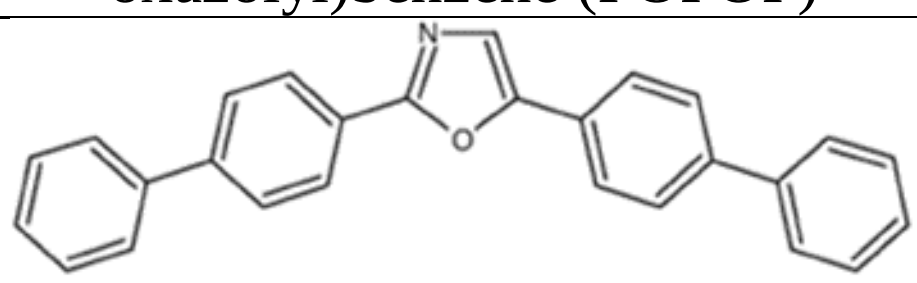 2,5-di(4-biphenylo)oxazole (BBO) | 434 (toluene) [146] | 1.4 (toluene) [147] | 3.1 (in toluene at 20 °C) [147] |

Resistance to ionizing radiation is also an important factor influencing the selection of a potential phosphor. It can be described using the parameter G – radiation chemical yield, or a number of damaged molecules under the action of a γ-photon with an energy of 100 eV. PPO and POPOP molecules have G values of 55 and 320 damages per 100 eV, respectively. The larger value for POPOP might be due to the large size of its molecule [148].

Due to the effect of quantum confinement, nanocrystals of some semiconductors acquire unique properties, including high quantum efficiency and ultrafast decay times [51]. In addition, such quantum dots are characterized by a wide absorption band and a narrow luminescence band. By varying the sizes of QDs, it possible to shift the maximum of the emission band. All these properties make QDs a promising candidate for the role of an activator or a shifter in composite scintillators [149]. In [150], transparent nanocomposites based on PMMA containing 0.5 wt.% CdTe QDs, as well as thin films based on PVA and CdTe with emission maxima at 565nm and 650 nm respectively (see Figure 17) were obtained.

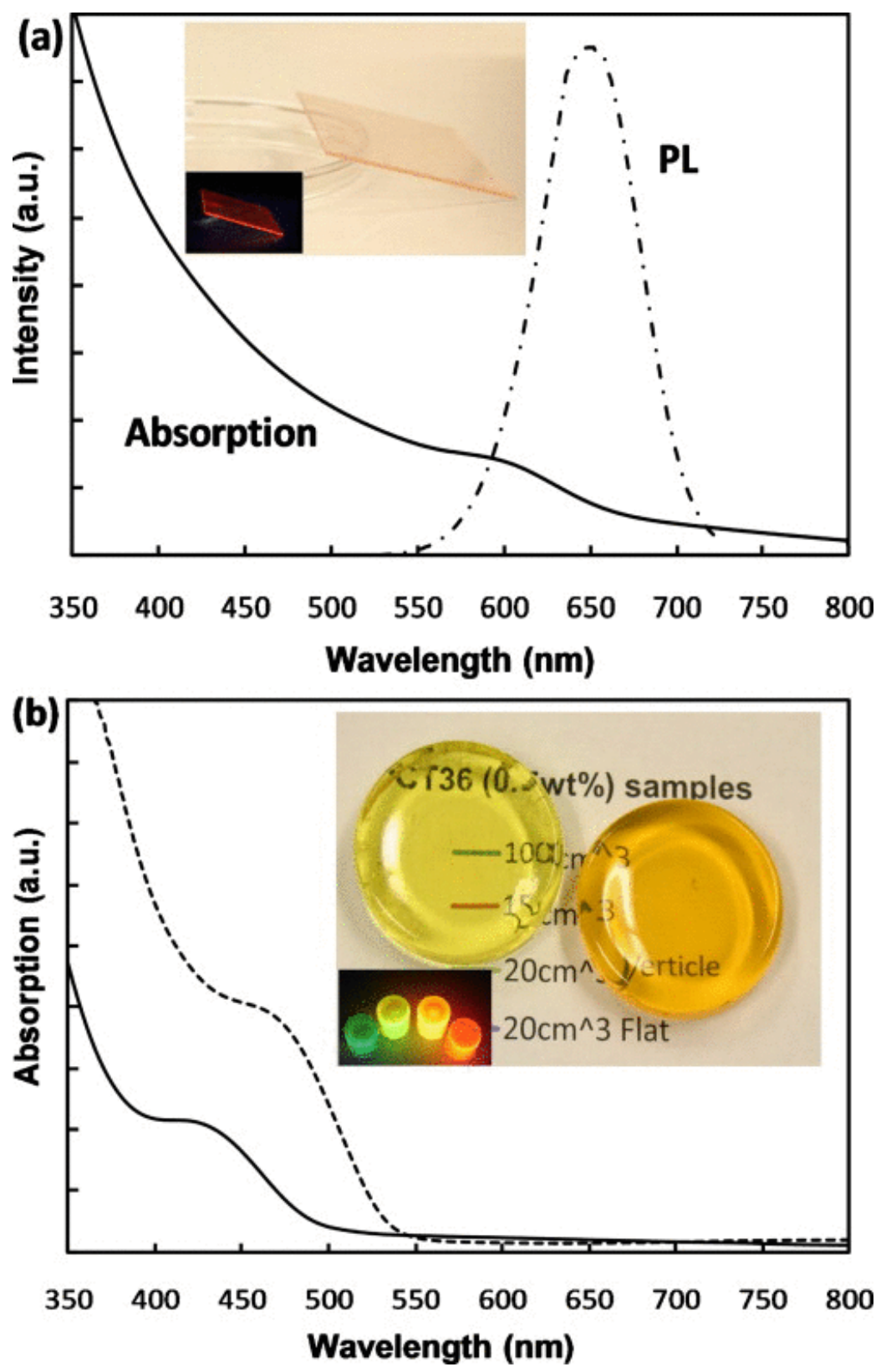

Figure 17. (a) Absorption and photoluminescence spectra of a CdTe/PVA nanocomposite film deposited on a glass substrate. The inset is a photo of a sample under room illumination and under UV irradiation with a wavelength of 365 nm. (b) Absorption spectra of composites based on PMMA with a filler CdTe (510 nm) 0.2wt% and filler CdTe (565 nm) 0.5 wt%. The inset shows photographs of the same samples under room light, as well as under UV illumination. (Reprinted with permission from Ref. [147]. Copyright with license Id: 4964100886753.)

CdSe/CdS core-crown nanoplatelets (ccNPLs) [151], [69] and CdSe/CdS/CdZnS core/crown/shell NPLs (ccsNPLs) as light emitters can be also proposed due to their very attractive properties. First, the bandgap of ccNPLs equals ~2.4 eV, while ccsNPLs display a slightly red shifted bandgap of 2.1 eV. When ccNPLs are used as a shifter, such bandgap is smaller than the energy of emission of most activators in plastics, facilitating one-way energy transfer from activator to ccNPLs.

Second, the ccNPLs have subnanosecond fluorescence lifetimes, while the emission decay of biexcitons efficiently formed via direct excitation by secondary electrons at higher excitation densities is even faster (<200ps), [69]. Third, fluorescence quantum efficiency (QE) of ccNPLs can exceed 60% [152], or, for ccsNPLs, even 90% due to encapsulation by a protective CdZnS shell [153], leading to low energy losses on the luminescence stage.

Organic molecules can also be used to increase the scintillation efficiency of composites with QDs. For example, a nanocomposite based on CdSe QDs and a polystyrene matrix demonstrates a higher scintillation efficiency when PPO is added [138]. The main problem in working with QDs as well as with nanoparticles of inorganic fillers is particle agglomeration [154]. In this regard, active research is underway on the dispersion of QDs inside a polymer matrix [120], [68], [155]. In [120], nanocomposite samples based on a polyurethane matrix and 1-thioglycerol stabilized QDs are presented (Figure 18).

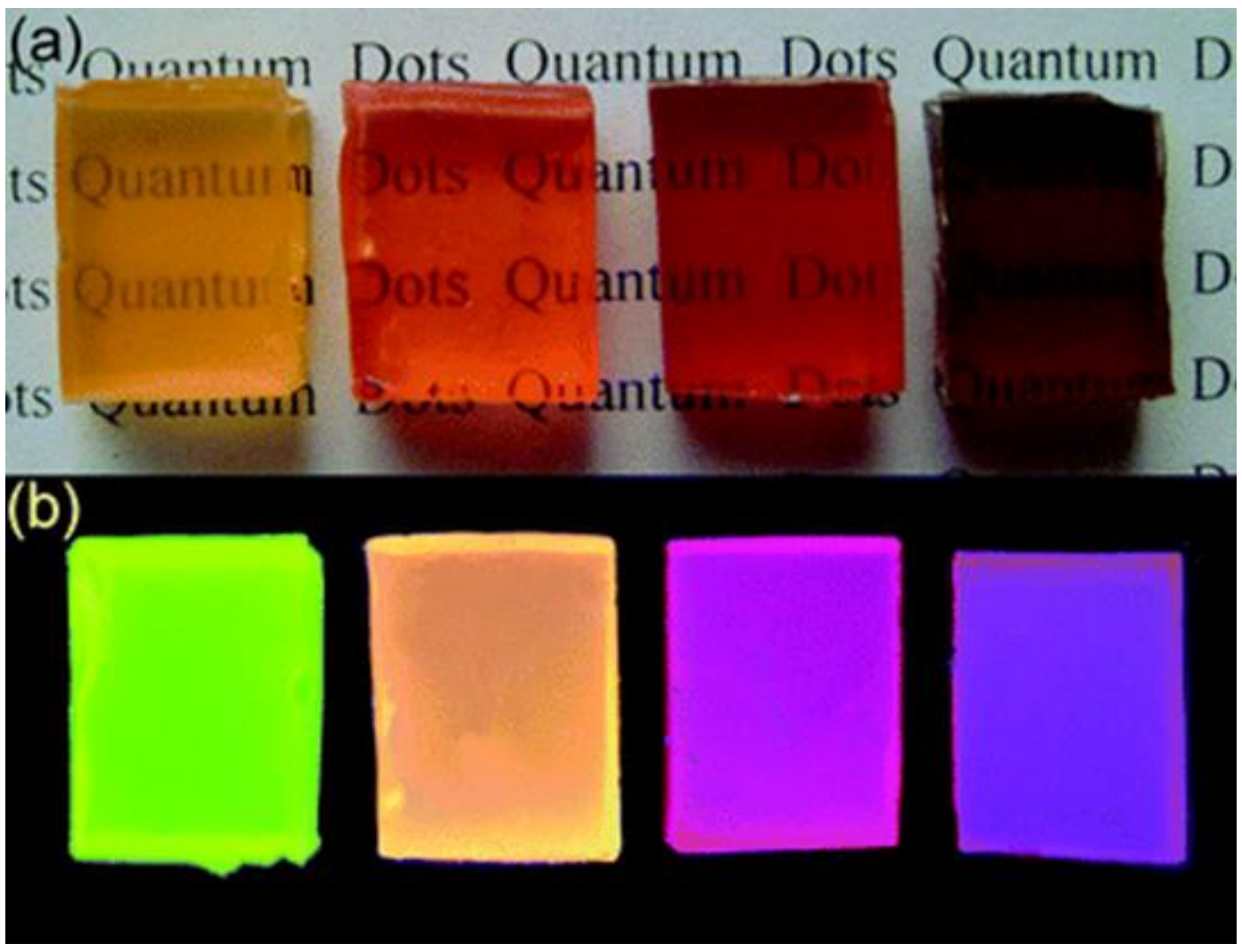

Figure 18. Polyurethane-based nanocomposites containing CdTe QDs with different size of particles. (Reprinted with permission from Ref. [120]. Copyright with license Id: 1122811-3.)

Figure 18 shows a scheme of possible energy transfer processes in scintillation nanocomposite materials with QD as ultrafast light emitters and inorganic nanoparticles as heavy absorber material. The bandgap ($E_g$) of the absorber should be larger than the bandgap of the plastic host, which facilitates one-way energy transfer. After gamma ray absorption, only a small fraction of its energy is released in the absorber nanoparticle due to its small size, while most of it is transferred to the plastics mainly via secondary electrons. Some of the secondary electrons may be absorbed by QDs directly, which happens at the sub-picosecond scale and is therefore very fast. A second energy transfer mechanism occurs via low-energy electronic excitations (excitons) in plastics. Exciton transfer takes place within a few picoseconds, essentially introducing no delays to the scintillation process. The rest of the energy is released in the plastic host itself and collected by the luminescent organic activator molecules, which are traditionally used as light emitters in conventional plastic scintillators. For the organic activator emission, QDs may act as an ultrafast wavelength shifter, increasing effective Stokes shift and minimizing re-absorption.

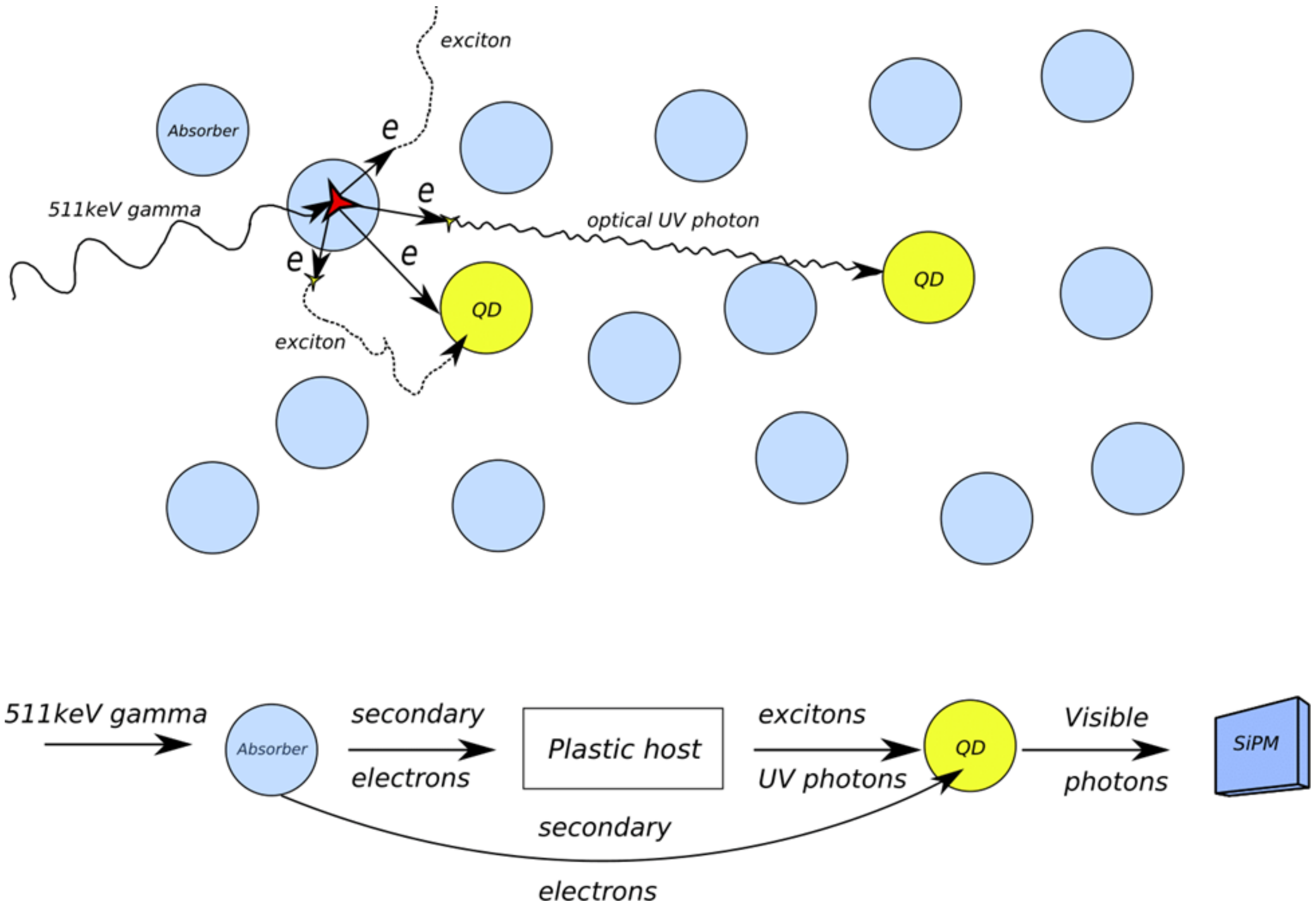

Figure 19. The scheme of hybrid scintillation nanocomposite material with QDs (top) and the energy transfer stages in the scintillation process (bottom).

The literature review on phosphors for organic scintillators showed that the technology of introducing a phosphor into a matrix and the mechanisms of energy transfer between organic phosphors and a polymer have been well studied. However, currently there is no information in the literature concerning systematic studies of the interaction of these components with inorganic particles possessing a high $Z_{eff}$ value (fillers). In this regard, it can be concluded that for the successful creation of nanocomposite scintillation materials with ultrafast decay, more detailed studies of the processes of energy exchange and conversion in heterogeneous structures containing two or more organic and inorganic components with different particles sizes are required.

**Conclusions**

Further development of diagnostics and therapy of diseases by nuclear medicine methods and research in the field of high-energy physics became almost impossible without the creation of conceptually new scintillation materials. These materials are subjects to increased requirements, unattainable for single-component systems. In particular, such requirements as fast timing response, high luminescence intensity, high stopping properties in relation to X-ray and gamma radiation, high optical transparency, radiation resistance, low cost of production and operation of scintillators have to be taken into consideration. The fulfilling of all these requirements for scintillation material will lead to a significant increase in the time and spatial resolution of the devices using TOF methods.

Research in this area has been intensified in recent years. New single-crystal and polycrystalline materials are being developed along with various structures which combine the properties of inorganic and organic substances. The most promising candidates are materials based on polymer matrices containing an inorganic filler with a high $Z_{eff}$ value and a dopant with a short decay time. Fast rise and decay times of the emission of QD and organic activators allow one to expect for nanocomposite scintillators CTR values of the order of tens picoseconds, that are characteristic for plastic scintillators. In addition, loading with a heavy inorganic filler can increase the efficiency of energy absorption as it has been shown for a number of compounds.

Despite the listed advantages, nanocomposite scintillators have also a number of drawbacks that limit their application. The $Z_{eff}$ value cannot exceed those of inorganic crystals. There is also a limitation of possible concentration of different additivities, which is set by matrix properties. At high concentrations, the polymerization reaction may be incomplete, which leads to a decrease in mechanical and temperature stability as well as radiation hardness. Besides, due to light scattering optical losses of such compounds are higher than in plastic scintillators and bulk inorganic crystals. As shown above, systematic studies of the processes occurring in heterogeneous nanocomposite structures are very scarce. Many problems associated with the synthesis of composite materials and understanding of the processes of creation and migration of electronic excitations in the nanocomposite materials remain unsolved. Nevertheless, the search for the most efficient combinations of components for nanocomposite scintillation detectors is at full throttle.

**Acknowledgement**

Authors thank Minobrnauki project FEUZ-2020-0059 and Estonian Research Council (grants PRG629 and PRG111) for financial support. Authors are also grateful for partial support from the European Regional Development Fund (DoRA Pluss program) and the ERDF funding in Estonia granted to the Center of Excellence TK141 “Advanced materials and high-technology devices for sustainable energetics, sensorics and nanoelectronics” (project No. 2014-2020.4.01.15-0011).

**References**

[1] J.B. Birks, the Scintillation Process in Organic Materials—I, Theory Pract. Scintill. Count. (1964) 39–67. https://doi.org/10.1016/b978-0-08-010472-0.50008-2.

[2] Y.N. Kharzheev, Scintillation counters in modern high-energy physics experiments, Phys. Part. Nucl. 46 (2015) 678–728. https://doi.org/10.1134/S1063779615040048.

[3] C.W.E. Van Eijk, Inorganic scintillators in medical imaging detectors, Nucl. Instruments Methods Phys. Res. Sect. A Accel. Spectrometers, Detect. Assoc. Equip. 509 (2003) 17–25. https://doi.org/10.1016/S0168-9002(03)01542-0.

[4] J. Glodo, Y. Wang, R. Shawgo, C. Brecher, R.H. Hawrami, J. Tower, K.S. Shah, New Developments in Scintillators for Security Applications, Phys. Procedia. 90 (2017) 285–290. https://doi.org/10.1016/j.phpro.2017.09.012.

[5] T. Jones, D. Townsend, History and future technical innovation in positron emission tomography, J. Med. Imaging. 4 (2017) 011013. https://doi.org/10.1117/1.jmi.4.1.011013.

[6] S. Vandenberghe, E. Mikhaylova, E. D'Hoe, P. Mollet, J.S. Karp, Recent developments in time-of-flight PET, EJNMMI Phys. 3 (2016). https://doi.org/10.1186/s40658-016-0138-3.

[7] I. Rausch, A. Ruiz, I. Valverde-Pascual, J. Cal-González, T. Beyer, I. Carrio, Performance evaluation of the VereoS PET/CT system according to the NEMA NU2-2012 standard, J. Nucl. Med. 60 (2019) 561–567. https://doi.org/10.2967/jnumed.118.215541.

[8] A.M. Grant, T.W. Deller, M.M. Khalighi, S.H. Maramraju, G. Delso, C.S. Levin, NEMA NU 2-2012 performance studies for the SiPM-based ToF-PET component of the GE SIGNA PET/MR system, Med. Phys. 43 (2016) 2334–2343. https://doi.org/10.1118/1.4945416.

[9] J. Van Sluis, J. De Jong, J. Schaar, W. Noordzij, P. Van Snick, R. Dierckx, R. Borra, A. Willemsen, R. Boellaard, Performance characteristics of the digital biograph vision PET/CT system, J. Nucl. Med. 60 (2019) 1031–1036. https://doi.org/10.2967/jnumed.118.215418.

[10] P. Lecoq, Pushing the Limits in Time-of-Flight PET Imaging, IEEE Trans. Radiat. Plasma Med. Sci. 1 (2017) 473–485. https://doi.org/10.1109/TRPMS.2017.2756674.

[11] The 10 ps Challenge website, (2020). https://the10ps-challenge.org/ (accessed May 31, 2021).

[12] S.R. Cherry, T. Jones, J.S. Karp, J. Qi, W.W. Moses, R.D. Badawi, Total-Body PET: Maximizing Sensitivity to Create New Opportunities for Clinical Research and Patient Care, J. Nucl. Med. 59 (2018) 3–12. https://doi.org/10.2967/jnumed.116.184028.

[13] P. Lecoq, C. Morel, J.O. Prior, D. Visvikis, S. Gundacker, E. Auffray, P. Križan, R.M. Turtos, D. Thers, E. Charbon, J. Varela, C. De La Taille, A. Rivetti, D. Breton, J.F. Pratte, J. Nuyts, S. Surti, S. Vandenberghe, P. Marsden, K. Parodi, J.M. Benlloch, M. Benoit, Roadmap toward the 10 ps time-of-flight PET challenge, Phys. Med. Biol. 65 (2020). https://doi.org/10.1088/1361-6560/ab9500.

[14] D.M. Binkley, Optimization Of Scintillation-Detector Timing Systems Using Monte Carlo Analysis, IEEE Trans. Nucl. Sci. 41 (1994) 386–393. https://doi.org/10.1109/23.281528.

[15] S. Gundacker, R. Martinez Turtos, N. Kratochwil, R.H. Pots, M. Paganoni, P. Lecoq, E. Auffray,

Experimental time resolution limits of modern SiPMs and TOF-PET detectors exploring different scintillators and Cherenkov emission, Phys. Med. Biol. 65 (2020) 025001. https://doi.org/10.1088/1361-6560/ab63b4.

[16] Y. Shao, A new timing model for calculating the intrinsic timing resolution of a scintillator detector, Phys. Med. Biol. 52 (2007) 1103–1117. https://doi.org/10.1088/0031-9155/52/4/016.

[17] C. Cuesta, M.A. Oliván, J. Amaré, S. Cebrián, E. García, C. Ginestra, M. Martínez, Y. Ortigoza, A. Ortiz De Solórzano, C. Pobes, J. Puimedón, M.L. Sarsa, J.A. Villar, P. Villar, Slow scintillation time constants in NaI(Tl) for different interacting particles, Opt. Mater. (Amst). 36 (2013) 316–320. https://doi.org/10.1016/j.optmat.2013.09.015.

[18] O.H. Nestor, C.Y. Huang, Bismuth germanate: A high-Z gamma-ray and charged particle detector, IEEE Trans. Nucl. Sci. 22 (1975) 68–71. https://doi.org/10.1109/TNS.1975.4327617.

[19] I. Holl, E. Lorenz, G. Mageras, A measurement of the light yield of common inorganic scintillators, IEEE Trans. Nucl. Sci. 35 (1988) 105–109.

[20] P. Lecoq, A. Gektin, M. Korzhik, Inorganic Scintillators for Detector Systems, Springer International Publishing, Cham, 2017. https://doi.org/10.1007/978-3-319-45522-8.

[21] S.E. Derenzo, Measurements of the intrinsic rise times of common inorganic scintillators, IEEE Trans. Nucl. Sci. 47 (2000) 860–864. https://doi.org/10.1109/23.856531.

[22] M. Moszyński, C. Gresset, J. Vacher, R. Odru, Timing properties of BGO scintillator, Nucl. Instruments Methods. 188 (1981) 403–409. https://doi.org/10.1016/0029-554X(81)90521-8.

[23] W. Chewpraditkul, L. Swiderski, M. Moszynski, T. Szczesniak, A. Syntfeld-Kazuch, C. Wanarak, P. Limsuwan, Scintillation properties of LuAG:Ce, YAG:Ce and LYSO:Ce crystals for gamma-ray detection, IEEE Trans. Nucl. Sci. 56 (2009) 3800–3805. https://doi.org/10.1109/TNS.2009.2033994.

[24] P. Dorenbos, C. Van Eijk, Proceedings of the international conference on inorganic scintillators and their applications, Delft, The Netherlands, 28 August-September 1, 1995: SCINT 95, in: Cryst. Growth Scintill. Prop. Rare Earth Oxyorthosilicates, 1996: p. 309.

[25] L. Pidol, A. Kahn-Harari, B. Viana, E. Virey, B. Ferrand, P. Dorenbos, J.T.M. De Haas, C.W.E. Van Eijk, High efficiency of lutetium silicate scintillators, Ce-doped LPS and LYSO crystals, IEEE Nucl. Sci. Symp. Conf. Rec. 2 (2003) 886–890. https://doi.org/10.1109/nssmic.2003.1351838.

[26] J. Glodo, W.W. Moses, W.M. Higgins, E. V.D. Van Loef, P. Wong, S.E. Derenzo, M.J. Weber, K.S. Shah, Effects of Ce concentration on scintillation properties of LaBr 3:Ce, IEEE Trans. Nucl. Sci. 52 (2005) 1805–1808. https://doi.org/10.1109/TNS.2005.856906.

[27] A. Kuhn, S. Surti, K.S. Shah, J.S. Karp, Investigation of LaBr3 detector timing resolution, IEEE Nucl. Sci. Symp. Conf. Rec. 4 (2005) 2022–2026.

https://doi.org/10.1109/NSSMIC.2005.1596730.

[28] E. V.D. Van Loef, P. Dorenbos, C.W.E. Van Eijk, K.W. Krämer, H.U. Güdel, Scintillation properties of LaBr3:Ce3+ crystals: Fast, efficient and high-energy-resolution scintillators, Nucl. Instruments Methods Phys. Res. Sect. A Accel. Spectrometers, Detect. Assoc. Equip. 486 (2002) 254–258. https://doi.org/10.1016/S0168-9002(02)00712-X.

[29] S. Aldawood, I. Castelhano, R. Gernhäuser, H. Van Der Kolff, C. Lang, S. Liprandi, R. Lutter, L. Maier, T. Marinšek, D.R. Schaart, K. Parodi, P.G. Thirolf, Comparative characterization study of a LaBr3(Ce) scintillation crystal in two surface wrapping scenarios: Absorptive and reflective, Front. Oncol. 5 (2015) 1–9. https://doi.org/10.3389/fonc.2015.00270.

[30] M. Laval, M. Moszyński, R. Allemand, E. Cormoreche, P. Guinet, R. Odru, J. Vacher, Barium fluoride - Inorganic scintillator for subnanosecond timing, Nucl. Instruments Methods Phys. Res. 206 (1983) 169–176. https://doi.org/10.1016/0167-5087(83)91254-1.

[31] T.K. Lewellen, Recent developments in PET detector technology, Phys. Med. Biol. 53 (2008). https://doi.org/10.1088/0031-9155/53/17/R01.

[32] T. Rostomyan, E. Cline, I. Lavrukhin, H. Atac, A. Atencio, J.C. Bernauer, W.J. Briscoe, D. Cohen, E.O. Cohen, C. Collicott, K. Deiters, S. Dogra, E. Downie, W. Erni, I.P. Fernando, A. Flannery, T. Gautam, D. Ghosal, R. Gilman, A. Golossanov, J. Hirschman, M. Kim, M. Kohl, B. Krusche, L. Li, W. Lin, A. Liyanage, W. Lorenzon, P. Mohanmurthy, J. Nazeer, P. Or, T. Patel, E. Piasetzky, N. Pilip, H. Reid, P.E. Reimer, G. Ron, E. Rooney, Y. Shamai, P. Solazzo, S. Strauch, D. Vidne, N. Wuerfel, Timing detectors with SiPM read-out for the MUSE experiment at PSI, Nucl. Instruments Methods Phys. Res. Sect. A Accel. Spectrometers, Detect. Assoc. Equip. 986 (2021). https://doi.org/10.1016/j.nima.2020.164801.

[33] X. Wen, J.P. Hayward, Time Resolution Measurements of EJ-232Q with Single- And Dual-Sided Readouts, IEEE Trans. Nucl. Sci. 67 (2020) 2081–2088. https://doi.org/10.1109/TNS.2020.3010469.

[34] S. Gundacker, E. Auffray, K. Pauwels, P. Lecoq, Measurement of intrinsic rise times for various L(Y)SO and LuAG scintillators with a general study of prompt photons to achieve 10 ps in TOF-PET, Phys. Med. Biol. 61 (2016) 2802–2837. https://doi.org/10.1088/0031-9155/61/7/2802.

[35] A. Artikov, J. Budagov, I. Chirikov-Zorin, D. Chokheli, M. Lyablin, G. Bellettini, A. Menzione, S. Tokar, N. Giokaris, A. Manousakis-Katsikakis, Properties of the Ukraine polystyrene-based plastic scintillator UPS 923A, Nucl. Instruments Methods Phys. Res. Sect. A Accel. Spectrometers, Detect. Assoc. Equip. 555 (2005) 125–131. https://doi.org/10.1016/j.nima.2005.09.021.

[36] R.M. Turtos, S. Gundacker, E. Auffray, P. Lecoq, Towards a metamaterial approach for fast timing in PET: Experimental proof-of-concept, Phys. Med. Biol. 64 (2019).

https://doi.org/10.1088/1361-6560/ab18b3.

[37] J.B. Birks, Applications of Organic Scintillators, Theory Pract. Scintill. Count. (1964) 354–430. https://doi.org/10.1016/b978-0-08-010472-0.50015-x.

[38] P. Adamson, K.V. Alexandrov, W.W.M. Allison, G.J. Alner, I. Ambats, B. Anderson, D.F. Anderson, Y. Antipov, C. Arroyo, D.S. Ayres, B. Baller, B. Barish, P.D. Barnes Jr., W.L. Barrett, R.H. Bernstein, R.E. Blair, V. Bocean, D.J. Boehnlein, D. Bogert, P.M. Border, C. Bower, J. Byrne, T. Chase, S. Chernichenko, S. Childress, B.C. Choudhary, J.H. Cobb, J.D. Cossairt, H. Courant, P. Cushman, J.W. Dawson, A. Denisov, P.J. Dervan, N. Diaczenko, G. Drake, M. Drew, A. Durum, R. Edgecock, V.K. Ermilova, O. Eroshin, R. Fahrutdinov, G.J. Feldman, T.H. Fields, D. Fujino, H.R. Gallagher, M. Gebhard, Y. Gilitsky, M.C. Goodman, Y. Gorin, Y. Gornushkin, M. Graham, N. Grossman, V.J. Guarino, Y. Gutnikov, R. Halsall, J. Hanson, P.G. Harris, E. Hartouni, R. Hatcher, R. Heinz, K. Heller, N. Hill, T. Hu, Y. Huang, J. Hylen, M. Ignatenko, G. Irwin, C. James, T. Joffe-Minor, T. Kafka, T. Kirichenko, J. Kilmer, H. Kim, V. Kochetkov, G. Koizumi, Z. Krumstein, N.N. Kundu, A. Ladran, Y.F. Lai, K. Lang, C. Laughton, P.J. Litchfield, N.P. Longley, P. Lucas, S. Madani, V. Makeev, W.A. Mann, H.S. Mao, M.L. Marshak, D. Maxam, E.N. May, J.R. Meier, E. Melnikov, G.I. Merzon, D.G. Michael, R.H. Milburn, L. Miller, W.H. Miller, S.J. Mishra, J. Morfin, L. Mualem, S. Mufson, J. Musser, A. Napier, J.K. Nelson, H. Newman, A. Nozdrin, S. O'Day, W. Oliver, A. Olshevski, V. Onuchin, A. Para, G.F. Pearce, C.W. Peck, C. Perry, E.A. Peterson, A. Petruhin, D.A. Petyt, R.. Plunkett, L.E. Price, M. Proga, D.R. Pushka, R.A. Rameika, A.L. Read, K. Ruddick, R. Rusack, V.A. Ryabov, A. Sadovski, M. Sanchez, J. Schneps, P.V. Schoessow, R. Schwienhorst, V. Semenov, B.W. Shen, P.D. Shield, R. Shivane, A. Sisakian, W. Smart, V. Smirnitsky, V. Smotriaev, T. Soesbe, A. Soldatov, R. Soltz, N.I. Starkov, R. Talaga, J. Thomas, J.L. Thron, D. Tovee, J. Trevor, I. Trostin, V.A. Tsarev, L.R. Turner, N. Tyurin, M. Vakili, L. Wai, D. Wall, R.C. Webb, A. Wehmann, N. West, R.F. White, S.G. Wojcicki, D. Wright, X.M. Xia, W.G. Yan, J.C. Yun, Technical Design Report, 1998.

[39] L. Aliaga, L. Bagby, B. Baldin, A. Baumbaugh, A. Bodek, R. Bradford, W.K. Brooks, D. Boehnlein, S. Boyd, H. Budd, A. Butkevich, D.A. Martinez Caicedo, C.M. Castromonte, M.E. Christy, J. Chvojka, H. Da Motta, D.S. Damiani, I. Danko, M. Datta, J. Devan, E. Draeger, S.A. Dytman, G.A. Díaz, B. Eberly, D.A. Edmondson, J. Felix, L. Fields, G.A. Fiorentini, R.S. Flight, A.M. Gago, H. Gallagher, C.A. George, J.A. Gielata, C. Gingu, R. Gran, J. Grange, N. Grossman, D.A. Harris, J. Heaton, A. Higuera, J.A. Hobbs, I.J. Howley, K. Hurtado, M. Jerkins, T. Kafka, M.O. Kantner, C. Keppel, J. Kilmer, M. Kordosky, A.H. Krajeski, G.J. Kumbartzki, H. Lee, A.G. Leister, G. Locke, G. Maggi, E. Maher, S. Manly, W.A. Mann, C.M. Marshall, K.S. McFarland, C.L. McGivern, A.M. McGowan, A. Mislivec, J.G. Morfín, J. Mousseau, D. Naples, J.K. Nelson, G. Niculescu, I. Niculescu, C.D. O'Connor, N. Ochoa, J. Olsen, B. Osmanov, J. Osta, J.L.

Palomino, V. Paolone, J. Park, G.N. Perdue, C. Peña, A. Pla-Dalmau, L. Rakotondravohitra, R.D. Ransome, H. Ray, L. Ren, P. Rubinov, C. Rude, K.E. Sassin, H. Schellman, D.W. Schmitz, R.M. Schneider, E.C. Schulte, C. Simon, F.D. Snider, M.C. Snyder, C.J. Solano Salinas, N. Tagg, B.G. Tice, R.N. Tilden, J.P. Velásquez, T. Walton, A. Westerberg, J. Wolcott, B.A. Wolthuis, N. Woodward, T. Wytock, G. Zavala, H.B. Zeng, D. Zhang, L.Y. Zhu, B.P. Ziemer, Design, calibration, and performance of the MINERvA detector, Nucl. Instruments Methods Phys. Res. Sect. A Accel. Spectrometers, Detect. Assoc. Equip. 743 (2014) 130–159. https://doi.org/10.1016/j.nima.2013.12.053.

[40] A. Lucotte, S. Bondil, K. Borer, J.E. Campagne, A. Cazes, M. Hess, C. De La Taille, G. Martin-Chassard, L. Raux, J.P. Repellin, A front-end read out chip for the OPERA scintillator tracker, Nucl. Instruments Methods Phys. Res. Sect. A Accel. Spectrometers, Detect. Assoc. Equip. 521 (2004) 378–392. https://doi.org/10.1016/j.nima.2003.10.104.

[41] V.M. Abazov, B. Abbott, M. Abolins, B.S. Acharya, D.L. Adams, M. Adams, T. Adams, M. Agelou, J.L. Agram, S.N. Ahmed, S.H. Ahn, M. Ahsan, G.D. Alexeev, G. Alkhazov, A. Alton, G. Alverson, G.A. Alves, M. Anastasoaie, T. Andeen, J.T. Anderson, S. Anderson, B. Andrieu, R. Angstadt, V. Anosov, Y. Arnoud, M. Arov, A. Askew, B. Åsman, A.C.S. Assis Jesus, O. Atramentov, C. Autermann, C. Avila, L. Babukhadia, T.C. Bacon, F. Badaud, A. Baden, S. Baffioni, L. Bagby, B. Baldin, P.W. Balm, P. Banerjee, S. Banerjee, E. Barberis, O. Bardon, W. Barg, P. Bargassa, P. Baringer, C. Barnes, J. Barreto, J.F. Bartlett, U. Bassler, M. Bhattacharjee, M.A. Baturitsky, D. Bauer, A. Bean, B. Baumbaugh, S. Beauceron, M. Begalli, F. Beaudette, M. Begel, A. Bellavance, S.B. Beri, G. Bernardi, R. Bernhard, I. Bertram, M. Besançon, A. Besson, R. Beuselinck, D. Beutel, V.A. Bezzubov, P.C. Bhat, V. Bhatnagar, M. Binder, C. Biscarat, A. Bishoff, K.M. Black, I. Blackler, G. Blazey, F. Blekman, S. Blessing, D. Bloch, U. Blumenschein, E. Bockenthien, V. Bodyagin, A. Boehnlein, O. Boeriu, T.A. Bolton, P. Bonamy, D. Bonifas, F. Borcherding, G. Borissov, K. Bos, T. Bose, C. Boswell, M. Bowden, A. Brandt, G. Briskin, R. Brock, G. Brooijmans, A. Bross, N.J. Buchanan, D. Buchholz, M. Buehler, V. Buescher, S. Burdin, S. Burke, T.H. Burnett, E. Busato, C.P. Buszello, D. Butler, J.M. Butler, J. Cammin, S. Caron, J. Bystricky, L. Canal, F. Canelli, W. Carvalho, B.C.K. Casey, D. Casey, N.M. Cason, H. Castilla-Valdez, S. Chakrabarti, D. Chakraborty, K.M. Chan, A. Chandra, D. Chapin, F. Charles, E. Cheu, L. Chevalier, E. Chi, R. Chiche, D.K. Cho, R. Choate, S. Choi, B. Choudhary, S. Chopra, J.H. Christenson, T. Christiansen, L. Christofek, I. Churin, G. Cisko, D. Claes, A.R. Clark, B. Clément, C. Clément, Y. Coadou, D.J. Colling, L. Coney, B. Connolly, M. Cooke, W.E. Cooper, D. Coppage, M. Corcoran, J. Coss, A. Cothenet, M.C. Cousinou, B. Cox, S. Crépé-Renaudin, M. Cristetiu, M.A.C. Cummings, D. Cutts, H. da Motta, M. Das, B. Davies, G. Davies, G.A. Davis, W. Davis, K. De, P. de Jong, S.J. de Jong, E. De La Cruz-Burelo, C. De La Taille, C. De Oliveira

Martins, S. Dean, J.D. Degenhardt, F. Déliot, P.A. Delsart, K. Del Signore, R. DeMaat, M. Demarteau, R. Demina, P. Demine, D. Denisov, S.P. Denisov, S. Desai, H.T. Diehl, M. Diesburg, M. Doets, M. Doidge, H. Dong, S. Doulas, L. V. Dudko, L. Duflot, S.R. Dugad, A. Duperrin, O. Dvornikov, J. Dyer, A. Dyshkant, M. Eads, D. Edmunds, T. Edwards, J. Ellison, J. Elmsheuser, J.T. Eltzroth, V.D. Elvira, S. Eno, P. Ermolov, O. V. Eroshin, J. Estrada, D. Evans, H. Evans, A. Evdokimov, V.N. Evdokimov, J. Fagan, J. Fast, S.N. Fatakia, D. Fein, L. Feligioni, A. V. Ferapontov, T. Ferbel, M.J. Ferreira, F. Fiedler, F. Filthaut, W. Fisher, H.E. Fisk, I. Fleck, T. Fitzpatrick, E. Flattum, F. Fleuret, R. Flores, J. Foglesong, M. Fortner, H. Fox, C. Franklin, W. Freeman, S. Fu, S. Fuess, T. Gadfort, C.F. Galea, E. Gallas, E. Galyaev, M. Gao, C. Garcia, A. Garcia-Bellido, J. Gardner, V. Gavrilov, A. Gay, P. Gay, D. Gelé, R. Gelhaus, K. Genser, C.E. Gerber, Y. Gershtein, D. Gillberg, G. Geurkov, G. Ginther, B. Gobbi, K. Goldmann, T. Golling, N. Gollub, V. Golovtsov, B. Gómez, G. Gomez, R. Gomez, R. Goodwin, Y. Gornushkin, K. Gounder, A. Goussiou, D. Graham, G. Graham, P.D. Grannis, K. Gray, S. Greder, D.R. Green, J. Green, J.A. Green, H. Greenlee, Z.D. Greenwood, E.M. Gregores, S. Grinstein, P. Gris, J.F. Grivaz, L. Groer, S. Grünendahl, M.W. Grünewald, W. Gu, J. Guglielmo, A. Gupta, S.N. Gurzhiev, G. Gutierrez, P. Gutierrez, A. Haas, N.J. Hadley, E. Haggard, H. Haggerty, S. Hagopian, I. Hall, R.E. Hall, C. Han, L. Han, R. Hance, K. Hanagaki, P. Hanlet, S. Hansen, K. Harder, A. Harel, R. Harrington, J.M. Hauptman, R. Hauser, C. Hays, J. Hays, E. Hazen, T. Hebbeker, C. Hebert, D. Hedin, J.M. Heinmiller, A.P. Heinson, U. Heintz, C. Hensel, G. Hesketh, M.D. Hildreth, R. Hirosky, J.D. Hobbs, B. Hoeneisen, M. Hohlfeld, S.J. Hong, R. Hooper, S. Hou, P. Houben, Y. Hu, J. Huang, Y. Huang, V. Hynek, D. Huffman, I. Iashvili, R. Illingworth, A.S. Ito, S. Jabeen, Y. Jacquier, M. Jaffré, S. Jain, V. Jain, K. Jakobs, R. Jayanti, A. Jenkins, R. Jesik, Y. Jiang, K. Johns, M. Johnson, P. Johnson, A. Jonckheere, P. Jonsson, H. Jöstlein, N. Jouravlev, M. Juarez, A. Juste, A.P. Kaan, M.M. Kado, D. Käfer, W. Kahl, S. Kahn, E. Kajfasz, A.M. Kalinin, J. Kalk, S.D. Kalmani, D. Karmanov, J. Kasper, I. Katsanos, D. Kau, R. Kaur, Z. Ke, R. Kehoe, S. Kermiche, S. Kesisoglou, A. Khanov, A. Kharchilava, Y.M. Kharzheev, H. Kim, K.H. Kim, T.J. Kim, N. Kirsch, B. Klima, M. Klute, J.M. Kohli, J.P. Konrath, E. V. Komissarov, M. Kopal, V.M. Korablev, A. Kostritski, J. Kotcher, B. Kothari, A. V. Kotwal, A. Koubarovsky, A. V. Kozelov, J. Kozminski, A. Kryemadhi, O. Kouznetsov, J. Krane, N. Kravchuk, K. Krempetz, J. Krider, M.R. Krishnaswamy, S. Krzywdzinski, M. Kubantsev, R. Kubinski, N. Kuchinsky, S. Kuleshov, Y. Kulik, A. Kumar, S. Kunori, A. Kupco, T. Kurča, J. Kvita, V.E. Kuznetsov, R. Kwarciany, S. Lager, N. Lahrichi, G. Landsberg, M. Larwill, P. Laurens, B. Lavigne, J. Lazoflores, A.C. Le Bihan, G. Le Meur, P. Lebrun, S.W. Lee, W.M. Lee, A. Leflat, C. Leggett, F. Lehner, R. Leitner, C. Leonidopoulos, J. Leveque, P. Lewis, J. Li, Q.Z. Li, X. Li, J.G.R. Lima, D. Lincoln, C. Lindenmeyer, S.L. Linn, J. Linnemann, V. V. Lipaev, R. Lipton, M.

Litmaath, J. Lizarazo, L. Lobo, A. Lobodenko, M. Lokajicek, A. Lounis, P. Love, J. Lu, H.J. Lubatti, A. Lucotte, L. Lueking, C. Luo, M. Lynker, A.L. Lyon, E. Machado, A.K.A. Maciel, R.J. Madaras, P. Mättig, C. Magass, A. Magerkurth, A.M. Magnan, M. Maity, N. Makovec, P.K. Mal, H.B. Malbouisson, S. Malik, V.L. Malyshev, V. Manakov, H.S. Mao, Y. Maravin, D. Markley, M. Markus, T. Marshall, M. Martens, M. Martin, G. Martin-Chassard, S.E.K. Mattingly, M. Matulik, A.A. Mayorov, R. McCarthy, R. McCroskey, M. McKenna, T. McMahon, D. Meder, H.L. Melanson, A. Melnitchouk, A. Mendes, D. Mendoza, L. Mendoza, X. Meng, Y.P. Merekov, M. Merkin, K.W. Merritt, A. Meyer, J. Meyer, M. Michaut, C. Miao, H. Miettinen, D. Mihalcea, V. Mikhailov, D. Miller, J. Mitrevski, N. Mokhov, J. Molina, N.K. Mondal, H.E. Montgomery, R.W. Moore, T. Moulik, G.S. Muanza, M. Mostafa, S. Moua, M. Mulders, L. Mundim, Y.D. Mutaf, P. Nagaraj, E. Nagy, M. Naimuddin, F. Nang, M. Narain, V.S. Narasimhan, A. Narayanan, N.A. Naumann, H.A. Neal, J.P. Negret, S. Nelson, R.T. Neuenschwander, P. Neustroev, C. Noeding, A. Nomerotski, S.F. Novaes, A. Nozdrin, T. Nunnemann, A. Nurczyk, E. Nurse, V. O’Dell, D.C. O’Neil, V. Oguri, D. Olis, N. Oliveira, B. Olivier, J. Olsen, N. Oshima, B.O. Oshinowo, G.J. Otero y Garzón, P. Padley, K. Papageorgiou, N. Parashar, J. Park, S.K. Park, J. Parsons, R. Partridge, N. Parua, A. Patwa, G. Pawloski, P.M. Perea, E. Perez, O. Peters, P. Pétroff, M. Petteni, L. Phaf, R. Piegaia, M.A. Pleier, P.L.M. Podesta-Lerma, V.M. Podstavkov, Y. Pogorelov, M.E. Pol, A. Pompoš, P. Polosov, B.G. Pope, E. Popkov, S. Porokhovoy, W.L. Prado da Silva, W. Pritchard, I. Prokhorov, H.B. Prosper, S. Protopopescu, M.B. Przybycien, J. Qian, A. Quadt, B. Quinn, E. Ramberg, R. Ramirez-Gomez, K.J. Rani, K. Ranjan, M.V.S. Rao, P.A. Rapidis, S. Rapisarda, J. Raskowski, P.N. Ratoff, R.E. Ray, N.W. Reay, R. Rechenmacher, L. V. Reddy, T. Regan, J.F. Renardy, S. Reucroft, J. Rha, M. Ridel, M. Rijssenbeek, I. Ripp-Baudot, F. Rizatdinova, S. Robinson, R.F. Rodrigues, M. Roco, C. Rotolo, C. Royon, P. Rubinov, R. Ruchti, R. Rucinski, V.I. Rud, N. Russakovich, P. Russo, B. Sabirov, G. Sajot, A. Sánchez-Hernández, M.P. Sanders, A. Santoro, B. Satyanarayana, G. Savage, L. Sawyer, T. Scanlon, D. Schaile, R.D. Schamberger, Y. Scheglov, H. Schellman, P. Schieferdecker, C. Schmitt, C. Schwanenberger, A.A. Schukin, A. Schwartzman, R. Schwienhorst, S. Sengupta, H. Severini, E. Shabalina, M. Shamim, H.C. Shankar, V. Shary, A.A. Shchukin, P. Sheahan, W.D. Shephard, R.K. Shivpuri, A.A. Shishkin, D. Shpakov, M. Shupe, R.A. Sidwell, V. Simak, V. Sirotenko, D. Skow, P. Skubic, P. Slattery, D.E. Smith, R.P. Smith, K. Smolek, G.R. Snow, J. Snow, S. Snyder, S. Söldner-Rembold, X. Song, Y. Song, L. Sonnenschein, A. Sopczak, V. Sorín, M. Sosebee, K. Soustruznik, M. Souza, N. Spartana, B. Spurlock, N.R. Stanton, J. Stark, J. Steele, A. Stefanik, J. Steinberg, G. Steinbrück, K. Stevenson, V. Stolin, A. Stone, D.A. Stoyanova, J. Strandberg, M.A. Strang, M. Strauss, R. Ströhmer, D. Strom, M. Strovink, L. Stutte, S. Sumowidagdo, A. Sznajder, M. Talby, S. Tentindo-Repond, P. Tamburello, W. Taylor, P. Telford, J. Temple, N. Terentyev, V.

Teterin, E. Thomas, J. Thompson, B. Thooris, M. Titov, D. Toback, V. V. Tokmenin, C. Tolian, M. Tomoto, D. Tompkins, T. Toole, J. Torborg, F. Touze, S. Towers, T. Trefzger, S. Trincaz-Duvoid, T.G. Trippe, D. Tsybychev, B. Tuchming, C. Tully, A.S. Turcot, P.M. Tuts, M. Utes, L. Uvarov, S. Uvarov, S. Uzunyan, B. Vachon, P.J. van den Berg, P. van Gemmeren, R. Van Kooten, W.M. van Leeuwen, N. Varelas, E.W. Varnes, A. Vartapetian, I.A. Vasilyev, M. Vaupel, M. Vaz, P. Verdier, L.S. Vertogradov, M. Verzocchi, M. Vigneault, F. Villeneuve-Seguier, P.R. Vishwanath, J.R. Vlimant, E. Von Toerne, A. Vorobyov, M. Vreeswijk, T. Vu Anh, V. Vysotsky, H.D. Wahl, R. Walker, N. Wallace, L. Wang, Z.M. Wang, J. Warchol, M. Warsinsky, G. Watts, M. Wayne, M. Weber, H. Weerts, M. Wegner, N. Wermes, M. Wetstein, A. White, V. White, D. Whiteson, D. Wicke, T. Wijnen, D.A. Wijngaarden, N. Wilcer, H. Willutzki, G.W. Wilson, S.J. Wimpenny, J. Wittlin, T. Wlodek, M. Wobisch, J. Womersley, D.R. Wood, T.R. Wyatt, Z. Wu, Y. Xie, Q. Xu, N. Xuan, S. Yacoob, R. Yamada, M. Yan, R. Yarema, T. Yasuda, Y.A. Yatsunenko, Y. Yen, K. Yip, H.D. Yoo, F. Yoffe, S.W. Youn, J. Yu, A. Yurkewicz, A. Zabi, M. Zanabria, A. Zatserklyaniy, M. Zdrazil, C. Zeitnitz, B. Zhang, D. Zhang, X. Zhang, T. Zhao, Z. Zhao, H. Zheng, B. Zhou, B. Zhou, J. Zhu, M. Zielinski, D. Zieminska, A. Zieminski, R. Zitoun, T. Zmuda, V. Zutshi, S. Zviagintsev, E.G. Zverev, A. Zylberstejn, The upgraded DØ detector, Nucl. Instruments Methods Phys. Res. Sect. A Accel. Spectrometers, Detect. Assoc. Equip. 565 (2006) 463–537. https://doi.org/10.1016/j.nima.2006.05.248.

[42] S. Cabrera, J. Fernández, G. Gómez, J. Piedra, T. Rodrigo, A. Ruiz, I. Vila, R. Vilar, C. Grozis, R. Kephart, R. Stanek, D.H. Kim, M.S. Kim, Y. Oh, Y.K. Kim, G. Veramendi, K. Anikeev, G. Bauer, I.K. Furic, A. Korn, I. Kravchenko, M. Mulhearn, C. Paus, S. Pavlon, K. Sumorok, C. Chen, M. Jones, W. Kononenko, J. Kroll, G.M. Mayers, F.M. Newcomer, R.G.C. Oldeman, D. Usynin, R. Van Berg, G. Bellettini, C. Cerri, A. Menzione, F. Spinella, E. Vataga, S. De Cecco, D. De Pedis, C. Dionisi, S. Giagu, A. De Girolamo, M. Rescigno, L. Zanello, M. Ahn, B.J. Kim, S.B. Kim, I. Cho, J. Lee, I. Yu, H. Kaneko, A. Kazama, S. Kim, K. Sato, K. Sato, F. Ukegawa, The CDF-II time-of-flight detector, Nucl. Instruments Methods Phys. Res. Sect. A Accel. Spectrometers, Detect. Assoc. Equip. 494 (2002) 416–423. https://doi.org/10.1016/S0168-9002(02)01512-7.

[43] T.A. Collaboration, G. Aad, E. Abat, J. Abdallah, A.A. Abdelalim, A. Abdesselam, O. Abdinov, B.A. Abi, M. Abolins, H. Abramowicz, E. Acerbi, B.S. Acharya, R. Achenbach, M. Ackers, D.L. Adams, F. Adamyan, T.N. Addy, M. Aderholz, C. Adorisio, P. Adragna, M. Aharrouche, S.P. Ahlen, F. Ahles, A. Ahmad, H. Ahmed, G. Aielli, P.F. Åkesson, T.P.A. Åkesson, A. V Akimov, S.M. Alam, J. Albert, S. Albrand, M. Aleksa, I.N. Aleksandrov, M. Aleppo, F. Alessandria, C. Alexa, G. Alexander, T. Alexopoulos, G. Alimonti, M. Aliyev, P.P. Allport, S.E. Allwood-Spiers, A. Aloisio, J. Alonso, R. Alves, M.G. Alviggi, K. Amako, P. Amaral, S.P. Amaral, G. Ambrosini,

G. Ambrosio, C. Amelung, V. V Ammosov, A. Amorim, N. Amram, C. Anastopoulos, B. Anderson, K.J. Anderson, E.C. Anderssen, A. Andreazza, V. Andrei, L. Andricek, M.-L. Andrieux, X.S. Anduaga, F. Anghinolfi, A. Antonaki, M. Antonelli, S. Antonelli, R. Apsimon, G. Arabidze, I. Aracena, Y. Arai, A.T.H. Arce, J.P. Archambault, J.-F. Arguin, E. Arik, M. Arik, K.E. Arms, S.R. Armstrong, M. Arnaud, C. Arnault, A. Artamonov, S. Asai, S. Ask, B. Åsman, D. Asner, L. Asquith, K. Assamagan, A. Astbury, B. Athar, T. Atkinson, B. Aubert, B. Auerbach, E. Auge, K. Augsten, V.M. Aulchenko, N. Austin, G. Avolio, R. Avramidou, A. Axen, C. Ay, G. Azuelos, G. Baccaglioni, C. Bacci, H. Bachacou, K. Bachas, G. Bachy, E. Badescu, P. Bagnaia, D.C. Bailey, J.T. Baines, O.K. Baker, F. Ballester, F.B.D.S. Pedrosa, E. Banas, D. Banfi, A. Bangert, V. Bansal, S.P. Baranov, S. Baranov, A. Barashkou, E.L. Barberio, D. Barberis, G. Barbier, P. Barclay, D.Y. Bardin, P. Bargassa, T. Barillari, M. Barisonzi, B.M. Barnett, R.M. Barnett, S. Baron, A. Baroncelli, M. Barone, A.J. Barr, F. Barreiro, J.B.G. da Costa, P. Barrillon, A.B. Poy, N. Barros, V. Bartheld, H. Bartko, R. Bartoldus, S. Basiladze, J. Bastos, L.E. Batchelor, R.L. Bates, J.R. Batley, S. Batraneanu, M. Battistin, G. Battistoni, V. Batusov, F. Bauer, B. Bauss, D.E. Baynham, M. Bazalova, A. Bazan, P.H. Beauchemin, B. Beaugiraud, R.B. Beccherle, G.A. Beck, H.P. Beck, K.H. Becks, I. Bedajanek, A.J. Beddall, A. Beddall, P. Bednár, V.A. Bednyakov, C. Bee, S.B. Harpaz, G.A.N. Belanger, C. Belanger-Champagne, B. Belhorma, P.J. Bell, W.H. Bell, G. Bella, F. Bellachia, L. Bellagamba, F. Bellina, G. Bellomo, M. Bellomo, O. Beltramello, A. Belymam, S. Ben Ami, M. Ben Moshe, O. Benary, D. Benchekroun, C. Benchouk, M. Bendel, B.H. Benedict, N. Benekos, J. Benes, Y. Benhammou, G.P. Benincasa, D.P. Benjamin, J.R. Bensinger, K. Benslama, S. Bentvelsen, M. Beretta, D. Berge, E. Bergeaas, N. Berger, F. Berghaus, S. Berglund, F. Bergsma, J. Beringer, J. Bernabéu, K. Bernardet, C. Berriaud, T. Berry, H. Bertelsen, A. Bertin, F. Bertinelli, S. Bertolucci, N. Besson, A. Beteille, S. Bethke, W. Bialas, R.M. Bianchi, M. Bianco, O. Biebel, M. Bieri, M. Biglietti, H. Bilokon, M. Binder, S. Binet, N. Bingefors, A. Bingul, C. Bini, C. Biscarat, R. Bischof, M. Bischofberger, A. Bitadze, J.P. Bizzell, K.M. Black, R.E. Blair, J.J. Blaising, O. Blanch, G. Blanchot, C. Blocker, J. Blocki, A. Blondel, W. Blum, U. Blumenschein, C. Boaretto, G.J. Bobbink, A. Bocci, D. Bocian, R. Bock, M. Boehm, J. Boek, J.A. Bogaerts, A. Bogouch, C. Bohm, J. Bohm, V. Boisvert, T. Bold, V. Boldea, V.G. Bondarenko, R. Bonino, J. Bonis, W. Bonivento, P. Bonneau, M. Boonekamp, G. Boorman, M. Boosten, C.N. Booth, P.S.L. Booth, P. Booth, J.R.A. Booth, K. Borer, A. Borisov, I. Borjanovic, K. Bos, D. Boscherini, F. Bosi, M. Bosman, M. Bosteels, B. Botchev, H. Boterenbrood, D. Botterill, J. Boudreau, E. V Bouhova-Thacker, C. Boulahouache, C. Bourdarios, M. Boutemeur, K. Bouzakis, G.R. Boyd, J. Boyd, B.H. Boyer, I.R. Boyko, N.I. Bozhko, S. Braccini, A. Braem, P. Branchini, G.W. Brandenburg, A. Brandt, O. Brandt, U. Bratzler, H.M. Braun, S. Bravo, I.P. Brawn, B. Brelier, J. Bremer, R. Brenner, S. Bressler, D.

Breton, N.D. Brett, P. Breugnon, P.G. Bright-Thomas, F.M. Brochu, I. Brock, R. Brock, T.J. Brodbeck, E. Brodet, F. Broggi, Z. Broklova, C. Bromberg, G. Brooijmans, G. Brouwer, J. Broz, E. Brubaker, P.A.B. de Renstrom, D. Bruncko, A. Bruni, G. Bruni, M. Bruschi, T. Buanes, N.J. Buchanan, P. Buchholz, I.A. Budagov, V. Büscher, L. Bugge, D. Buira-Clark, E.J. Buis, F. Bujor, T. Buran, H. Burckhart, D. Burckhart-Chromek, S. Burdin, R. Burns, E. Busato, J.J.F. Buskop, K.P. Buszello, F. Butin, J.M. Butler, C.M. Buttar, J. Butterworth, J.M. Butterworth, T. Byatt, S.C. Urbán, E.C. Casas, M. Caccia, D. Caforio, O. Cakir, P. Calafiura, G. Calderini, D.C. Terol, J. Callahan, L.P. Caloba, R. Caloi, D. Calvet, A. Camard, F. Camarena, P. Camarri, M. Cambiaghi, D. Cameron, J. Cammin, F.C. Segura, S. Campana, V. Canale, J. Cantero, M.D.M.C. Garrido, I. Caprini, M. Caprini, M. Caprio, D. Caracinha, C. Caramarcu, Y. Carcagno, R. Cardarelli, C. Cardeira, L.C. Sas, A. Cardini, T. Carli, G. Carlino, L. Carminati, B. Caron, S. Caron, C. Carpentieri, F.S. Carr, A.A. Carter, J.R. Carter, J. Carvalho, D. Casadei, M.P. Casado, M. Cascella, C. Caso, J. Castelo, V.C. Gimenez, N. Castro, F. Castrovillari, G. Cataldi, F. Cataneo, A. Catinaccio, J.R. Catmore, A. Cattai, S. Caughron, D. Cauz, A. Cavallari, P. Cavalleri, D. Cavalli, M. Cavalli-Sforza, V. Cavasinni, F. Ceradini, C. Cerna, C. Cernoch, A.S. Cerqueira, A. Cerri, F. Cerutti, M. Cervetto, S.A. Cetin, F. Cevenini, M. Chalifour, M.C. Llatas, A. Chan, J.W. Chapman, D.G. Charlton, S. Charron, S. V Chekulaev, G.A. Chelkov, H. Chen, L. Chen, T. Chen, X. Chen, S. Cheng, T.L. Cheng, A. Cheplakov, V.F. Chepurnov, R.C. El Moursli, D. Chesneanu, E. Cheu, L. Chevalier, J.L. Chevalley, F. Chevallier, V. Chiarella, G. Chiefari, L. Chikovani, A. Chilingarov, G. Chiodini, S. Chouridou, D. Chren, T. Christiansen, I.A. Christidi, A. Christov, M.L. Chu, J. Chudoba, A.G. Chuguev, G. Ciapetti, E. Cicalini, A.K. Ciftci, V. Cindro, M.D. Ciobotaru, A. Ciocio, M. Cirilli, M. Citterio, M. Ciubancan, J. V Civera, A. Clark, W. Cleland, J.C. Clemens, B.C. Clement, C. Clément, D. Clements, R.W. Clifft, M. Cobal, A. Coccaro, J. Cochran, R. Coco, P. Coe, S. Coelli, E. Cogneras, C.D. Cojocaru, J. Colas, A.P. Colijn, C. Collard, C. Collins-Tooth, J. Collot, R. Coluccia, G. Comune, P.C. Muiño, E. Coniavitis, M. Consonni, S. Constantinescu, C. Conta, F.A. Conventi, J. Cook, M. Cooke, N.J. Cooper-Smith, T. Cornelissen, M. Corradi, S. Correard, A. Corso-Radu, J. Coss, G. Costa, M.J. Costa, D. Costanzo, T. Costin, R.C. Torres, L. Courneyea, C. Couyoumtzelis, G. Cowan, B.E. Cox, J. Cox, D.A. Cragg, K. Cranmer, J. Cranshaw, M. Cristinziani, G. Crosetti, C.C. Almenar, S. Cuneo, A. Cunha, M. Curatolo, C.J. Curtis, P. Cwetanski, Z. Czyczula, S. D’Auria, M. D’Onofrio, A.D.R.G. Mello, P.V.M. Da Silva, R. Da Silva, W. Dabrowski, A. Dael, A. Dahlhoff, T. Dai, C. Dallapiccola, S.J. Dallison, J. Dalmau, C.H. Daly, M. Dam, D. Damazio, M. Dameri, K.M. Danielsen, H.O. Danielsson, R. Dankers, D. Dannheim, G. Darbo, P. Dargent, C. Daum, J.P. Dauvergne, M. David, T. Davidek, N. Davidson, R. Davidson, I. Dawson, J.W. Dawson, R.K. Daya, K. De, R. de Asmundis, R. de Boer, S. De Castro, N. De Groot, P. de Jong, X. de La Broise, E.D. La Cruz-

Burelo, C.D. La Taille, B. De Lotto, M.D.O. Branco, D. De Pedis, P. de Saintignon, A. De Salvo, U. De Sanctis, A. De Santo, J.B.D.V. De Regie, G. De Zorzi, S. Dean, G. Dedes, D. V Dedovich, P.O. Defay, R. Degele, M. Dehchar, M. Deile, C. Del Papa, J. Del Peso, T. Del Prete, E. Delagnes, P. Delebecque, A. Dell'Acqua, M. Della Pietra, D. della Volpe, M. Delmastro, P. Delpierre, N. Delruelle, P.A. Delsart, C.D. Silberberg, S. Demers, M. Demichev, P. Demierre, B. Demirköz, W. Deng, S.P. Denisov, C. Dennis, C.J. Densham, M. Dentan, J.E. Derkaoui, F. Derue, P. Dervan, K.K. Desch, A. Dewhurst, A. Di Ciaccio, L. Di Ciaccio, A. Di Domenico, A. Di Girolamo, B. Di Girolamo, S. Di Luise, A. Di Mattia, A. Di Simone, M.M.D. Gomez, E.B. Diehl, H. Dietl, J. Dietrich, W. Dietsche, S. Diglio, M. Dima, K. Dindar, B. Dinkespiler, C. Dionisi, R. Dipanjan, P. Dita, S. Dita, F. Dittus, S.D. Dixon, F. Djama, R. Djilkibaev, T. Djobava, M.A.B. do Vale, M. Dobbs, R. Dobinson, D. Dobos, E. Dobson, M. Dobson, J. Dodd, O.B. Dogan, T. Doherty, Y. Doi, J. Dolejsi, I. Dolenc, Z. Dolezal, B.A. Dolgoshein, E. Domingo, M. Donega, J. Dopke, D.E. Dorfan, O. Dorholt, A. Doria, A. Dos Anjos, M. Dosil, A. Dotti, M.T. Dova, J.D. Dowell, A.T. Doyle, G. Drake, D. Drakoulakos, Z. Drasal, J. Drees, N. Dressnandt, H. Drevermann, C. Driouichi, M. Dris, J.G. Drohan, J. Dubbert, T. Dubbs, E. Duchovni, G. Duckeck, A. Dudarev, M. Dührssen, H. Dür, I.P. Duerdoth, S. Duffin, L. Duflot, M.-A. Dufour, N.D. Dayot, H.D. Yildiz, D. Durand, A. Dushkin, R. Duxfield, M. Dwuznik, F. Dydak, D. Dzahini, S.D. Cornell, M. Düren, W.L. Ebenstein, S. Eckert, S. Eckweiler, P. Eerola, I. Efthymiopoulos, U. Egede, K. Egorov, W. Ehrenfeld, T. Eifert, G. Eigen, K. Einsweiler, E. Eisenhandler, T. Ekelof, L.M. Eklund, M. El Kacimi, M. Ellert, S. Elles, N. Ellis, J. Elmsheuser, M. Elsing, R. Ely, D. Emeliyanov, R. Engelmann, M. Engström, P. Ennes, B. Epp, A. Eppig, V.S. Epshteyn, A. Ereditato, V. Eremin, D. Eriksson, I. Ermoline, J. Ernwein, D. Errede, S. Errede, M. Escalier, C. Escobar, X.E. Curull, B. Esposito, F. Esteves, F. Etienne, A.I. Etienvre, E. Etzion, H. Evans, V.N. Evdokimov, P. Evtoukhovitch, A. Eyring, L. Fabbri, C.W. Fabjan, C. Fabre, P. Faccioli, K. Facius, V. Fadeyev, R.M. Fakhrutdinov, S. Falciano, I. Falleau, A.C. Falou, Y. Fang, M. Fanti, A. Farbin, A. Farilla, J. Farrell, P. Farthouat, D. Fasching, F. Fassi, P. Fassnacht, D. Fassouliotis, F. Fawzi, L. Fayard, F. Fayette, R. Febbraro, O.L. Fedin, I. Fedorko, L. Feld, G. Feldman, L. Feligioni, C. Feng, E.J. Feng, J. Fent, A.B. Fenyuk, J. Ferencei, D. Ferguson, J. Ferland, W. Fernando, S. Ferrag, A. Ferrari, P. Ferrari, R. Ferrari, A. Ferrer, M.L. Ferrer, D. Ferrere, C. Ferretti, F. Ferro, M. Fiascaris, S. Fichet, F. Fiedler, V. Filimonov, A. Filipčič, A. Filippas, F. Filthaut, M. Fincke-Keeler, G. Finocchiaro, L. Fiorini, A. Firan, P. Fischer, M.J. Fisher, S.M. Fisher, V. Flaminio, J. Flammer, M. Flechl, I. Fleck, W. Flegel, P. Fleischmann, S. Fleischmann, C.M.F. Corral, F. Fleuret, T. Flick, J. Flix, L.R.F. Castillo, M.J. Flowerdew, F. Föhlisch, M. Fokitis, T.M.F. Martin, J. Fopma, D.A. Forbush, A. Formica, J.M. Foster, D. Fournier, A. Foussat, A.J. Fowler, H. Fox, P. Francavilla, D. Francis, S. Franz, J.T. Fraser, M. Fraternali, S. Fratianni, J. Freestone, R.S. French,

K. Fritsch, D. Froidevaux, J.A. Frost, C. Fukunaga, J. Fulachier, E.F. Torregrosa, J. Fuster, C. Gabaldon, S. Gadomski, G. Gagliardi, P. Gagnon, E.J. Gallas, M. V Gallas, B.J. Gallop, K.K. Gan, F.C. Gannaway, Y.S. Gao, V.A. Gapienko, A. Gaponenko, C. Garciá, M. Garcia-Sciveres, J.E.G. Navarro, V. Garde, R.W. Gardner, N. Garelli, H. Garitaonandia, V.G. Garonne, J. Garvey, C. Gatti, G. Gaudio, O. Gaumer, V. Gautard, P. Gauzzi, I.L. Gavrilenko, C. Gay, J.-C. Gayde, E.N. Gazis, E. Gazo, C.N.P. Gee, C. Geich-Gimbel, K. Gellerstedt, C. Gemme, M.H. Genest, S. Gentile, M.A. George, S. George, P. Gerlach, Y. Gernizky, C. Geweniger, H. Ghazlane, V.M. Ghete, P. Ghez, N. Ghodbane, B. Giacobbe, S. Giagu, V. Giakoumopoulou, V. Giangiobbe, F. Gianotti, B. Gibbard, A. Gibson, M.D. Gibson, S.M. Gibson, G.F. Gieraltowski, I.G. Botella, L.M. Gilbert, M. Gilchriese, O. Gildemeister, V. Gilewsky, A.R. Gillman, D.M. Gingrich, J. Ginzburg, N. Giokaris, M.P. Giordani, C.G. Girard, P.F. Giraud, P. Girtler, D. Giugni, P. Giusti, B.K. Gjelsten, C. Glasman, A. Glazov, K.W. Glitza, G.L. Glonti, K.G. Gnanvo, J. Godlewski, T. Göpfert, C. Gössling, T. Göttfert, S. Goldfarb, D. Goldin, N. Goldschmidt, T. Golling, N.P. Gollub, P.J. Golonka, S.N. Golovnia, A. Gomes, J. Gomes, R. Gonçalo, A. Gongadze, A. Gonidec, S. Gonzalez, S.G. de la Hoz, V.G. Millán, M.L.G. Silva, B. Gonzalez-Pineiro, S. González-Sevilla, M.J. Goodrick, J.J. Goodson, L. Goossens, P.A. Gorbounov, A. Gordeev, H. Gordon, I. Gorelov, G. Gorfine, B. Gorini, E. Gorini, A. Gorišek, E. Gornicki, S.A. Gorokhov, B.T. Gorski, S. V Goryachev, V.N. Goryachev, M. Gosselink, M.I. Gostkin, M. Gouanère, I.G. Eschrich, D. Goujdami, M. Goulette, I. Gousakov, J. Gouveia, S. Gowdy, C. Goy, I. Grabowska-Bold, V. Grabski, P. Grafström, C. Grah, K.-J. Grahn, F. Grancagnolo, S. Grancagnolo, H. Grassmann, V. Gratchev, H.M. Gray, E. Graziani, B. Green, A. Greenall, D. Greenfield, D. Greenwood, I.M. Gregor, A. Grewal, E. Griesmayer, N. Grigalashvili, C. Grigson, A.A. Grillo, F. Grimaldi, K. Grimm, P.L.Y. Gris, Y. Grishkevich, H. Groenstege, L.S. Groer, J. Grognuz, M. Groh, E. Gross, J. Grosse-Knetter, M.E.M. Grothe, J. Grudzinski, C. Gruse, M. Gruwe, K. Grybel, P. Grybos, E.M. Gschwendtner, V.J. Guarino, C.J. Guicheney, G. Guilhem, T. Guillemin, J. Gunther, B. Guo, A. Gupta, L. Gurriana, V.N. Gushchin, P. Gutierrez, L. Guy, C. Guyot, C. Gwenlan, C.B. Gwilliam, A. Haas, S. Haas, C. Haber, G. Haboubi, R. Hackenburg, E. Hadash, H.K. Hadavand, C. Haeberli, R. Härtel, R. Haggerty, F. Hahn, S. Haider, Z. Hajduk, M. Hakimi, H. Hakobyan, H. Hakobyan, J. Haller, G.D. Hallewell, B. Hallgren, K. Hamacher, A. Hamilton, H. Han, L. Han, K. Hanagaki, M. Hance, P. Hanke, C.J. Hansen, F.H. Hansen, J.R. Hansen, J.B. Hansen, J.D. Hansen, P.H. Hansen, T. Hansl-Kozanecka, G. Hanson, P. Hansson, K. Hara, S. Harder, A. Harel, T. Harenberg, R. Harper, J.C. Hart, R.G.G. Hart, F. Hartjes, N. Hartman, T. Haruyama, A. Harvey, Y. Hasegawa, K. Hashemi, S. Hassani, M. Hatch, R.W. Hatley, T.G. Haubold, D. Hauff, F. Haug, S. Haug, M. Hauschild, R. Hauser, C. Hauviller, M. Havranek, B.M. Hawes, R.J. Hawkings, D. Hawkins, T. Hayler, H.S. Hayward, S.J. Haywood, E. Hazen, M. He,

Y.P. He, S.J. Head, V. Hedberg, L. Heelan, F.E.W. Heinemann, M. Heldmann, S. Hellman, C. Helsens, R.C.W. Henderson, P.J. Hendriks, A.M.H. Correia, S. Henrot-Versille, F. Henry-Couannier, T. Henß, G. Herten, R. Hertenberger, L. Hervas, M. Hess, N.P. Hessey, A. Hicheur, A. Hidvegi, E. Higón-Rodriguez, D. Hill, J. Hill, J.C. Hill, N. Hill, S.J. Hillier, I. Hinchliffe, D. Hindson, C. Hinkelbein, T.A. Hodges, M.C. Hodgkinson, P. Hodgson, A. Hoecker, M.R. Hoeferkamp, J. Hoffman, A.E. Hoffmann, D. Hoffmann, H.F. Hoffmann, M. Holder, T.I. Hollins, G. Hollyman, A. Holmes, S.O. Holmgren, R. Holt, E. Holtom, T. Holy, R.J. Homer, Y. Homma, P. Homola, W. Honerbach, A. Honma, I. Hooton, T. Horazdovsky, C. Horn, S. Horvat, J.-Y. Hostachy, T. Hott, S. Hou, M.A. Houlden, A. Hoummada, J. Hover, D.F. Howell, J. Hrivnac, I. Hruska, T. Hryn'ova, G.S. Huang, Z. Hubacek, F. Hubaut, F. Huegging, B.T. Huffman, E. Hughes, G. Hughes, R.E. Hughes-Jones, W. Hulsbergen, P. Hurst, M. Hurwitz, T. Huse, N. Huseynov, J. Huston, J. Huth, G. Iacobucci, M. Ibbotson, I. Ibragimov, R. Ichimiya, L. Iconomidou-Fayard, J. Idarraga, M. Idzik, P. Iengo, M.C.I. Escudero, O. Igonkina, Y. Ikegami, M. Ikeno, Y. Ilchenko, Y. Ilyushenka, D. Imbault, P. Imbert, M. Imhaeuser, M. Imori, T. Ince, J. Inigo-Golfin, K. Inoue, P. Ioannou, M. Iodice, G. Ionescu, K. Ishii, M. Ishino, Y. Ishizawa, R. Ishmukhametov, C. Issever, H. Ito, A. V Ivashin, W. Iwanski, H. Iwasaki, J.M. Izen, V. Izzo, J. Jackson, J.N. Jackson, M. Jaekel, S. Jagielski, M. Jahoda, V. Jain, K. Jakobs, J. Jakubek, E. Jansen, P.P.M. Jansweijer, R.C. Jared, G. Jarlskog, S. Jarp, P. Jarron, K. Jelen, I.J.-L. Plante, P. Jenni, A. Jeremie, P. Jez, S. Jézéquel, Y. Jiang, G. Jin, S. Jin, O. Jinnouchi, D. Joffe, L.G. Johansen, M. Johansen, K.E. Johansson, P. Johansson, K.A. Johns, K. Jon-And, M. Jones, R. Jones, R.W.L. Jones, T.W. Jones, T.J. Jones, A. Jones, O. Jonsson, K.K. Joo, D. Joos, M. Joos, C. Joram, S. Jorgensen, J. Joseph, P. Jovanovic, S.S. Junnarkar, V. Juranek, P. Jussel, V. V Kabachenko, S. Kabana, M. Kaci, A. Kaczmarska, M. Kado, H. Kagan, S. Kagawa, S. Kaiser, E. Kajomovitz, S. Kakurin, L. V Kalinovskaya, S. Kama, H. Kambara, N. Kanaya, A. Kandasamy, S. Kandasamy, M. Kaneda, V.A. Kantserov, J. Kanzaki, B. Kaplan, A. Kapliy, J. Kaplon, M. Karagounis, M.K. Unel, K. Karr, P. Karst, V. Kartvelishvili, A.N. Karyukhin, L. Kashif, A. Kasmi, R.D. Kass, A. Kastanas, M. Kataoka, Y. Kataoka, E. Katsoufis, S. Katunin, K. Kawagoe, M. Kawai, T. Kawamoto, F. Kayumov, V.A. Kazanin, M.Y. Kazarinov, A. Kazarov, S.I. Kazi, J.R. Keates, R. Keeler, P.T. Keener, R. Kehoe, M. Keil, G.D. Kekelidze, M. Kelly, J. Kennedy, M. Kenyon, O. Kepka, N. Kerschen, B.P. Kerševan, S. Kersten, C. Ketterer, M. Khakzad, F. Khalilzade, H. Khandanyan, A. Khanov, D. Kharchenko, A. Khodinov, A.G. Kholodenko, A. Khomich, V.P. Khomutnikov, G. Khoriauli, N. Khovanskiy, V. Khovanskiy, E. Khramov, J. Khubua, G. Kieft, J.A. Kierstead, G. Kilvington, H. Kim, H. Kim, S.H. Kim, P. Kind, B.T. King, J. Kirk, G.P. Kirsch, L.E. Kirsch, A.E. Kiryunin, D. Kisielewska, B. Kisielewski, T. Kittelmann, A.M. Kiver, H. Kiyamura, E. Kladiva, J. Klaiber-Lodewigs, K. Kleinknecht, A. Klier, A.

Klimentov, C.R. Kline, R. Klingenberg, E.B. Klinkby, T. Klioutchnikova, P.F. Klok, S. Klous, E.-E. Kluge, P. Kluit, M. Klute, S. Kluth, N.K. Knecht, E. Kneringer, E. Knezo, J. Knobloch, B.R. Ko, T. Kobayashi, M. Kobel, P. Kodys, A.C. König, S. König, L. Köpke, F. Koetsveld, T. Koffas, E. Koffeman, Z. Kohout, T. Kohriki, T. Kokott, G.M. Kolachev, H. Kolanoski, V. Kolesnikov, I. Koletsou, M. Kollefrath, S. Kolos, S.D. Kolya, A.A. Komar, J.R. Komaragiri, T. Kondo, Y. Kondo, N. V Kondratyeva, T. Kono, A.I. Kononov, R. Konoplich, S.P. Konovalov, N. Konstantinidis, A. Kootz, S. Koperny, S. V Kopikov, K. Korcyl, K. Kordas, V. Koreshev, A. Korn, I. Korolkov, V.A. Korotkov, H. Korsmo, O. Kortner, M.E. Kostrikov, V. V Kostyukhin, M.J. Kotamäki, D. Kotchetkov, S. Kotov, V.M. Kotov, K.Y. Kotov, C. Kourkoumelis, A. Koutsman, S. Kovalenko, R. Kowalewski, H. Kowalski, T.Z. Kowalski, W. Kozanecki, A.S. Kozhin, V. Kral, V. Kramarenko, G. Kramberger, A. Kramer, O. Krasel, M.W. Krasny, A. Krasznahorkay, A. Krepouri, P. Krieger, P. Krivkova, G. Krobath, H. Kroha, J. Krstic, U. Kruchonak, H. Krüger, K. Kruger, Z. V Krumshteyn, P. Kubik, W. Kubischta, T. Kubota, L.G. Kudin, J. Kudlaty, A. Kugel, T. Kuhl, D. Kuhn, V. Kukhtin, Y. Kulchitsky, N. Kundu, A. Kupco, M. Kupper, H. Kurashige, L.L. Kurchaninov, Y.A. Kurochkin, V. Kus, W. Kuykendall, P. Kuzhir, E.K. Kuznetsova, O. Kvasnicka, R. Kwee, D. La Marra, M. La Rosa, L. La Rotonda, L. Labarga, J.A. Labbe, C. Lacasta, F. Lacava, H. Lacker, D. Lacour, V.R. Lacuesta, E. Ladygin, R. Lafaye, B. Laforge, T. Lagouri, S. Lai, E. Lamanna, M. Lambacher, F. Lambert, W. Lampl, E. Lancon, U. Landgraf, M.P.J. Landon, H. Landsman, R.R. Langstaff, A.J. Lankford, F. Lanni, K. Lantzsch, A. Lanza, V. V Lapin, S. Laplace, J.F. Laporte, V. Lara, T. Lari, A. V Larionov, C. Lasseur, W. Lau, P. Laurelli, A. Lavorato, W. Lavrijsen, A.B. Lazarev, A.-C. Le Bihan, O. Le Dortz, C. Le Maner, M. Le Vine, L. Leahu, M. Leahu, C. Lebel, M. Lechowski, T. LeCompte, F. Ledroit-Guillon, H. Lee, J.S.H. Lee, S.C. Lee, M. Lefebvre, R.P. Lefevre, M. Legendre, A. Leger, B.C. LeGeyt, C. Leggett, M. Lehmacher, G.L. Miotto, M. Lehto, R. Leitner, D. Lelas, D. Lellouch, M. Leltchouk, V. Lendermann, K.J.C. Leney, T. Lenz, G. Lenzen, J. Lepidis, C. Leroy, J.-R. Lessard, J. Lesser, C.G. Lester, M. Letheren, A.L.F. Cheong, J. Levêque, D. Levin, L.J. Levinson, M.S. Levitski, M. Lewandowska, M. Leyton, J. Li, W. Li, M. Liabline, Z. Liang, Z. Liang, B. Liberti, P. Lichard, W. Liebig, R. Lifshitz, D. Liko, H. Lim, M. Limper, S.C. Lin, A. Lindahl, F. Linde, L. Lindquist, S.W. Lindsay, V. Linhart, A.J. Lintern, A. Liolios, A. Lipniacka, T.M. Liss, A. Lissauer, J. List, A.M. Litke, S. Liu, T. Liu, Y. Liu, M. Livan, A. Lleres, G.L. Llácer, S.L. Lloyd, F. Lobkowicz, P. Loch, W.S. Lockman, T. Loddenkoetter, F.K. Loebinger, A. Loginov, C.W. Loh, T. Lohse, K. Lohwasser, M. Lokajicek, J. Loken, S. Lokwitz, M.C. Long, L. Lopes, D.L. Mateos, M.J. Losty, X. Lou, K.F. Loureiro, L. Lovas, J. Love, A. Lowe, M.L. Fantoba, F. Lu, J. Lu, L. Lu, H.J. Lubatti, S. Lucas, C. Luci, A. Lucotte, A. Ludwig, I. Ludwig, J. Ludwig, F. Luehring, D. Lüke, G. Luijckx, L. Luisa, D. Lumb, L. Luminari, E. Lund, B. Lund-Jensen, B.

Lundberg, J. Lundquist, A. Lupi, N. Lupu, G. Lutz, D. Lynn, J. Lynn, J. Lys, V. Lysan, E. Lytken, J.M. López-Amengual, H. Ma, L.L. Ma, M.M. En, G. Maccarrone, G.G.R. Mace, D. Macina, R. Mackeprang, A. Macpherson, D. MacQueen, C. Macwaters, R.J. Madaras, W.F. Mader, R. Maenner, T. Maeno, P. Mättig, S. Mättig, C.A. Magrath, Y. Mahalalel, K. Mahboubi, G. Mahout, C. Maidantchik, A. Maio, G.M. Mair, K. Mair, Y. Makida, D. Makowiecki, P. Malecki, V.P. Maleev, F. Malek, D. Malon, S. Maltezos, V. Malychev, S. Malyukov, M. Mambelli, R. Mameghani, J. Mamuzic, A. Manabe, A. Manara, G. Manca, L. Mandelli, I. Mandić, M. Mandl, J. Maneira, M. Maneira, P.S. Mangeard, M. Mangin-Brinet, I.D. Manjavidze, W.A. Mann, S. Manolopoulos, A. Manousakis-Katsikakis, B. Mansoulie, A. Manz, A. Mapelli, L. Mapelli, L. March, J.F. Marchand, M. Marchesotti, M. Marcisovsky, A. Marin, C.N. Marques, F. Marroquim, R. Marshall, Z. Marshall, F.K. Martens, S.M. i Garcia, A.J. Martin, B. Martin, B. Martin, F.F. Martin, J.P. Martin, P. Martin, G. Martinez, C.M. Lacambra, V.M. Outschoorn, A. Martini, J. Martins, T. Maruyama, F. Marzano, T. Mashimo, R. Mashinistov, J. Masik, A.L. Maslennikov, M. Maß, I. Massa, G. Massaro, N. Massol, M. Mathes, J. Matheson, P. Matricon, H. Matsumoto, H. Matsunaga, J.M. Maugain, S.J. Maxfield, E.N. May, J.K. Mayer, C. Mayri, R. Mazini, M. Mazzanti, P. Mazzanti, E. Mazzoni, F. Mazzucato, S.P.M. Kee, R.L. McCarthy, C. McCormick, N.A. McCubbin, J. McDonald, K.W. McFarlane, S. McGarvie, H. McGlone, R.A. McLaren, S.J. McMahon, T.R. McMahon, T.J. McMahon, R.A. McPherson, M. Mechtel, D. Meder-Marouelli, M. Medinnis, R. Meera-Lebbai, C. Meessen, R. Mehdiyev, A. Mehta, K. Meier, H. Meinhard, J. Meinhardt, C. Meirosu, F. Meisel, A. Melamed-Katz, B.R.M. Garcia, P.M. Jorge, P. Mendez, S. Menke, C. Menot, E. Meoni, D. Merkl, L. Merola, C. Meroni, F.S. Merritt, I. Messmer, J. Metcalfe, S. Meuser, J.-P. Meyer, T.C. Meyer, W.T. Meyer, V. Mialkovski, M. Michelotto, L. Micu, R. Middleton, P. Miele, A. Migliaccio, L. Mijović, G. Mikenberg, M. Mikestikova, M. Mikestikova, B. Mikulec, M. Mikuž, D.W. Miller, R.J. Miller, W. Miller, M. Milosavljevic, D.A. Milstead, S. Mima, A.A. Minaenko, M. Minano, I.A. Minashvili, A.I. Mincer, B. Mindur, M. Mineev, L.M. Mir, G. Mirabelli, L.M. Verge, S. Misawa, S. Miscetti, A. Misiejuk, A. Mitra, G.Y. Mitrofanov, V.A. Mitsou, P.S. Miyagawa, Y. Miyazaki, J.U. Mjörnmark, S. Mkrtchyan, D. Mladenov, T. Moa, M. Moch, A. Mochizuki, P. Mockett, P. Modesto, S. Moed, K. Mönig, N. Möser, B. Mohn, W. Mohr, S. Mohrdieck-Möck, A.M. Moisseev, R.M.M. Valls, J. Molina-Perez, A. Moll, G. Moloney, R. Mommsen, L. Moneta, E. Monnier, G. Montarou, S. Montesano, F. Monticelli, R.W. Moore, T.B. Moore, G.F. Moorhead, A. Moraes, J. Morel, A. Moreno, D. Moreno, P. Morettini, D. Morgan, M. Morii, J. Morin, A.K. Morley, G. Mornacchi, M.-C. Morone, S. V Morozov, E.J. Morris, J. Morris, M.C. Morrissey, H.G. Moser, M. Mosidze, A. Moszczynski, S. V Mouraviev, T. Mouthuy, T.H. Moye, E.J.W. Moyse, J. Mueller, M. Müller, A. Muijs, T.R. Muller, A. Munar, D.J. Munday, K. Murakami, R.M. Garcia, W.J. Murray, A.G.

Myagkov, M. Myska, K. Nagai, Y. Nagai, K. Nagano, Y. Nagasaka, A.M. Nairz, D. Naito, K. Nakamura, Y. Nakamura, I. Nakano, G. Nanava, A. Napier, M. Nassiakou, I. Nasteva, N.R. Nation, T. Naumann, F. Nauyock, S.K. Nderitu, H.A. Neal, E. Nebot, P. Nechaeva, A. Neganov, A. Negri, S. Negroni, C. Nelson, S. Nemecek, P. Nemethy, A.A. Nepomuceno, M. Nessi, S.Y. Nesterov, L. Neukermans, P. Nevski, F.M. Newcomer, A. Nichols, C. Nicholson, R. Nicholson, R.B. Nickerson, R. Nicolaidou, G. Nicoletti, B. Nicquevert, M. Niculescu, J. Nielsen, T. Niinikoski, M.J. Niinimaki, N. Nikitin, K. Nikolaev, I. Nikolic-Audit, K. Nikolopoulos, H. Nilsen, B.S. Nilsson, P. Nilsson, A. Nisati, R. Nisius, L.J. Nodulman, M. Nomachi, H. Nomoto, J.-M. Noppe, M. Nordberg, O.N. Francisco, P.R. Norton, J. Novakova, M. Nowak, M. Nozaki, R. Nunes, G.N. Hanninger, T. Nunnemann, T. Nyman, P. O'Connor, S.W. O'Neale, D.C. O'Neil, M. O'Neill, V. O'Shea, F.G. Oakham, H. Oberlack, M. Obermaier, P. Oberson, A. Ochi, W. Ockenfels, S. Odaka, I. Odenthal, G.A. Odino, H. Ogren, S.H. Oh, T. Ohshima, H. Ohshita, H. Okawa, M. Olcese, A.G. Olchevski, C. Oliver, J. Oliver, M.O. Gomez, A. Olszewski, J. Olszowska, C. Omachi, A. Onea, A. Onofre, C.J. Oram, G. Ordonez, M.J. Oreglia, F. Orellana, Y. Oren, D. Orestano, I.O. Orlov, R.S. Orr, F. Orsini, L.S. Osborne, B. Osculati, C. Osuna, R. Otec, R. Othegraven, B. Ottewell, F. Ould-Saada, A. Ouraou, Q. Ouyang, O.K. Øye, V.E. Ozcan, K. Ozone, N. Ozturk, A.P. Pages, S. Padhi, C.P. Aranda, E. Paganis, F. Paige, P.M. Pailler, K. Pajchel, S. Palestini, J. Palla, D. Pallin, M.J. Palmer, Y.B. Pan, N. Panikashvili, V.N. Panin, S. Panitkin, D. Pantea, M. Panuskova, V. Paolone, A. Paoloni, I. Papadopoulos, T. Papadopoulou, I. Park, W. Park, M.A. Parker, S. Parker, C. Parkman, F. Parodi, J.A. Parsons, U. Parzefall, E. Pasqualucci, G. Passardi, A. Passeri, M.S. Passmore, F. Pastore, F. Pastore, S. Pataraia, D. Pate, J.R. Pater, S. Patricelli, T. Pauly, E. Pauna, L.S. Peak, S.J.M. Peeters, M. Peez, E. Pei, S. V Peleganchuk, G. Pellegrini, R. Pengo, J. Pequenao, M. Perantoni, A. Perazzo, A. Pereira, E. Perepelkin, V.J.O. Perera, E.P. Codina, V.P. Reale, I. Peric, L. Perini, H. Pernegger, E. Perrin, R. Perrino, P. Perrodo, G. Perrot, P. Perus, V.D. Peshekhonov, E. Petereit, J. Petersen, T.C. Petersen, P.J.F. Petit, C. Petridou, E. Petrolo, F. Petrucci, R. Petti, M. Pezzetti, B. Pfeifer, A. Phan, A.W. Phillips, P.W. Phillips, The ATLAS Experiment at the CERN Large Hadron Collider, J. Instrum. 3 (2008) S08003–S08003. https://doi.org/10.1088/1748-0221/3/08/S08003.

[44] G. Petrucciani, G. Petrucciani, The CMS experiment at the CERN LHC, Search Higgs Boson C. (2013) 15–58. https://doi.org/10.1007/978-88-7642-482-3_2.

[45] A. Schopper, Overview of the LHCb calorimeter system, Nucl. Instruments Methods Phys. Res. Sect. A Accel. Spectrometers, Detect. Assoc. Equip. 623 (2010) 219–221. https://doi.org/10.1016/j.nima.2010.02.201.

[46] S. Niedźwiecki, P. Białas, C. Curceanu, E. Czerwiński, K. Dulski, A. Gajos, B. Głowacz, M. Gorgol, B.C. Hiesmayr, B. Jasińska, Kapłon, D. Kisielewska-Kamińska, G. Korcyl, P. Kowalski,

T. Kozik, N. Krawczyk, W. Krzemień, E. Kubicz, M. Mohammed, M. Pawlik-Niedźwiecka, M. Pałka, L. Raczynski, Z. Rudy, N.G. Sharma, S. Sharma, R.Y. Shopa, M. Silarski, M. Skurzok, A. Wieczorek, W. Wiślicki, B. Zgardzińsk, M. Zieliński, P. Moskal, J-PET: A new technology for the whole-body PET imaging, Acta Phys. Pol. B. 48 (2017) 1567–1576. https://doi.org/10.5506/APhysPolB.48.1567.

[47] A.N. Vasil'ev, Relaxation of hot electronic excitations in scintillators : account for scattering , track effects , complicated electronic structure, in: Proc. 5th Int. Conf. Inorg. Scintill. Their Appl., 1999: pp. 43–52.

[48] P. Martin, S. Guizard, P. Daguzan, G. Petite, P. D'Oliveira, P. Meynadier, M. Perdrix, Subpicosecond study of carrier trapping dynamics in wide-band-gap crystals, Phys. Rev. B - Condens. Matter Mater. Phys. 55 (1997) 5799–5810. https://doi.org/10.1103/PhysRevB.55.5799.

[49] S. Guizard, P. Martin, P. Daguzan, G. Petite, A. Dos Santos, A. Antonnetti, Contrasted behaviour of an electron gas in mgo, al2o3 and sio2, Epl. 29 (1995) 401–406. https://doi.org/10.1209/0295-5075/29/5/009.

[50] V. Nagirnyi, G. Geoffroy, S. Guizard, M. Kirm, A. Kotlov, Relaxation of electronic excitations in wide-gap crystals studied by femtosecond interferometry technique, Phys. Solid State. 50 (2008) 1784–1788. https://doi.org/10.1134/S1063783408090382.

[51] R.M. Turtos, S. Gundacker, A. Polovitsyn, S. Christodoulou, M. Salomoni, E. Auffray, I. Moreels, P. Lecoq, J.Q. Grim, Ultrafast emission from colloidal nanocrystals under pulsed X-ray excitation, J. Instrum. 11 (2016). https://doi.org/10.1088/1748-0221/11/10/P10015.

[52] S.I. Omelkov, V. Nagirnyi, S. Gundacker, D.A. Spassky, E. Auffray, P. Lecoq, M. Kirm, Scintillation yield of hot intraband luminescence, J. Lumin. 198 (2018) 260–271. https://doi.org/10.1016/j.jlumin.2018.02.027.

[53] S.I. Omelkov, V. Nagirnyi, M. Kirm, New Properties and Prospects of Hot Intraband Luminescence for Fast timing, in: 2019: pp. 41–53. https://doi.org/10.1007/978-3-030-21970-3_4.

[54] R.G. Deich, M.S. Abdrakhmanov, Intraband Luminescence in Ionic Crystals, Phys. Status Solidi. 171 (1992) 263–273. https://doi.org/10.1002/pssb.2221710129.

[55] V.N. Makhov, N.M. Khaidukov, Cross-luminescence peculiarities of complex KF-based fluorides, Nucl. Inst. Methods Phys. Res. A. 308 (1991) 205–207. https://doi.org/10.1016/0168-9002(91)90627-3.

[56] P.A. Rodnyi, Core-valence luminescence in scintillators, Radiat. Meas. 38 (2004) 343–352. https://doi.org/10.1016/j.radmeas.2003.11.003.

[57] T. Sekikawa, T. Yamazaki, Y. Nabekawa, S. Watanabe, Femtosecond lattice relaxation induced by inner-shell excitation, Conf. Lasers Electro-Optics Eur. - Tech. Dig. 19 (2001) 25–26.

https://doi.org/10.1109/cleo.2001.947417.
[58] T. Shimizu, T. Sekikawa, T. Kanai, S. Watanabe, M. Itoh, Time-Resolved Auger Decay in CsBr Using High Harmonics, Phys. Rev. Lett. 91 (2003) 6–9. https://doi.org/10.1103/PhysRevLett.91.017401.
[59] H. Nishimura, S. Shionoya, Dynamical Aspects of Self-Trapping of 1s Excitons in RbI and KI, J. Phys. Soc. Japan. 52 (1983) 4277–4282. https://doi.org/10.1143/JPSJ.52.4277.
[60] C. Luschick, A. Luschick, Decay of electronic excitations with defect formation in solids, Nauka, Moscow, 1989.
[61] K.S. Song, Williams, T. Richard, Self-Trapped Excitons, Springer-Verlag Berlin Heidelberg, 1996. https://doi.org/10.1007/978-3-642-85236-7.
[62] H. Nishimura, Luminescence And Self-Trapping Of Excitons In Alkali Halides, in: Defect Process. Induc. by Electron. Excit. Insul., 1989: p. 288. https://doi.org/https://doi.org/10.1142/9789814415736_0003.
[63] S.I. Omelkov, J. Bogdanov, J. Saaring, A. Jurgilaitis, D. Kroon, V.-T. Pham, J. Larsson, S. Burri, C. Bruschini, E. Charbon, M. Kirm, Time-correlated multiphoton counting technique for luminescence decay measurement with low repetition rate sources, J. Lumin. (In Press) (2021).
[64] S.E. Brunner, D.R. Schaart, BGO as a hybrid scintillator / Cherenkov radiator for cost-effective time-of-flight PET, Phys. Med. Biol. 62 (2017) 4421–4439. https://doi.org/10.1088/1361-6560/aa6a49.
[65] M. Kasha, Characterization of electronic transitions in complex molecules, Discuss. Faraday Soc. 9 (1950) 14–19.
[66] T.O. White, Scintillating fibres, Nucl. Inst. Methods Phys. Res. A. 273 (1988) 820–825. https://doi.org/10.1016/0168-9002(88)90102-7.
[67] P. Peng, M.S. Judenhofer, S.R. Cherry, Compton PET: A layered structure PET detector with high performance, Phys. Med. Biol. 64 (2019). https://doi.org/10.1088/1361-6560/ab1ba0.
[68] C. Liu, Z. Li, T.J. Hajagos, D. Kishpaugh, D.Y. Chen, Q. Pei, Transparent Ultra-High-Loading Quantum Dot/Polymer Nanocomposite Monolith for Gamma Scintillation, ACS Nano. 11 (2017) 6422–6430. https://doi.org/10.1021/acsnano.7b02923.
[69] R.M. Turtos, S. Gundacker, S. Omelkov, B. Mahler, A.H. Khan, J. Saaring, Z. Meng, A. Vasil'ev, C. Dujardin, M. Kirm, I. Moreels, E. Auffray, P. Lecoq, On the use of CdSe scintillating nanoplatelets as time taggers for high-energy gamma detection, Npj 2D Mater. Appl. 3 (2019) 1–10. https://doi.org/10.1038/s41699-019-0120-8.
[70] E.A. McKigney, R.E. Del Sesto, L.G. Jacobsohn, P.A. Santi, R.E. Muenchausen, K.C. Ott, T. Mark McCleskey, B.L. Bennett, J.F. Smith, D. Wayne Cooke, Nanocomposite scintillators for radiation detection and nuclear spectroscopy, Nucl. Instruments Methods Phys. Res. Sect. A

Accel. Spectrometers, Detect. Assoc. Equip. 579 (2007) 15–18. https://doi.org/10.1016/j.nima.2007.04.004.

[71] W.F. Hornyak, A fast neutron detector, Rev. Sci. Instrum. 23 (1952) 264–267. https://doi.org/10.1063/1.1746248.

[72] D.N. Kelkar, P. V. Joshi, A rapid method for estimating radium and radon in water, Health Phys. 17 (1969) 253–257. https://doi.org/10.1097/00004032-196908000-00007.

[73] L.E. Grudskaya, V.A. Vlasov, V.G. Poduzhailo, S.A. Malinovskaya, Phoswich for neutron detection, Instrum Exp Tech. (1970) 1006–1007.

[74] I. Petr, J.B. Birks, A. Adams, The composite directional γ-ray scintillation detector, Nucl. Instruments Methods. 99 (1972) 285–293. https://doi.org/10.1016/0029-554X(72)90788-4.

[75] G.F. Knoll, T.F. Knoll, M.H. Timothy, Light collection in scintillation detector composites for neutron detection, IEEE Trans. Nucl. Sci. 35 (1988) 872–875. https://doi.org/10.1109/23.12850.

[76] T. Shimizu, S. Kubota, T. Motobayashi, J. Ruan, F. Shiraishi, Y. Takami, Tokyo, 103, IEEE Trans. Nucl. Sci. 33 (1986) 370–373. https://doi.org/10.1109/TNS.1986.4337121.

[77] V.A. Bumazhnov, V.I. Kochetkov, V.A. Onuchin, V.K. Semenov, A.P. Soldatov, Composite Scintillators for Presice Calorimetery, IHEP 98-14. (1998) 11.

[78] P. Lecoq, Metamaterials for novel X- or gamma-ray detector designs, in: 2008 IEEE Nucl. Sci. Symp. Conf. Rec., IEEE, 2008: pp. 1405–1409. https://doi.org/10.1109/NSSMIC.2008.4774678.

[79] P. Lecoq, Development of new scintillators for medical applications, Nucl. Instruments Methods Phys. Res. Sect. A Accel. Spectrometers, Detect. Assoc. Equip. 809 (2016) 130–139. https://doi.org/10.1016/j.nima.2015.08.041.

[80] G. Konstantinou, P. Lecoq, J.M. Benlloch, A.J. Gonzalez, Metascintillators for ultra-fast gamma detectors: a review of current state and future perspectives, IEEE Trans. Radiat. Plasma Med. Sci. (2021) 1–1. https://doi.org/10.1109/trpms.2021.3069624.

[81] T.J. Hajagos, C. Liu, N.J. Cherepy, Q. Pei, High-Z Sensitized Plastic Scintillators: A Review, Adv. Mater. 30 (2018) 1–13. https://doi.org/10.1002/adma.201706956.

[82] Y.N. Kharzheev, Radiation Hardness of Scintillation Detectors Based on Organic Plastic Scintillators and Optical Fibers, Phys. Part. Nucl. 50 (2019) 42–76. https://doi.org/10.1134/S1063779619010027.

[83] G.H. V. Bertrand, M. Hamel, F. Sguerra, Current Status on Plastic Scintillators Modifications, Chem. - A Eur. J. 20 (2014) 15660–15685. https://doi.org/10.1002/chem.201404093.

[84] C. Zorn, M. Bowen, S. Majewski, J. Walker, R. Wojcik, C. Hurlbut, W. Moser, Pilot study of new radiation-resistant plastic scintillators doped with 3-hydroxyflavone, Nucl. Inst. Methods Phys. Res. A. 273 (1988) 108–116. https://doi.org/10.1016/0168-9002(88)90804-2.

[85] C. Liu, T.J. Hajagos, D. Kishpaugh, Y. Jin, W. Hu, Q. Chen, Q. Pei, Synthesis of transparent

nanocomposite monoliths for gamma scintillation, Hard X-Ray, Gamma-Ray, Neutron Detect. Phys. XVII. 9593 (2015) 959312. https://doi.org/10.1117/12.2189664.

[86] Y. Jin, D. Kishpaugh, C. Liu, T.J. Hajagos, Q. Chen, L. Li, Y. Chen, Q. Pei, Partial ligand exchange as a critical approach to the synthesis of transparent ytterbium fluoride-polymer nanocomposite monoliths for gamma ray scintillation, J. Mater. Chem. C. 4 (2016) 3654–3660. https://doi.org/10.1039/c6tc00447d.

[87] C.A. Harper, Modern Plastics Handbook, 1st editio, McGraw-Hill Professional, New York, 2000.

[88] F.W. Markley, Plastic Scintillators from Cross-Linked Epoxy Resins, Mol. Cryst. 4 (1968) 303–317. https://doi.org/10.1080/15421406808082920.

[89] V.N. Salimgareeva, S. V. Kolesov, Plastic scintillators based on polymethyl methacrylate: A review, Instruments Exp. Tech. 48 (2005) 273–282. https://doi.org/10.1007/s10786-005-0052-8.

[90] J. Harmon, J. Gaynor, V. Feygelman, J. Walker, Linear polydiorganosiloxanes as plastic bases for radiation hard scintillators, Nucl. Inst. Methods Phys. Res. B. 53 (1991) 309–314. https://doi.org/10.1016/0168-583X(91)95619-O.

[91] T. Zolper, Z. Li, M. Jungk, A. Stammer, H. Stoegbauer, T. Marks, Y.W. Chung, Q. Wang, Traction characteristics of siloxanes with aryl and cyclohexyl branches, Tribol. Lett. 49 (2013) 301–311. https://doi.org/10.1007/s11249-012-0066-x.

[92] M. Bowen, S. Majewski, D. Pettey, J. Walker, R. Wojcik, C. Zorn, A new radiation-resistant plastic scintillator, IEEE Trans. Nucl. Sci. 36 (1989) 562–566. https://doi.org/10.1109/23.34501.

[93] Z. Kang, M. Barta, J. Nadler, B. Wagner, R. Rosson, B. Kahn, Synthesis of BaF2:Ce nanophosphor and epoxy encapsulated transparent nanocomposite, J. Lumin. 131 (2011) 2140–2143. https://doi.org/10.1016/j.jlumin.2011.05.030.

[94] H. Wang, T. Ritter, W. Cao, K.K. Shung, High frequency properties of passive materials for ultrasonic transducers, IEEE Trans. Ultrason. Ferroelectr. Freq. Control. 48 (2001) 78–84. https://doi.org/10.1109/58.895911.

[95] M.G. Schorr, F.L. Torney, Solid Non-Crystalline Scintillation Phosphors, Phys. Rev. Journals Arch. 80 (1950) 474. https://doi.org/https://doi.org/10.1103/PhysRev.80.474.

[96] T. Forster, 10Th Spiers Memorial Lecture, Discuss. Faraday Soc. 27 (1959) 7–17. https://doi.org/10.1039/DF9592700007.

[97] V.K. Milinchuk, N.M. Bolbit, E.R. Klinshpont, V.I. Tupikov, G.S. Zhdanov, S.B. Taraban, I.P. Shelukhov, A.S. Smoljanskii, Radiation-induced chemical processes in polystyrene scintillators, Nucl. Instruments Methods Phys. Res. Sect. B Beam Interact. with Mater. Atoms. 151 (1999) 457–461. https://doi.org/10.1016/S0168-583X(99)00096-8.

[98] T.M. Demkiv, O.O. Halyatkin, V. V. Vistovskyy, A. V. Gektin, A.S. Voloshinovskii, X-ray excited luminescence of polystyrene-based scintillator loaded with LaPO 4 -Pr nanoparticles, J.

Appl. Phys. 120 (2016). https://doi.org/10.1063/1.4964334.

[99] T.M. Demkiv, O.O. Halyatkin, V. V. Vistovskyy, A. V. Gektin, A.S. Voloshinovskii, Luminescent and kinetic properties of the polystyrene composites based on BaF2 nanoparticles, Nucl. Instruments Methods Phys. Res. Sect. A Accel. Spectrometers, Detect. Assoc. Equip. 810 (2016) 1–5. https://doi.org/10.1016/j.nima.2015.11.130.

[100] A. Quaranta, S.M. Carturan, T. Marchi, V.L. Kravchuk, F. Gramegna, G. Maggioni, M. Degerlier, Optical and scintillation properties of polydimethyl-diphenylsiloxane based organic scintillators, IEEE Trans. Nucl. Sci. 57 (2010) 891–900. https://doi.org/10.1109/TNS.2010.2042817.

[101] M.S. Skorotetcky, O. V. Borshchev, N.M. Surin, I.B. Meshkov, A.M. Muzafarov, S.A. Ponomarenko, Novel Cross-Linked Luminescent Silicone Composites Based on Reactive Nanostructured Organosilicon Luminophores, Silicon. 7 (2015) 191–200. https://doi.org/10.1007/s12633-014-9256-5.

[102] D. Sun, H.J. Sue, N. Miyatake, Optical properties of ZnO quantum dots in epoxy with controlled dispersion, J. Phys. Chem. C. 112 (2008) 16002–16010. https://doi.org/10.1021/jp805104h.

[103] T. Rogers, C. Han, B. Wagner, J. Nadler, Z. Kang, Synthesis of luminescent nanoparticle embedded polymer nanocomposites for scintillation applications, Mater. Res. Soc. Symp. Proc. 1312 (2011) 355–360. https://doi.org/10.1557/opl.2011.123.

[104] J.J. Bai, G.S. Hu, J.T. Zhang, B.X. Liu, J.J. Cui, X.R. Hou, F. Yu, Z.Z. Li, Preparation and Rheology of Isocyanate Functionalized Graphene Oxide/Thermoplastic Polyurethane Elastomer Nanocomposites, J. Macromol. Sci. Part B Phys. 58 (2019) 425–441. https://doi.org/10.1080/00222348.2019.1565102.

[105] L. Kan, H. Cheng, B. Li, X. Zhang, Q. Wang, H. Wei, N. Ma, Anthracene dimer crosslinked polyurethanes as mechanoluminescent polymeric materials, New J. Chem. 43 (2019) 2658–2664. https://doi.org/10.1039/C8NJ06005C.

[106] M. Li, X. Qiang, W. Xu, H. Zhang, Synthesis, characterization and application of AFC-based waterborne polyurethane, Prog. Org. Coatings. 84 (2015) 35–41. https://doi.org/10.1016/j.porgcoat.2015.02.009.

[107] Z. Zhou, Q. Wang, Z. Zeng, L. Yang, X. Ding, N. Lin, Z. Cheng, Polyurethane-based Eu(iii) luminescent foam as a sensor for recognizing Cu2+ in water, Anal. Methods. 5 (2013) 6045–6050. https://doi.org/10.1039/c3ay41365a.

[108] J.T. Haponiuk, K. Formela, PU Polymers, Their Composites, and Nanocomposites: State of the Art and New Challenges, Elsevier Inc., 2017. https://doi.org/10.1016/B978-0-12-804065-2.00001-2.

[109] V.S. Shevelev, A.V. Ishchenko, S.Y. Sokovnin, V.G. Il'Ves, B.V. Shulgin, Influence of luminescent additives on the optical and luminescent properties of organic polymers, in: AIP

Conf. Proc., 2019. https://doi.org/10.1063/1.5134404.
[110] A.K. Barick, Micro- and Nanomechanics of PU Polymer-Based Composites and Nanocomposites, Elsevier Inc., 2017. https://doi.org/10.1016/B978-0-12-804065-2.00002-4.
[111] C. Tao, X. Han, J. Bao, Q. Chen, Y. Huang, G. Xu, Preparation of waterborne polyurethane with outstanding fluorescence properties and programmable emission intensity, Polym. Int. 66 (2017) 770–778. https://doi.org/10.1002/pi.5310.
[112] V.I. Bezrodnyi, M.C. Stratilat, L.F. Kosyanchuk, A.M. Negriyko, G. V. Klishevich, T.T. Todosiichuk, Spectral and photophysical properties of phenalenone dyes in aliphatic polyurethane matrix, Funct. Mater. 22 (2015) 212–218. https://doi.org/10.15407/fm22.02.212.
[113] Y.C. Chung, J.W. Choi, S.H. Lee, B.C. Chun, Investigation of fluorescent shape memory polyurethanes grafted with various dyes, Bull. Korean Chem. Soc. 32 (2011) 2988–2996. https://doi.org/10.5012/bkcs.2011.32.8.2988.
[114] A. Ishchenko, Molecular engineering of dye-doped polymers for optoelectronics, Polym. Adv. Technol. 13 (2002) 744–752. https://doi.org/10.1002/pat.269.
[115] M. Hilder, P.C. Junk, M.M. Lezhnina, M. Warzala, U.H. Kynast, Rare earth functionalized polymers, J. Alloys Compd. 451 (2008) 530–533. https://doi.org/10.1016/j.jallcom.2007.04.114.
[116] T. Gbur, M. Vlk, V. Čuba, A. Beitlerová, M. Nikl, Preparation and luminescent properties of ZnO:Ga(La)/polymer nanocomposite, Radiat. Meas. 56 (2013) 102–106. https://doi.org/10.1016/j.radmeas.2013.01.053.
[117] T. Romaskevic, M. Sedlevicius, S. Budriene, A. Ramanavicius, N. Ryskevic, S. Miasojedovas, A. Ramanaviciene, Assembly and characterization of polyurethane-gold nanoparticle conjugates, Macromol. Chem. Phys. 212 (2011) 2291–2299. https://doi.org/10.1002/macp.201100390.
[118] S. Chen, J. Zhu, Y. Shen, C. Hu, L. Chen, Synthesis of nanocrystal - Polymer transparent hybrids via polyurethane matrix grafted onto functionalized CdS nanocrystals, Langmuir. 23 (2007) 850–854. https://doi.org/10.1021/la062210g.
[119] S. Gogoi, M. Kumar, B.B. Mandal, N. Karak, High performance luminescent thermosetting waterborne hyperbranched polyurethane/carbon quantum dot nanocomposite with in vitro cytocompatibility, Compos. Sci. Technol. 118 (2015) 39–46. https://doi.org/10.1016/j.compscitech.2015.08.010.
[120] X. Luo, J. Han, Y. Ning, Z. Lin, H. Zhang, B. Yang, Polyurethane-based bulk nanocomposites from 1-thioglycerol-stabilized CdTe quantum dots with enhanced luminescence, J. Mater. Chem. 21 (2011) 6569–6575. https://doi.org/10.1039/c0jm04425c.
[121] J. Tan, R. Zou, J. Zhang, W. Li, L. Zhang, D. Yue, Large-scale synthesis of N-doped carbon quantum dots and their phosphorescence properties in a polyurethane matrix, Nanoscale. 8 (2016) 4742–4747. https://doi.org/10.1039/c5nr08516k.

[122] J. Bai, W. Ren, Y. Wang, X. Li, C. Zhang, Z. Li, Z. Xie, High-performance thermoplastic polyurethane elastomer/carbon dots bulk nanocomposites with strong luminescence, High Perform. Polym. 32 (2020) 857–867. https://doi.org/10.1177/0954008320907123.

[123] J. Ryszkowska, Quantitative description of the microstructure of polyurethane nanocomposites with YAG including Tb3+, Mater. Sci. Eng. B Solid-State Mater. Adv. Technol. 146 (2008) 54–58. https://doi.org/10.1016/j.mseb.2007.07.044.

[124] J. Ryszkowska, E.A. Zawadzak, W. Łojkowski, A. Opalińska, K.J. Kurzydłowski, Structure and properties of polyurethane nanocomposites with zirconium oxide including Eu, Mater. Sci. Eng. C. 27 (2007) 994–997. https://doi.org/10.1016/j.msec.2006.09.046.

[125] V.S. Shevelev, A. V. Ishchenko, V. V. Platonov, S.Y. Sokovnin, V.G. Il'Ves, E. V. Tikhonov, O.I. Karzhenkov, B. V. Shulgin, Radioluminescence properties of nanocomposite scintillators with BaF 2 fillers, in: J. Phys. Conf. Ser., 2018. https://doi.org/10.1088/1742-6596/1115/5/052009.

[126] C. Lü, B. Yang, High refractive index organic-inorganic nanocomposites: Design, synthesis and application, J. Mater. Chem. 19 (2009) 2884–2901. https://doi.org/10.1039/b816254a.

[127] L. Lei, D. Chen, Y. Yu, R. Zhang, H. Ling, J. Xu, F. Huang, Y. Wang, Growth of hexagonal NaGdF4 nanocrystals based on cubic Ln 3+: CaF2 precursors and the multi-color upconversion emissions, J. Alloys Compd. 591 (2014) 370–376. https://doi.org/10.1016/j.jallcom.2013.12.238.

[128] M. Pang, D. Liu, Y. Lei, S. Song, J. Feng, W. Fan, H. Zhang, Rare-earth-doped bifunctional alkaline-earth metal fluoride nanocrystals via a facile microwave-assisted process, Inorg. Chem. 50 (2011) 5327–5329. https://doi.org/10.1021/ic200551g.

[129] V. V. Osipov, Y.A. Kotov, M.G. Ivanov, O.M. Samatov, V. V. Lisenkov, V. V. Platonov, A.M. Murzakaev, A.I. Medvedev, E.I. Azarkevich, Laser synthesis of nanopowders, Laser Phys. 16 (2006) 116–125. https://doi.org/10.1134/S1054660X06010105.

[130] S.Y. Sokovnin, V.G. Il'Ves, Production of nanopowders using pulsed electron beam, Ferroelectrics. 436 (2012) 101–107. https://doi.org/10.1080/10584587.2012.730951.

[131] J.S. Taurozzi, V.A. Hackley, M.R. Wiesner, Ultrasonic dispersion of nanoparticles for environmental, health and safety assessment issues and recommendations, Nanotoxicology. 5 (2011) 711–729. https://doi.org/10.3109/17435390.2010.528846.

[132] J.D. Olson, G.P. Gray, S.A. Carter, Optimizing hybrid photovoltaics through annealing and ligand choice, Sol. Energy Mater. Sol. Cells. 93 (2009) 519–523. https://doi.org/10.1016/j.solmat.2008.11.022.

[133] Y. Dirix, C. Bastiaansen, W. Cased, P. Smith, Oriented pearl-necklace arrays of metallic nanoparticles in polymers: A new route toward polarization-dependent color filters, Adv. Mater. 11 (1999) 223–227. https://doi.org/10.1002/(SICI)1521-4095(199903)11:3<223::AID-

ADMA223>3.0.CO;2-J.

[134] T. Rajh, O.I. Mićić, A.J. Nozik, Synthesis and characterization of surface-modified colloidal CdTe quantum dots, J. Phys. Chem. 97 (1993) 11999–12003. https://doi.org/10.1021/j100148a026.

[135] N. Gaponik, D. V. Talapin, A.L. Rogach, A. Eychmüller, H. Weller, Efficient Phase Transfer of Luminescent Thiol-Capped Nanocrystals: From Water to Nonpolar Organic Solvents, Nano Lett. 2 (2002) 803–806. https://doi.org/10.1021/nl025662w.

[136] E. Tang, G. Cheng, X. Ma, X. Pang, Q. Zhao, Surface modification of zinc oxide nanoparticle by PMAA and its dispersion in aqueous system, Appl. Surf. Sci. 252 (2006) 5227–5232. https://doi.org/10.1016/j.apsusc.2005.08.004.

[137] D. Baram-Pinto, S. Shukla, N. Perkas, A. Gedanken, R. Sarid, Inhibition of herpes simplex virus type 1 infection by silver nanoparticles capped with mercaptoethane sulfonate, Bioconjug. Chem. 20 (2009) 1497–1502. https://doi.org/10.1021/bc900215b.

[138] C.B. Murray, D.J. Norris, M.G. Bawendi, Synthesis and Characterization of Nearly Monodisperse CdE (E = S, Se, Te) Semiconductor Nanocrystallites, J. Am. Chem. Soc. 115 (1993) 8706–8715. https://doi.org/10.1021/ja00072a025.

[139] H. Althues, J. Henle, S. Kaskel, Functional inorganic nanofillers for transparent polymers, Chem. Soc. Rev. 36 (2007) 1454–1465. https://doi.org/10.1039/b608177k.

[140] H. Althues, R. Palkovits, A. Rumplecker, P. Simon, W. Sigle, M. Bredol, U. Kynast, S. Kaskel, Synthesis and characterization of transparent luminescent ZnS:Mn/PMMA nanocomposites, Chem. Mater. 18 (2006) 1068–1072. https://doi.org/10.1021/cm0477422.

[141] M. Drofenik, V. Uskoković, Synthesis of materials within reverse micelles, Surf. Rev. Lett. 12 (2005) 239–277.

[142] R.K. Feller, G.M. Purdy, D. Ortiz-Acosta, S. Stange, A. Li, E.A. McKigney, E.I. Esch, R.E. Muenchausen, R. Gilbertson, M. Bacrania, B.L. Bennett, K.C. Ott, L. Brown, C.S. MacOmber, B.L. Scott, R.E. Del Sesto, Large-scale synthesis of CexLa1-xF3nanocomposite scintillator materials, J. Mater. Chem. 21 (2011) 5716–5722. https://doi.org/10.1039/c0jm04162a.

[143] S. Sahi, W. Chen, K. Jiang, Luminescence enhancement of PPO/PVT scintillators by CeF 3 nanoparticles, J. Lumin. 159 (2015) 105–109. https://doi.org/10.1016/j.jlumin.2014.11.004.

[144] C. Liu, T.J. Hajagos, D. Kishpaugh, Y. Jin, W. Hu, Q. Chen, Q. Pei, Facile Single-Precursor Synthesis and Surface Modification of Hafnium Oxide Nanoparticles for Nanocomposite γ-Ray Scintillators, Adv. Funct. Mater. 25 (2015) 4607–4616. https://doi.org/10.1002/adfm.201501439.

[145] W. Cai, Q. Chen, N. Cherepy, A. Dooraghi, D. Kishpaugh, A. Chatziioannou, S. Payne, W. Xiang, Q. Pei, Synthesis of bulk-size transparent gadolinium oxide-polymer nanocomposites for gamma ray spectroscopy, J. Mater. Chem. C. 1 (2013) 1970–1976. https://doi.org/10.1039/c2tc00245k.

[146] J.B. Birks, Organic Liquid Scintillators, (1964) 269–320. https://doi.org/10.1016/B978-0-08-010472-0.50013-6.

[147] A. Wieczorek, dr Andrzej Kochanowski, Development of novel plastic scintillators based on polyvinyltoluene for the hybrid J-PET/MR tomograph, (2017).

[148] C.H. Mesquita, C.L. Duarte, M.M. Hamada, Radiation physical chemistry effects on organic detectors, Nucl. Instruments Methods Phys. Res. Sect. A Accel. Spectrometers, Detect. Assoc. Equip. 505 (2003) 385–388. https://doi.org/10.1016/S0168-9002(03)01103-3.

[149] W.G. Lawrence, S. Thacker, S. Palamakumbura, K.J. Riley, V. V. Nagarkar, Quantum dot-organic polymer composite materials for radiation detection and imaging, IEEE Trans. Nucl. Sci. 59 (2012) 215–221. https://doi.org/10.1109/TNS.2011.2178861.

[150] Z. Kang, Y. Zhang, H. Menkara, B.K. Wagner, C.J. Summers, W. Lawrence, V. Nagarkar, CdTe quantum dots and polymer nanocomposites for x-ray scintillation and imaging, Appl. Phys. Lett. 98 (2011) 10–13. https://doi.org/10.1063/1.3589366.

[151] A. Prudnikau, A. Chuvilin, M. Artemyev, CdSe-CdS nanoheteroplatelets with efficient photoexcitation of central CdSe region through epitaxially grown CdS wings, J. Am. Chem. Soc. 135 (2013) 14476–14479. https://doi.org/10.1021/ja401737z.

[152] M.D. Tessier, P. Spinicelli, D. Dupont, G. Patriarche, S. Ithurria, B. Dubertret, Efficient exciton concentrators built from colloidal core/crown CdSe/CdS semiconductor nanoplatelets, Nano Lett. 14 (2014) 207–213. https://doi.org/10.1021/nl403746p.

[153] S. Shendre, S. Delikanli, M. Li, D. Dede, Z. Pan, S.T. Ha, Y.H. Fu, P.L. Hernández-Martínez, J. Yu, O. Erdem, A.I. Kuznetsov, C. Dang, T.C. Sum, H.V. Demir, Ultrahigh-efficiency aqueous flat nanocrystals of CdSe/CdS@Cd1−xZnxS colloidal core/crown@alloyed-shell quantum wells, Nanoscale. 11 (2019) 301–310. https://doi.org/10.1039/c8nr07879c.

[154] N. Tomczak, D. Jańczewski, M. Han, G.J. Vancso, Designer polymer-quantum dot architectures, Prog. Polym. Sci. 34 (2009) 393–430. https://doi.org/10.1016/j.progpolymsci.2008.11.004.

[155] S. Lee, Design Principle of Reactive Components for Dimethacrylate-Terminated Quantum Dots: Preserved Photoluminescent Quantum Yield, Excellent Pattern Uniformity, and Suppression of Aggregation in the Matrix, Macromol. Chem. Phys. 221 (2020) 1–8. https://doi.org/10.1002/macp.201900488.